\documentclass[reprint,aps,prb,twocolumn,groupedaddress,nobibnotes,superscriptaddress]{revtex4-1}
\setcitestyle{numbers,square}
\usepackage[english]{babel}
\usepackage{amsmath,amssymb,amsfonts} 
\usepackage{graphicx}
\usepackage{xcolor}
\usepackage{color} 
\usepackage{dcolumn}
\usepackage{chngcntr}
\usepackage{hyperref}
\usepackage{mathtools}
\usepackage{amsmath}
\usepackage{amsfonts}
\usepackage{amsthm}
\usepackage{amssymb}
\usepackage{graphicx}
\usepackage{float}
\usepackage{multirow}
\usepackage{float}
\usepackage{placeins}
\usepackage{caption}
\usepackage{ragged2e}
\usepackage[mathlines]{lineno}
\modulolinenumbers[5]
\usepackage{commath}
\usepackage{verbatim}
\usepackage{upgreek}
\usepackage{bm}
\usepackage{makecell}
\usepackage[makeroom]{cancel}

\newcommand\myundersetETA{\mathrel{\underset{\makebox[0pt]{\mbox{\normalfont\tiny\sffamily $\eta\to0$}}}{=}}}
\newcommand\myundersetDELTAZERO{\mathrel{\underset{\makebox[0pt]{\mbox{\normalfont\tiny\sffamily $\Delta_{\nu}=0$}}}{=}}}

\usepackage{comment}

\begin{document}

\title{Phonon collisional broadening and heat transport beyond the Boltzmann equation}

\author{Enrico Di Lucente}
\affiliation{Theory and Simulation of Materials (THEOS), École Polytechnique Fédérale de Lausanne, Lausanne 1015, Switzerland}
\affiliation{Department of Applied Physics and Applied Mathematics, Columbia University, New York (USA)}
\author{Nicola Marzari}
\affiliation{Theory and Simulation of Materials (THEOS), École Polytechnique Fédérale de Lausanne, Lausanne 1015, Switzerland}
\affiliation{Theory of Condensed Matter, Cavendish Laboratory, University of Cambridge, Cambridge CB3 0US, United Kingdom}
\author{Michele Simoncelli}
\affiliation{Department of Applied Physics and Applied Mathematics, Columbia University, New York (USA)}
\date{\today}

\begin{abstract}
In crystals, macroscopic technological properties such as thermal conductivity originate from the microscopic drift and scattering of phonons, which is commonly described by the Boltzmann Transport Equation (BTE). Despite its widespread use, the most general space-time non-local form of the BTE still lacks a rigorous derivation of its collisional part based on Fermi’s Golden Rule (FGR), and becomes inadequate in several regimes, including when the energy-variation scale set by the phonon dispersion steepness approaches that of collisional broadening. A common hallmark of this issue is the poor numerical convergence of conductivity with respect to the smearing used to evaluate FGR rates. This is often circumvented using adaptive schemes, which however violate detailed balance and allow unphysical negative eigenvalues in the collision operator. Here, we overcome these limitations by rigorously deriving the space–time-dependent BTE from the Kadanoff–Baym Equations (KBE), and introduce a linearized generalized BTE (LGBTE) that goes beyond the FGR framework, incorporating self-consistent (physically-derived, fully anharmonic, and mode-resolved) collisional broadening and energy-nonconserving scattering. More generally, we establish a hierarchy of ansätze on Green’s functions, enabling controlled extensions of the semiclassical BTE, and a roadmap toward quantum KBE accuracy. Finally, using first-principles simulations complemented by analytical arguments, we show that this approach formally addresses two long-standing problems of the FGR-based linearized BTE across crystal dimensionalities: (i) its lack of conductivity convergence, common to heat conductors like Diamond; (ii) its universal failure in all 2D systems, rooted in FGR universally predicting an unphysical overdamping for scattering channels involving flexural vibrations, which we showcase in the heat insulator $\alpha$-GeSe monolayer.
\end{abstract}

\maketitle

%\newpage
%\tableofcontents

\newpage
\section{Introduction}

Heat conduction in solids originates from drift and scattering of non-homogeneous and interacting quantized atomic vibrations (phonons), and is often described using the statistical framework of the phonon Boltzmann transport equation (BTE) introduced by Peierls \cite{peierls1929kinetischen,peierls1996quantum}. In particular, Peierls adopted a semiclassical approach, considering well-defined phonon quasiparticles (i.e., as having a non-overdamped Lorentzian spectral function with FWHM smaller than its central value\cite{landau1987statistical_part2}) that drift in space and time akin to the particles of a classical gas, and interact according to the quantum-mechanical Fermi golden rule (FGR) \cite{fermi1950nuclear}. The FGR states that the scattering rate is proportional to the squared matrix element of the interaction Hamiltonian that couples initial and final states, and contains a Dirac delta distribution that ensures exact energy conservation. Within the BTE, the FGR-based scattering operator is often linearized for computational tractability \cite{omini1997heat,broido_intrinsic_2007,donadio_thermal_2007,tian_phonon_2012,fugallo2013ab,phono3py,tadano_anharmonic_2014,ShengBTE_2014,lindsay2016first,carrete2017almabte} and accounts for the intrinsic scattering mechanisms—such as anharmonic phonon-phonon \cite{paulatto2013anharmonic,tadano_anharmonic_2014,feng_quantum_2016},
electron-phonon \cite{liao2015significant}, and phonon-mass impurity \cite{tamura_isotope_1983} interactions—that limit thermal transport in bulk pristine crystals (without these mechanisms, the crystal's conductivity would diverge \cite{ziman2001electrons}). 
Such a linearized version of the BTE (LBTE) is widely used in conjuction with first-principles simulations \cite{omini1997heat,broido_intrinsic_2007,donadio_thermal_2007,tian_phonon_2012,fugallo2013ab,phono3py,tadano_anharmonic_2014,ShengBTE_2014,lindsay2016first,carrete2017almabte} to predict the conductivity of bulk crystals without any empirical parameter. 
This framework yields robust predictions that agree with experiments in many common materials \cite{mcgaughey2004quantitative,ShengBTE_2014,togo2015distributions,lindsay_perspective_2019,hanus_thermal_2021,qian_phonon-engineered_2021}, but it also shows strong sensitivity to numerical parameters, requiring extensive convergence testing when applied to two-dimensional materials \cite{hanThermalConductivityMonolayer2023a,fugallo2014thermal} and extreme thermal conductors such as diamond \cite{fugallo2013ab}. As a result, a longstanding question in the community is whether the high numerical sensitivity of LBTE solutions in these cases originates from the approximations introduced in the derivation of the LBTE, and, if so, how it can be overcome.

A central hypothesis underlying the derivation of the LBTE is that phonons are well-defined quasiparticles, i.e., their lifetimes are longer than their oscillation periods\cite{landau1959theory,landau1957theory,simoncelli2022wigner}. In three-dimensional crystals, this assumption is violated in strongly anharmonic materials (e.g., Refs.\cite{renExtremePhononAnharmonicity2023,tadano2022first,dangic2024lattice}), whereas in two-dimensional crystals it is universally violated\cite{bonini2012acoustic}. Specifically, Ref.\cite{bonini2012acoustic} showed that a sufficient condition for overdamped phonons, whose lifetimes are shorter than their oscillation periods and therefore violate the quasiparticle picture, is the presence of  quadratic-dispersion modes that intereact with linear-dispersion modes. Under these conditions, the semiclassical LBTE is no longer applicable and a full spectral description is required. This issue is also connected to the use of FGR to describe phonon scattering in the LBTE. The FGR constraint of exact energy conservation at the level of each individual scattering event is sufficient\cite{spohn2006phonon}, but not strictly necessary, for global energy conservation over macroscopic timescales. In particular, over many scattering events the system's energy can be conserved on average, even if individual processes slighly violate exact energy conservation.\\
In addition to these formal problems, the FGR in the LBTE suffers from computational issues: to numerically evaluate the Dirac delta distribution appearing in the FGR, a finite-variance Gaussian approximation is employed  \cite{phono3py,ShengBTE_2014}; the variance of the Gaussian is treated as a numerical parameter that is expected not to influence the physical observables, like thermal conductivity, that results from the simulation when numerical convergence is reached.
In this context, systems such as diamond, molecular crystals and some crystalline polymers, and transition metal dichalcogenides do not exhibit clear convergence in thermal conductivity with respect to Gaussian smearing, rendering this approach unreliable. Several numerical methods have been proposed in the literature to address this limitation, including adaptive smearing \cite{ShengBTE_2014}. However, as we will see, such an approach does not respect the analytical property of the LBTE to exactly conserve energy, and also does not remove the necessity to perform a numerical convergence test.
These problems highlight the need to go beyond the semiclassical approximations used in deriving the LBTE, and to include physical mechanisms such as collisional broadening\cite{rossi2011theory}, which determine the shape and width of the energy-exchange distribution in scattering processes.
The role of collisional broadening in, e.g., charge transport, particularly in the context of hot-electron dynamics in semiconductors, has already been extensively studied \cite{reggiani1987quantum,haug2008quantum,levinson1970translational}. These studies have shown that a proper treatment of collisional broadening and intracollisional field effects is important for accurately describing carrier dynamics at high electric fields. However, a detailed treatment of collisional broadening in the context of phonon scattering is still lacking.
Finally, and most importantly, the semiclassical approach does not allow for a rigorous derivation of the LBTE scattering operator with full spatial and temporal dependence \cite{spohn2006phonon}—an essential feature when solving the LBTE beyond the steady-state limit or when accounting for spatial inhomogeneities in scattering. To date, such effects have typically been treated using empirical approximations \cite{vasko2006quantum}.\\

Here, we demonstrate that it is possible to go beyond the semiclassical and quasiparticle regime of phonon transport by introducing a physical collisional broadening that overcomes the formal and numerical limitations of the FGR. Our approach extends the applicability of quantum-kinetic heat-transport theory to regimes where spectral resolution is essential, addressing the failure of the phonon LBTE in all two dimensional materials, and more generally solving the long standing problem of accurately predicting the thermal conductivity of extreme three-dimensional conductors.

The article is organized as follows. We start from
the Dyson equation for the Green's function, from which we derive the quantum kinetic equation for the phonon Green's function, the 
Kadanoff–Baym equation (KBE) \cite{kadanoff2018quantum}. This equation describes a wide range of interactions and excitations,  it is highly dimensional and non-Markovian. Because solving the KBE exactly is computationally prohibitive, we formally coarse-grain it into a generalizes BTE (GBTE) that includes spectral broadening. We then rewrite the KBE in the Wigner's mixed representation, showing that this allows to simplify the formalism significantly, using the functional-method formalisms to express the self-energy in terms of the Green's function. Then we perform a mapping between the KBE and the GBTE, which we rigorously demonstrate via the generalized Kadanoff-Baym (GKB) ansatz \cite{haug2008quantum,reggiani1987quantum,vspivcka1995quasiparticle,lipavsky1986generalized,stefanucci2023and,stefanucci2024semiconductor}. Finally, we apply the gradient approximation, which is reminiscent of the generalized Moyal product formula \cite{moyal1949quantum,simoncelli2021thermal}, to both the driving and collision terms. This approximation simplifies the commutators in the driving term, resulting in gradient corrections that lead to the time and space derivatives of the GBTE. Importantly, we note that the collision term in the GBTE is generally nonlinear, providing a natural way to describe system in the far-from-equilibrium regime \cite{yamada1968nonlinear}. With the GKB ansatz, we establish a procedure that allows us to transition from the KBE to a solvable version of the GBTE, linearized GBTE (LGBTE), identifying the approximations made and discussing how energy is conserved over macroscopic timescales even if individual scattering  processes can slightly violate exact energy conservation. We quantitatively show with first-principles simulations that this framework solves the failures of the LBTE in two-dimensional materials and extreme three-dimensional conductors, with examples in monolayer $\alpha$-GeSe, and bulk diamond.

\section{Non-equilibrium phonon Green's function theory}

In this Section, we briefly review the fundamentals of the non-equilibrium Green's function (NEGF) formalism for phonons. We disucss the properties of the KBE for the phonon Green's function, and how it allows to predict macroscopic thermal properties. 
As we will see, this approach is alternative to the one based on the one-body density matrix \cite{simoncelli2021thermal} and is particularly convenient to understand and control the approximations performed in the description of quantum events between phonons.

\subsection{Phonon hamiltonian}

The Hamiltonian of a crystal lattice is given by \cite{ashcroft2022solid,ziman2001electrons}:
\begin{equation}
H=\sum_{\bm{R}b}\frac{\boldsymbol{p}_{\bm{R}b}^{2}}{2m_{b}}+V(\boldsymbol{r}_{\bm{R}_{1}b_{1}},\boldsymbol{r}_{\bm{R}_{2}b_{2}},\boldsymbol{r}_{\bm{R}_{3}b_{3}},...),
\end{equation}
where $\boldsymbol{r}_{\bm{R}b}=\boldsymbol{r}_{\bm{R}}+\boldsymbol{r}_{b}$ identifies the position of every atom ($\bm{R}$) including its basis ($b$) in the unit cell of the crystal, $V$ is the interatomic potential and $m_{b}$ is the mass of atom $b$.  Within the harmonic approximation, the potential energy is approximated quadratically around equilibrium positions as \cite{ashcroft2022solid,ziman2001electrons}
\begin{equation}
V_{\text{harm}}=\frac{1}{2}\sum_{\substack{\bm{R}b\alpha\\\bm{R}'b'\beta}}\upphi_{\bm{R}b\alpha,\bm{R}'b'\beta}{u}_{\bm{R}b\alpha}{u}_{\bm{R}'b'\beta},
\end{equation}
where we introduced the dynamical matrix $\upphi$ in real space,
\begin{equation}
\upphi_{\bm{R}b\alpha,\bm{R}'b'\beta}=\left.\frac{\partial V}{\partial {u}_{\bm{R}b\alpha}\partial {u}_{\bm{R}'b'\beta}}\right|_{\text{eq}},
\end{equation}
whose matrix elements are the so-called interatomic force constants of the second order. The displacement $\boldsymbol{u}(jb)$ quantifies the small oscillations around atomic equilibrium positions. By seeking solutions in the form of waves we move to a reciprocal space representation in terms of the momentum ($\boldsymbol{P}_{\bm{q}b}=\frac{1}{\sqrt{m_b}}\sum_{\bm{R}}\boldsymbol{p}_{\bm{R}b}e^{-i\bm{q}\cdot(r_{\bm{R}}+\bm{B}_b)}$) and position ($\boldsymbol{X}_{\bm{q}b}=\frac{1}{\sqrt{m_b}}\sum_{\bm{R}}\boldsymbol{u}_{\bm{R}b}e^{-i\bm{q}\cdot(r_{\bm{R}}+\bm{B}_b)}$) operators:
\begin{equation}
\begin{split}
H_{\text{harm}}=\,&\frac{1}{\mathcal{V}} \sum_{\bm{q}b} \frac{P_{\bm{q}b} P_{\bm{q}b}^{\dagger}}{2m_b}\\
&+\frac{1}{2\mathcal{V}} 
\sum_{\substack{\bm{q} \\ b\alpha \\ b'\beta}} 
\upphi_{b\alpha, b'\beta}(\bm{q}) \, 
X_{\bm{q} b\alpha}X_{\bm{q}b'\beta}^{\dagger}
\end{split}
\end{equation}
where $\mathcal{V}$ is the volume of the crystal and $\upphi_{b\alpha,b'\beta;\boldsymbol{q}}=\sum_{\bm{R}'}\upphi_{\bm{0}b\alpha,\bm{R}'b'\beta}e^{-i\boldsymbol{q}\cdot r_{\bm{R}'}}$ is the Fourier transform of the dynamical matrix in real space.
This matrix acts on the phonon (polarizations) eigenvectors $\varepsilon_{\alpha}(\boldsymbol{q}s,b)$ as 
\begin{equation}
\sum_{b'\beta}\upphi_{b\alpha,b'\beta;\boldsymbol{q}}\varepsilon_{b'\beta;\boldsymbol{q}s}=\omega_{\boldsymbol{q}s}^{2}\varepsilon_{b\alpha;\boldsymbol{q}s}
\end{equation}
and, once diagonalized at each $\boldsymbol{q}$-point, gives eigenvalues that correspond to phonon frequencies, $\omega_{\boldsymbol{q}s}$. Expressing the bosonic creation and annihilation operators in terms of $\boldsymbol{P}$ and $\boldsymbol{X}$ we get 
\begin{equation} \label{a_a+_function_of_X_and_P}
\begin{split}
&a_{\boldsymbol{q}s}=\sqrt{\frac{m_{b}\omega_{\boldsymbol{q}s}}{2\hbar}}\sum_{b}\left[\boldsymbol{\varepsilon}^{*}_{b;\boldsymbol{q}s}\cdot\boldsymbol{X}_{\boldsymbol{q}b}+i\frac{1}{m_{b}\omega_{\boldsymbol{q}s}}\boldsymbol{\varepsilon}_{b;\boldsymbol{q}s}\cdot\boldsymbol{P}_{\boldsymbol{q}b}\right],\\
&a_{\boldsymbol{q}s}^\dagger=\sqrt{\frac{m_{b}\omega_{\boldsymbol{q}s}}{2\hbar}}\sum_{b}\left[\boldsymbol{\varepsilon}^{*}_{b;\boldsymbol{q}s}\cdot\boldsymbol{X}_{\boldsymbol{q}b}-i\frac{1}{m_{b}\omega_{\boldsymbol{q}s}}\boldsymbol{\varepsilon}_{b;\boldsymbol{q}s}\cdot\boldsymbol{P}_{\boldsymbol{q}b}\right].
\end{split}
\end{equation}
This transformation is a linear, unitary, and canonical change of variables that diagonalizes the lattice Hamiltonian by mapping real-space operators onto normal mode operators. Inverting Eq. \eqref{a_a+_function_of_X_and_P} we have
\begin{equation} \label{X_as_function_of_a_a+}
\begin{split}
&\boldsymbol{X}_{\boldsymbol{q}b}=\sum_{s}\sqrt{\frac{\hbar}{2m_{b}\omega_{\boldsymbol{q}s}}}\boldsymbol{\varepsilon}_{b;\boldsymbol{q}s}(a_{\boldsymbol{q}s}+a^{\dagger}_{-\boldsymbol{q}s}),\\
&\boldsymbol{P}_{\boldsymbol{q}b}=-i\sum_{s}\sqrt{\frac{\hbar m_{b}\omega_{\boldsymbol{q}s}}{2}}\boldsymbol{\varepsilon}^{*}_{b;\boldsymbol{q}s}(a_{\boldsymbol{q}s}^{\dagger}-a_{\boldsymbol{q}s}),
\end{split}
\end{equation}
which allows to write the harmonic Hamiltonian in the form of an ensemble of harmonic oscillators in the second quantization
\begin{equation} \label{harmonic_hamiltonian}
H_{\text{harm}}=\sum_{\nu}\hbar\omega_{\nu}\left(\frac{1}{2}+a^{\dagger}_{\nu}a_{\nu}\right),
\end{equation}
where each operator $a_{\nu}$ and $a^{\dagger}_{\nu}$ respectively annihilates or creates one quantum of vibrational energy—i.e., a phonon. Throughout this work, we adopt the notation $\pm\nu=(\pm\boldsymbol{q},s)$. The anharmonic contribution is introduced by including a third-order term in the potential energy expansion, yielding
\begin{equation}
\begin{split}
H&=H_{\text{harm}}+\Delta H_{\text{3rd}}=\\
&=H_{\text{harm}}+\frac{1}{3!}\sum_{\substack{\bm{R}\bm{R}'\bm{R}''\\bb'b''\\ \alpha\beta\eta}}\uppsi_{\bm{R}b\alpha,\bm{R}'b'\beta,\bm{R}''b''\eta}\,u_{\bm{R}b\alpha}u_{\bm{R}'b'\beta}u_{\bm{R}''b''\eta},
\end{split}
\end{equation}
where the interatomic force constants of the third-order in real space are the matrix elements of
\begin{equation} \label{matrix_element_real_space}
\uppsi_{\bm{R}b\alpha,\bm{R}'b'\beta,\bm{R}''b''\eta}=\left.\frac{\partial V}{\partial u_{\bm{R}b\alpha}\partial u_{\bm{R}'b'\beta}\partial u_{\bm{R}''b''\eta}}\right|_{\text{eq}}.
\end{equation}
In reciprocal space we have 
\begin{equation} \label{phonon_hamiltonina_with_anharmonic_term}
\begin{split}
\Delta H_{\text{3rd}}=\frac{1}{3!}\sum_{\nu\nu'\nu''}&\uppsi_{\nu\nu'\nu''}\delta_{\boldsymbol{q}+\boldsymbol{q}'+\boldsymbol{q}'',\boldsymbol{K}}\cdot\\
\cdot&(a_{\nu}+a^{\dagger}_{-\nu})(a_{\nu'}+a^{\dagger}_{-\nu'})(a_{\nu''}+a^{\dagger}_{-\nu''})=\\
=\frac{1}{3!}\sum_{\nu\nu'\nu''}&\uppsi_{\nu\nu'\nu''}\delta_{\boldsymbol{q}+\boldsymbol{q}'+\boldsymbol{q}'',\boldsymbol{K}}A_{\nu}A_{\nu'}A_{\nu''},
\end{split}
\end{equation}
where $\boldsymbol{K}$ is a reciprocal lattice vector,
\begin{equation} \label{3rd_order_matrix_element}
\uppsi_{\nu\nu'\nu''}=\sqrt{\frac{\hbar^{3}}{\mathcal{V}}}\sum_{\substack{bb'b''\\ \alpha\beta\eta}}\frac{\varepsilon_{b\alpha;\nu}\varepsilon_{b'\beta;\nu'}\varepsilon_{b''\eta;\nu''}\uppsi_{b\alpha;\boldsymbol{q},b'\beta;\boldsymbol{q}',b''\eta;\boldsymbol{q}''}}{\sqrt{8m_{b}m_{b'}m_{b''}\omega_{\nu}\omega_{\nu'}\omega_{\nu''}}}
\end{equation}
and 
\begin{equation} \label{3rd_order_matrix_element_fourier_transform}
\uppsi_{b\alpha;\boldsymbol{q},b'\beta;\boldsymbol{q}',b''\eta;\boldsymbol{q}''}=\sum_{\bm{R},\bm{R}',\bm{R}''}\uppsi_{\bm{R}b\alpha,\bm{R}'b'\beta,\bm{R}''b''\eta}e^{i\boldsymbol{q}\cdot\bm{R}}e^{i\boldsymbol{q}'\cdot\bm{R}'}e^{i\boldsymbol{q}''\cdot\bm{R}''}
\end{equation}
It is instructive to look at the triplets of bosonic operators resulting from Eq. \eqref{phonon_hamiltonina_with_anharmonic_term}:
\begin{equation}
\begin{split}
&\,(a_{\nu}+a^{\dagger}_{-\nu})(a_{\nu'}+a^{\dagger}_{-\nu'})(a_{\nu''}+a^{\dagger}_{-\nu''})=\\
=&\,\,a_{\nu}a_{\nu'}a_{\nu''}+a_{\nu}a_{\nu'}a^{\dagger}_{-\nu''}+a_{\nu}a^{\dagger}_{-\nu'}a_{\nu''}+\\
&+a_{\nu}a^{\dagger}_{-\nu'}a^{\dagger}_{-\nu''}+a^{\dagger}_{-\nu}a_{\nu'}a_{\nu''}+a^{\dagger}_{-\nu}a_{\nu'}a^{\dagger}_{-\nu''}+\\
&+a^{\dagger}_{-\nu}a^{\dagger}_{-\nu'}a_{\nu''}+a^{\dagger}_{-\nu}a^{\dagger}_{-\nu'}a^{\dagger}_{-\nu''}.
\end{split}
\end{equation}
Each term in this expansion corresponds to a specific combination of phonon creation and annihilation operators. This expansion plays a crucial role in understanding how anharmonicity leads to scattering processes, such as phonon decays and coalescences, and also three-phonon annihilation/creation events, which we will discuss in more detail later in the text. To simplify this representation, we introduce the phonon operators in the vibronic formalism \cite{ryndyk2016theory}:
\begin{equation}
\begin{split}
&A_{\nu}=a_{\nu}+a^{\dagger}_{-\nu}\\
&A^{\dagger}_{\nu}=A_{-\nu}=a^{\dagger}_{\nu}+a_{-\nu}.
\end{split}
\end{equation}
These operators provide a convenient way to rewrite the anharmonic interactions in terms of collective phonon excitations as they naturally represent the mixed-mode interactions that occur in anharmonic systems, thus offering a clearer physical interpretation when constructing the phonon Green's function. In fact, as seen in Eq. \eqref{phonon_hamiltonina_with_anharmonic_term}, these are directly related to physical observables: they enter the expression of the anharmonic part of the phonon Hamiltonian, therefore they allow for a clear physical interpretation when used to construct the phonon Green's function.

\subsection{Bosonic Green's function} \label{bosonic_g_Section}

In contrast to the vibronic representation, given by the operators $A$ and $A^{\dagger}$ (related to the position operator, see Eq. \eqref{X_as_function_of_a_a+} above) the Green's function $g$ (which we indicate in lowercase to distinguish it from that of the vibronic representation) is constructed directly through the bosonic creation and annihilation operators. We start by defining a time-ordered (we set $\hbar=1$) single-phonon Green's function in the bosonic notation:
\begin{equation} \label{green's_function_boson}
\begin{split}
g_{\nu_{1}\nu_{2}}(t_{1},t_{2})&=-i{\rm{Tr}}\left(\hat{\rho}\hat{T}\Big(a_{\nu_{1}}(t_{1})a^{\dagger}_{\nu_{2}}(t_{2})\Big)\right)=\\
&=-i\langle\hat{T}a_{\nu_{1}}(t_{1})a^{\dagger}_{\nu_{2}}(t_{2})\rangle,
\end{split}
\end{equation}
where $\hat{T}$ is the time-ordering operator that always moves the operator with the earlier time argument to the right
\begin{equation}
\hat{T}\left[a(t_{1})b(t_{2})\right]=\theta(t_{1}-t_{2})a(t_{1})b(t_{2})+\theta(t_{2}-t_{1})b(t_{2})a(t_{1}).
\end{equation}
For future use we define the “lesser” and “greater”
Green’s functions as
\begin{equation} \label{G<_corr_function_boson}
g^{<}_{\nu_{1}\nu_{2}}(t_{1},t_{2})=-i{\rm{Tr}}\Big(\hat{\rho}a^{\dagger}_{\nu_{2}}(t_{2})a_{\nu_{1}}(t_{1})\Big)=-i\langle a^{\dagger}_{\nu_{2}}(t_{2})a_{\nu_{1}}(t_{1})\rangle,
\end{equation}
\begin{equation} \label{G>_corr_function_boson}
g^{>}_{\nu_{1}\nu_{2}}(t_{1},t_{2})=-i{\rm{Tr}}\Big(\hat{\rho}a_{\nu_{1}}(t_{1})a^{\dagger}_{\nu_{2}}(t_{2})\Big)=-i\langle a_{\nu_{1}}(t_{1})a^{\dagger}_{\nu_{2}}(t_{2})\rangle.
\end{equation}
Note that the vibronic representation is more commonly used in the literature \cite{volz2020quantum,klein1969derivation,wehner1966infra,wehner1967phonon,wehner1972scattering,niklasson1968theory,xu2008nonequilibrium} as it highlights the link with the physical observables (the operators $A$ enters the definition of the anharmonic part of the phonon Hamiltonian). The Green's function in the bosonic representation instead is a mathematical construct which helps to spot differences or analogies with i.g. its electronic counterpart. This is particularly convenient when dealing with heat flow's response functions \cite{caldarelli2022many}.

\subsection{Vibronic Green's function} \label{vibronic_g_Section}

The single-phonon Green's function in the vibronic notation (capital letters) reads
\begin{equation} \label{green's_function}
\begin{split}
G_{\nu_{1}\nu_{2}}(t_{1},t_{2})&=-i{\rm{Tr}}\left(\hat{\rho}\hat{T}\Big(A_{\nu_{1}}(t_{1})A^{\dagger}_{\nu_{2}}(t_{2})\Big)\right)=\\
&=-i\langle\hat{T}A_{\nu_{1}}(t_{1})A^{\dagger}_{\nu_{2}}(t_{2})\rangle.
\end{split}
\end{equation}
Similarly to Eqs. \eqref{G<_corr_function_boson} and \eqref{G>_corr_function_boson} the “lesser” and “greater” Green's functions can be written as
\begin{equation} \label{G<_corr_function}
G^{<}_{\nu_{1}\nu_{2}}(t_{1},t_{2})=-i{\rm{Tr}}\Big(\hat{\rho}A^{\dagger}_{\nu_{2}}(t_{2})A_{\nu_{1}}(t_{1})\Big)=-i\langle A^{\dagger}_{\nu_{2}}(t_{2})A_{\nu_{1}}(t_{1})\rangle,
\end{equation}
\begin{equation} \label{G>_corr_function}
G^{>}_{\nu_{1}\nu_{2}}(t_{1},t_{2})=-i{\rm{Tr}}\Big(\hat{\rho}A_{\nu_{1}}(t_{1})A^{\dagger}_{\nu_{2}}(t_{2})\Big)=-i\langle A_{\nu_{1}}(t_{1})A^{\dagger}_{\nu_{2}}(t_{2})\rangle.
\end{equation}
In this way Eq. \eqref{green's_function} can be also written as
\begin{equation} \label{green's_function_bis}
G_{\nu_{1}\nu_{2}}(t_{1},t_{2})=\theta(t_{2}-t_{1})G^{<}_{\nu_{1}\nu_{2}}(t_{1},t_{2})+\theta(t_{1}-t_{2})G^{>}_{\nu_{1}\nu_{2}}(t_{1},t_{2}).
\end{equation}
To have a convenient algebraic expression for dealing with unlimited time integrals, the set of Green's functions can be extended using the standard retarded and advanced functions,
\begin{equation} \label{GR_commutation}
\begin{split}
G^{R}_{\nu_{1}\nu_{2}}(t_{1},t_{2})&=-i\theta(t_{1}-t_{2})\langle[A_{\nu_{1}}(t_{1})\,,\,A^{\dagger}_{\nu_{2}}(t_{2})]\rangle=\\
&=\theta(t_{1}-t_{2})\left[G^{>}_{\nu_{1}\nu_{2}}(t_{1},t_{2})-G^{<}_{\nu_{1}\nu_{2}}(t_{1},t_{2})\right],
\end{split}
\end{equation}
\begin{equation} \label{GA_commutation}
\begin{split}
G^{A}_{\nu_{1}\nu_{2}}(t_{1},t_{2})&=i\theta(t_{2}-t_{1})\langle[A_{\nu_{1}}(t_{1})\,,\,A^{\dagger}_{\nu_{2}}(t_{2})]\rangle=\\
&=-\theta(t_{2}-t_{1})\left[G^{>}_{\nu_{1}\nu_{2}}(t_{1},t_{2})-G^{<}_{\nu_{1}\nu_{2}}(t_{1},t_{2})\right],
\end{split}
\end{equation}
where $[A,B]=AB-BA$ stems from phonons being bosons and obeying commutation relations \cite{mahan2000many}.
The retarded Green's function $G^{R}$ differs from zero only for times $t_{1}\geq t_{2}$, thus this function can be used to calculate the response at time $t_{1}$ to an earlier perturbation of the system at time $t_{2}$. 
The advanced Green's function $G^{A}$ is only finite for $t_{1}\leq t_{2}$. Importantly, in Appendix \ref{relation_g_bosonic_and_G_vibronic} we formally derive the following relation between vibronic and bosonic Green's functions:
\begin{equation} \label{relation_G_g_main}
\begin{split}
&G^{<}_{\nu_{1}\nu_{2}}(t_{1},t_{2})=g^{<}_{\nu_{1}\nu_{2}}(t_{1},t_{2})+g^{>}_{\nu_{1}\nu_{2}}(-t_{1},-t_{2}),\\
&G^{>}_{\nu_{1}\nu_{2}}(t_{1},t_{2})=g^{>}_{\nu_{1}\nu_{2}}(t_{1},t_{2})+g^{<}_{\nu_{1}\nu_{2}}(-t_{1},-t_{2}).
\end{split}
\end{equation}
The theoretical framework based on the phonon Green's function introduced in this Section can also be entirely reformulated within the one-body density matrix formalism, starting from the Liouville equation and following, for instance, the approaches presented in Refs. \cite{bonitz2016quantum} (Eqs. 7.4 and 7.81, in the context of electron transport) and \cite{stefanucci2024semiconductor} (Eq. 32 for electrons and Eq. 43 for phonons).

\section{Kadanoff-Baym equation}

Here we introduce the equations of motion of the NEGF in real time, which will serve as the foundation for all future discussions. There are two different but equivalent approaches to obtain such equations: the Kadanoff–Baym method \cite{kadanoff2018quantum} and the Keldysh method \cite{keldysh1964diagram}; here we will follow the first one.\\
The derivation begins with the differential form of the Dyson equation
\begin{equation} \label{dyson_eq}
\begin{split}
G_{\nu_{1}\nu_{2}}(t_{1},t_{2})=&\,G^{0}_{\nu_{1}\nu_{2}}(t_{1},t_{2})+\\
&+\int d\boldsymbol{q}'\int d\boldsymbol{q}''\int dt'\int dt''\\
&\,\,\,\,\,\,\,\,G^{0}_{\nu_{1}\nu'}(t_{1},t')\Sigma_{\nu'\nu''}(t',t'')G_{\nu''\nu_{2}}(t'',t_{2}),
\end{split}
\end{equation}
where $G^{0}$ is the non-interacting or unperturbed vibronic Green's function (see the SM \cite{supplementary}) and the interactions are contained in the self-energy $\Sigma$ which in turn can be expressed as a functional of the Green's function \cite{martin1959theory}. In the following, we focus on intraband phonon propagation ($s_{1}=s_{2}$), neglecting interband wave-like tunneling \cite{simoncelli2019unified,simoncelli2022wigner,di2023crossover}. Extending the theory developed in this work to include interband phonon effects is straightforward and involves keeping both band indices in the KBE (see later in the text). When considering the Dyson equation, one must handle products involving three or more terms. In this context, it can be shown that Langreth's theorem (regarding analytic continuation, see i.g. the discussion on page 71 of Ref. \cite{haug2008quantum} for more details) holds for products of the $C=AB$ type \cite{langreth1976linear,haug2008quantum}:
\begin{equation} \label{Langreth_theorem_1}
\begin{split}
C^{\lessgtr}_{\nu_{1}\nu_{2}}(t_{1},t_{2})=\int dt'\Big[&A^{R}_{\nu_{1}\nu'}(t_{1},t')B^{\lessgtr}_{\nu'\nu_{2}}(t',t_{2})+\\
&+A^{\lessgtr}_{\nu_{1}\nu'}(t_{1},t')B^{A}_{\nu'\nu_{2}}(t',t_{2})\Big].
\end{split}
\end{equation}
In what follows, it is more practical to adopt a notation where the multiplication of two terms is interpreted as an operator multiplication with respect to internal variables. 
With this the result \eqref{Langreth_theorem_1} can be easily generalized for a (matrix) product of three functions \cite{haug2008quantum}. If $D=ABC$, we have 
\begin{equation} \label{Langreth_theorem_2}
D^{\lessgtr}=A^{R}B^{R}C^{\lessgtr}+A^{R}B^{\lessgtr}C^{A}+A^{\lessgtr}B^{A}C^{A}.
\end{equation}
Finally, using such condensed notation, Eq. \eqref{dyson_eq} reads
\begin{equation} \label{dyson_eq_compact}
G=G^{0}+G^{0}\Sigma G,
\end{equation}
or equivalently
\begin{equation} \label{double_dyson_eq_for_KB}
\begin{split}
G^{0^{-1}}G=1+\Sigma G,\\
GG^{0^{-1}}=1+G\Sigma.
\end{split}
\end{equation}
By applying Langreth's theorem \eqref{Langreth_theorem_1} to Eqs. \eqref{double_dyson_eq_for_KB} we get
\begin{equation}
\begin{split}
G^{0^{-1}}G^{<}=\Sigma^{R}G^{<}+\Sigma^{<}G^{A},\\
G^{<}G^{0^{-1}}=G^{R}\Sigma^{<}+G^{<}\Sigma^{A}.
\end{split}
\end{equation}
Subtracting these two equations from each other we obtain
\begin{equation} \label{differential_KB_equation}
\Big[G^{0^{-1}}\,,\,G^{<}\Big]=\Sigma^{R}G^{<}+\Sigma^{<}G^{A}-G^{R}\Sigma^{<}-G^{<}\Sigma^{A},
\end{equation}
where $[A,B]=AB-BA$ is a commutator. This result is precisely the differential form of the KBE \cite{kadanoff2018quantum}. The structure of this equation resembles a lot that of the non-equilibrium BTE if the Green's function $G^{<}$ is mapped into a generalized quasiparticle distribution function (see later in the text). The first commutator on the left-hand side gives rise to a (generalized) driving term, while the terms on the right-hand side lead to a quantum collision term. In fact, the latter are integrals that can be interpreted as scattering-in (positive sign, repumping) and scattering-out (negative sign, depumping) terms. \\
The following NEGF relations also hold \cite{kadanoff2018quantum,ryndyk2016theory}
\begin{equation} \label{spectral_function_def}
\mathrm{A}=-2\text{Im}\big\{G^{R}\big\}=i\big(G^{R}-G^{A}\big)=i\big(G^{>}-G^{<}\big),
\end{equation}
\begin{equation} \label{gamma_def_NEGF}
\Gamma=-2\text{Im}\big\{\Sigma^{R}\big\}=i\big(\Sigma^{R}-\Sigma^{A}\big)=i\big(\Sigma^{>}-\Sigma^{<}\big),
\end{equation}
\begin{equation} \label{G_from_GR_and_GA}
G=\frac{1}{2}\big(G^{R}+G^{A}\big),
\end{equation}
\begin{equation} \label{Resigma_as_sum_sigma_R_A}
\text{Re}\{\Sigma_{\nu}(\omega)\}=\frac{1}{2}\big(\Sigma_{\nu}^{R}(\omega)+\Sigma_{\nu}^{A}(\omega)\big),
\end{equation}
where $\mathrm{A}$ is the spectral function and $\Gamma$ is the phonon linewidth (reciprocal of twice the phonon lifetime). Importantly, if the self-energy is weakly frequency-dependent near the phonon pole, and the linewidth is sufficiently small compared to the phonon energy, then the spectral function is Lorentzian. This is easily proved first by explicitly writing the retarded Green's function in terms of the self-energy written as a sum of real and imaginary parts and then using it in Eq. \eqref{spectral_function_def}. Using the following identities to symmetrize retarded and advanced functions $f^{R}=\frac{1}{2}(f^{R}+f^{A})+\frac{1}{2}(f^{R}-f^{A})$ and $f^{A}=\frac{1}{2}(f^{A}+f^{R})+\frac{1}{2}(f^{A}-f^{R})$, Eq. \eqref{differential_KB_equation} becomes
\begin{equation} \label{KB_equation_ongoing}
\begin{split}
\Big[G^{0^{-1}}\,,\,G^{<}\Big]=&\,\frac{1}{2}\left(\Sigma^{R}+\Sigma^{A}\right)G^{<}+\frac{1}{2}\left(\Sigma^{R}-\Sigma^{A}\right)G^{<}+\\
&+\frac{1}{2}\Sigma^{<}\left(G^{A}+G^{R}\right)+\frac{1}{2}\Sigma^{<}\left(G^{A}-G^{R}\right)-\\
&-\frac{1}{2}\left(G^{R}+G^{A}\right)-\frac{1}{2}\left(G^{R}-G^{A}\right)\Sigma^{<}-\\
&-\frac{1}{2}G^{<}\left(\Sigma^{A}+\Sigma^{R}\right)-\frac{1}{2}G^{<}\left(\Sigma^{A}-\Sigma^{R}\right).
\end{split}
\end{equation}
Factorizing and exploiting Eqs. \eqref{G_from_GR_and_GA} and \eqref{Resigma_as_sum_sigma_R_A} we get
\begin{equation}
\begin{split}
&\Big[G^{0^{-1}}\,,\,G^{<}\Big]=\\
=&\Big[\text{Re}\{\Sigma\},G^{<}\Big]+\Big[\Sigma^{<},G\Big]+\\
&+\frac{1}{2}\Big\{\left(\Sigma^{R}-\Sigma^{A}\right),G^{<}\Big\}-\frac{1}{2}\Big\{\left(G^{R}-G^{A}\right),\Sigma^{<}\Big\},
\end{split}
\end{equation}
where $\lbrace A,B \rbrace=AB+BA$ are anticommutators. Using Eqs. \eqref{spectral_function_def} and \eqref{gamma_def_NEGF} we finally have
\begin{equation} \label{KB_equation}
\begin{split}
&\big[G^{0^{-1}}-\text{Re}\{\Sigma\},G^{<}\big]-\big[\Sigma^{<},G\big]=\\
=&\frac{1}{2i}\big\lbrace\Gamma,G^{<}\big\rbrace-\frac{1}{2i}\big\lbrace \mathrm{A},\Sigma^{<}\big\rbrace=\\
=&\frac{1}{2}\big\lbrace\Sigma^{>},G^{<}\big\rbrace-\frac{1}{2}\big\lbrace G^{>},\Sigma^{<}\big\rbrace.
\end{split}
\end{equation}
In the KBE, we can distinguish the scattering-out given by linewidths $\Gamma$, the scattering-in given by $\lbrace \mathrm{A},\Sigma^{<}\rbrace$ and the self-energy renormalization of the single-particle energy described by $G^{0^{-1}}$, $-\text{Re}\{\Sigma\}$ (contributing to frequency lineshifts \cite{maradudin1962scattering}). This interpretation is sufficient only for highlighting the presence of a connection with the BTE, but we can already spot the generalization: in the standard BTE the scattering rates and the energy renormalization (when taken into account) are related to an energy conserving picture of each phonon collision, while the KBE has full spectral resolution, $\mathrm{A}$, through which such energy conservation could also be relaxed. Moreover, the KBE \eqref{KB_equation} makes evident the presence of the term $[\Sigma^{<},G]$, which does not have any natural interpretation within the semiclassical regime. This term is associated with the out-of-pole (off-shell in quantum field theory) or off-resonant propagation \cite{vspivcka1994quasiparticle,lipavsky2001kinetic,di2025theory}, which accounts for transport mediated by the satellite structures of the spectral function \cite{lipavsky2001kinetic}. In the regime of low scattering rates (the interactions are assumed to be weak as also prescribed by the Landau's theory \cite{landau1987statistical_part2}), it can be shown that this term does not affect the driving term of the quantum kinetic equation \cite{vspivcka1995quasiparticle}. Consequently, for a spectral function with a Lorentzian profile, the scattering term is also unaffected by the off-resonant propagation. As such, this term naturally lies beyond the standard BTE framework, which is constructed entirely on the energy pole. A rigorous and comprehensive treatment of how the off-resonant term should be incorporated, along with its implications for phonon transport, will be presented in detail in a separate publication \cite{di2025theory}.\\
For future use, it is useful to rewrite the KBE neglecting the off-resonant contribution \cite{di2025theory}, obtaining \cite{reggiani1987quantum}:
\begin{equation} \label{differential_KB_equation_final}
\left[G^{0^{-1}},G^{<}\right]=\Sigma^{>}G^{<}+G^{<}\Sigma^{>}-\Sigma^{<}G^{>}-G^{>}\Sigma^{<},
\end{equation}
which in compact notation can be read as
\begin{equation} \label{KBE_skeleton}
{\rm{DT}}=\left[G^{0^{-1}},G^{<}\right]=\left\lbrace\Sigma^{>},G^{<}\right\rbrace-\left\lbrace\Sigma^{<},G^{>}\right\rbrace={\rm{CT}},
\end{equation}
where we label “DT” the driving term and “CT” the collision term. As seen in Eqs. \eqref{differential_KB_equation_final} and \eqref{KBE_skeleton}, we neglect the effects of frequency lineshifts in this work, and will discuss them further later in the text.

\section{Extended linerized Boltzmann transport equation}

Having established the foundational structure of the transport equation (the KBE), the next step is to derive a mapping between the Green's function entering the KBE and the quasiparticle (phonon) distribution. This mapping, known as the Kadanoff-Baym ansatz, is first proven to exist and to be exact at thermal equilibrium. We then extend its validity to a general out-of-equilibrium phonon distribution, provided it is described via the so-called Wigner distribution function \cite{wigner1932quantum,moyal1949quantum}. Under this condition, we can derive the BTE from the KBE and further generalize it to incorporate collisional broadening and individual non-energy-conserving scattering events. To accomplish this objective, it is essential to go beyond the semiclassical BTE limit by preserving spectral resolution (at least Lorentzian) inherent in the KBE.

\subsection{Gradient approximation}

To derive a BTE-like equation that surpasses its semiclassical form, a robust approach is to formulate the theory using Wigner's mixed representation \cite{reggiani1987quantum,vspivcka1995quasiparticle}. In this context, we demonstrate how to obtain the extended BTE from the KBE through a gradient approximation, which is valid under the condition that the variation of the center-of-mass coordinates is slow relative to that of the relative coordinates. The BTE itself is indeed expected to be applicable in regimes of slow spatial and temporal variations. Therefore, it is crucial to separate the “fast” quantum fluctuations from the “slow” macroscopic variations. This separation is accomplished by introducing the Wigner coordinates and then systematically performing a gradient approximation, retaining the lowest-order terms. The Wigner's mixed representation associated with the phonon Green's function $G_{\nu_{1}\nu_{2}}(t_{1},t_{2})$
is defined by \cite{vspivcka1995quasiparticle}: \\
\begin{equation} \label{Wigner's_mixed_representation}
\begin{split}
&\tilde{G}_{\nu}^{\lessgtr}(\omega;\boldsymbol{R},t)=\\
=&\iint d\tau d\boldsymbol{q}'\,e^{i\omega\tau+i\boldsymbol{q}'\cdot\boldsymbol{R}}G_{s}^{\lessgtr}\left(\boldsymbol{q}+\frac{\boldsymbol{q}'}{2},t+\frac{\tau}{2};\boldsymbol{q}-\frac{\boldsymbol{q}'}{2},t-\frac{\tau}{2}\right),
\end{split}
\end{equation}
with the integration domain in reciprocal space corresponding to the Brillouin zone (BZ), where, in the near-equilibrium regime, BZ boundary effects are negligible \cite{simoncelli2022wigner}. Eq. \eqref{Wigner's_mixed_representation} relies on the center-of-mass and relative variables:
\begin{equation} \label{center-of-mass_difference_variables}
\begin{split}
&\boldsymbol{q}'=\boldsymbol{q}_1-\boldsymbol{q}_2,\hspace{1cm}\tau=t_{1}-t_{2},\\
&\boldsymbol{q}=\frac{1}{2}(\boldsymbol{q}_1+\boldsymbol{q}_2),\hspace{1cm}t=\frac{1}{2}(t_{1}+t_{2}).
\end{split}
\end{equation}
Hereafter, we omit the tilde to simplify the notation, as the center-of-mass variables dependence makes the Wigner's mixed representation evident. Importantly, the variables $\boldsymbol{q}'$ and $\tau$ vary on a fast, microscopic scale, and must be treated exactly (after Fourier transforms, $\boldsymbol{q}'\leftrightarrow\boldsymbol{R}$, $\tau\leftrightarrow\omega$), while $\boldsymbol{q}$ and $t$ (after Fourier transforms, $\boldsymbol{q}\leftrightarrow\boldsymbol{r}$, $t\leftrightarrow\Omega$) are macroscopic, slow variables with small gradients, and hence are treated approximately. \\
Clearly, inverting the center-of-mass and relative
variables we get
\begin{equation} \label{times_expressions}
\begin{split}
&\boldsymbol{q}_1=\boldsymbol{q}+\frac{\boldsymbol{q}'}{2},\hspace{1cm}t_{1}=t+\frac{\tau}{2},\\
&\boldsymbol{q}_2=\boldsymbol{q}-\frac{\boldsymbol{q}'}{2},\hspace{1cm}t_{2}=t-\frac{\tau}{2},
\end{split}
\end{equation}
and definition \eqref{green's_function} gives
\begin{equation} \label{Wigner's_mixed_representation_bis}
G^{<}_{\nu_{1}\nu_{2}}(t_{1},t_{2})=G_{s}^{<}\left(\boldsymbol{q}+\frac{\boldsymbol{q}'}{2},t+\frac{\tau}{2};\boldsymbol{q}-\frac{\boldsymbol{q}'}{2},t-\frac{\tau}{2}\right).
\end{equation}
In the Wigner mixed representation \eqref{Wigner's_mixed_representation} the role of the nonequilibrium distribution of phonon populations is played by the Wigner distribution function (see Appendix \ref{wigner_distr_section} to see how its definition naturally emerges from the Wigner mixed representation of the Green’s functions):
\begin{equation} \label{wigner_distr_function_main}
\begin{split}
\tilde{n}_{s}(\boldsymbol{q},\boldsymbol{R},t)=\frac{1}{2\pi}\int d\boldsymbol{q}'\,e^{i\boldsymbol{q}'\cdot\boldsymbol{R}}\left\langle a_{s}^{\dagger}\left(\boldsymbol{q}-\frac{\boldsymbol{q}'}{2},t\right) a_{s}\left(\boldsymbol{q}+\frac{\boldsymbol{q}'}{2},t\right)\right\rangle,
\end{split}
\end{equation}
Moreover, within this formalism, the KBE \eqref{KBE_skeleton} results in a complicated form of matrix products \cite{haug2008quantum,lipavsky2001kinetic,vspivcka1995quasiparticle}, $C=AB$ (see Eq. \eqref{KBE_to_use_for_derivation} below):
\begin{equation} 
C_{\nu_{1}\nu_{2}}(t_{1},t_{2})=\int d\boldsymbol{q}'dt'\,A_{\nu_{1}\nu'}(t_{1},t')B_{\nu'\nu_{2}}(t',t_{2}).
\end{equation}
Expressed in terms of the new variables $(\boldsymbol{q},\omega,\boldsymbol{R},t)$, this becomes \cite{haug2008quantum}
\begin{equation} 
C_{s}(\boldsymbol{q},\omega,\boldsymbol{R},t)=A_{s}(\boldsymbol{q},\omega,\boldsymbol{R},t)\mathsf{G}(\boldsymbol{q},\omega,\boldsymbol{R},t)B_{s}(\boldsymbol{q},\omega,\boldsymbol{R},t),
\end{equation}
where the gradient operator $\mathsf{G}$ is defined as (see the SM \cite{supplementary} for a detailed derivation)
\begin{equation} \label{gradient_operator}
\mathsf{G}=\text{exp}\left[\frac{1}{2i}\left(\frac{\partial^{A}}{\partial t}\frac{\partial^{B}}{\partial\omega}-\frac{\partial^{A}}{\partial\omega}\frac{\partial^{B}}{\partial t}-\frac{\partial^{A}}{\partial\boldsymbol{R}}\cdot\frac{\partial^{B}}{\partial\boldsymbol{q}}+\frac{\partial^{A}}{\partial\boldsymbol{q}}\cdot\frac{\partial^{B}}{\partial\boldsymbol{R}}\right)\right],
\end{equation}
where the superscripts ($A$, $B$) indicate which term in the product is to be differentiated. The lowest nonvanishing order constitutes the gradient approximation and in this way the commutators and anticommutators of the KBE \eqref{KB_equation} assume the following form \cite{haug2008quantum,vspivcka1995quasiparticle}:
\begin{equation} \label{gradient_expansion_1} 
\left\lbrace A_{s},B_{s}\right\rbrace_{(\boldsymbol{q},\omega,\boldsymbol{R},t)}=2A_{s}(\boldsymbol{q},\omega,\boldsymbol{R},t)B_{s}(\boldsymbol{q},\omega,\boldsymbol{R},t),
\end{equation}
\begin{equation} \label{gradient_expansion_2}
\begin{split}
&\left[A_{s},B_{s}\right]_{(\boldsymbol{q},\omega,\boldsymbol{R},t)}=\\
=&-i\left(\frac{\partial A}{\partial t}\frac{\partial B}{\partial\omega}-\frac{\partial A}{\partial\omega}\frac{\partial B}{\partial t}-\frac{\partial A}{\partial\boldsymbol{R}}\cdot\frac{\partial B}{\partial\boldsymbol{q}}+\frac{\partial A}{\partial\boldsymbol{q}}\cdot\frac{\partial B}{\partial\boldsymbol{R}}\right),
\end{split}
\end{equation}
where, for the sake of brevity, dependencies in the commutator are left implicit. A direct generalization to higher-order derivatives in Eq. \eqref{gradient_operator} illustrates the strength of the KBE formalism in extending the transport theory beyond linear response approaches. 
It is essential to clarify that the KBE is fundamentally notation-independent, meaning it applies universally to any type of Green's function $G$—whether bosonic, vibronic, electronic, etc.—and is valid for any set of phase-space variables. This includes representations in $(\boldsymbol{r},t)$, $(\boldsymbol{q},t)$, or even the mixed representation of $(\boldsymbol{q},\omega,\boldsymbol{R},t)$ as described by Wigner’s mixed representation. Here, we rewrite the KBE \eqref{KBE_skeleton} in Wigner’s mixed representation:
\begin{equation} \label{KBE_to_use_for_derivation}
\begin{split}
&\,\text{DT}_{\nu}(\omega,\boldsymbol{R},t)=\left[G^{0^{-1}}_{\nu}(\omega)\,,\,G^{<}_{\nu}(\omega,\boldsymbol{R},t)\right]=\\
=&\,2\Big[\Sigma_{\nu}^{>}(\omega,\boldsymbol{R},t)G_{\nu}^{<}(\omega,\boldsymbol{R},t)-\Sigma_{\nu}^{<}(\omega,\boldsymbol{R},t)G_{\nu}^{>}(\omega,\boldsymbol{R},t)\Big]=\\
=&\,\text{CT}_{\nu}(\omega,\boldsymbol{R},t),
\end{split}
\end{equation}
where we already exploited the gradient approximation \eqref{gradient_expansion_1} in the collision term.

\subsection{Generalized Kadanoff-Baym ansatz}

The final step we need to perform in transitioning from the KBE for the Green's function, $G$, to a BTE-like transport equation for the (phonon) Wigner distribution function, $n$, involves establishing a mapping between these two quantities. This is known in the literature as the GKB ansatz, which was first introduced in the context of electron transport \cite{haug2008quantum,reggiani1987quantum,vspivcka1995quasiparticle,lipavsky1986generalized} through the relations $g^{<}=iAf$ and $g^{>}=-iA[1-f]$, where $g$ and $f$ represent the non-equilibrium electron Green's function and distribution function, respectively. This was part of the early development of NEGF theory to describe intracollisional field effects under strong electric fields \cite{haug2008quantum,jauho1982rigorous,ciancio2004gauge}. The phonon analogue of such ansatz has also been explored \cite{niklasson1968theory,kwok1966unified,meier1969green,horie1964boltzmann,stefanucci2023and},
primarily using bosonic notation. Consequently, one can show that it is necessary to adapt the self-energy definition (commonly written in vibronic notation for phonons) to this framework. Indeed, it is no coincidence that the BTE derived using an ansatz involving bosonic quantities yields an incomplete collision term (see, for instance, Eqs. 3.3$'$, 3.12, and 3.13 of Ref. \cite{kwok1966unified}, where the resulting collision term includes only half of the contributions present in the standard BTE). This is because the bosonic self-energy used was based on a partial diagrammatic expansion that omitted terms arising from different combinations of phonon operators, each with its corresponding multiplicity. In this study, we consider the Green's function, spectral function, and self-energy in vibronic notation for two main reasons. First, the vibronic operators naturally appear in the perturbative term of the phonon Hamiltonian. Second, this approach ensures consistency with the literature, where the functional form of the self-energy is typically expressed in vibronic notation. The GKB ansatz in bosonic notation keeps the same structure of the solution of the unperturbed phonon (bosonic) Green's function (see SI \cite{supplementary}) and reads
\begin{equation} \label{GKB_ansatz_equation_g_piccola}
\begin{split}
g^{<}_{\nu}(\omega,\boldsymbol{R},t)&=-2\pi i\,n_{\nu}(\omega,\boldsymbol{R},t)\mathrm{d}_{\nu}(\omega)\\
g^{>}_{\nu}(\omega,\boldsymbol{R},t)&=-2\pi i\big[n_{\nu}(\omega,\boldsymbol{R},t)+1\big]\mathrm{d}_{\nu}(\omega).
\end{split}
\end{equation}
where the spectral function is $\mathrm{a}_{\nu}(\omega)=2\pi\mathrm{d}_{\nu}(\omega)$ with
\begin{equation} \label{spectral_function_d_bosonic_main}
\begin{split}
\mathrm{d}_{\nu}(\omega)=\frac{1}{\pi}\frac{\gamma_{\nu}(\omega)}{(\omega-\omega_{\nu}-\text{\scriptsize{$\Delta$}}_{\nu}(\omega))^{2}+\gamma_{\nu}^{2}(\omega)},
\end{split}
\end{equation}
where frequency lineshifts $\text{\scriptsize{$\Delta$}}_{\nu}(\omega)=\text{Re}\{\sigma_{\nu}(\omega)\}$ will not be the focus of this work and we decide to neglect them for the moment. This ansatz is entirely general, as it only replaces one unknown function $g^{\lessgtr}$ by another function $n$, which indeed plays the role of the phonon distribution function, and it satisfies the exact relation $\mathrm{a}=i(g^{>}-g^{<})$. Using the property \eqref{relation_G_g_in_wigner_space} we can finally write the GKB ansatz in the vibronic representation
\begin{equation} \label{GKB_ansatz_equation}
\resizebox{0.5\textwidth}{!}{$
\begin{split}
G^{<}_{\nu}(\omega,\boldsymbol{R},t)&=-2\pi i\Big[n_{\nu}(\omega,\boldsymbol{R},t)\mathrm{d}_{\nu}(\omega)+\big[n_{\nu}(-\omega,\boldsymbol{R},-t)+1\big]\mathrm{d}_{\nu}(-\omega)\Big],\\
G^{>}_{\nu}(\omega,\boldsymbol{R},t)&=-2\pi i\Big[n_{\nu}(-\omega,\boldsymbol{R},-t)\mathrm{d}_{\nu}(-\omega)+\big[n_{\nu}(\omega,\boldsymbol{R},t)+1\big]\mathrm{d}_{\nu}(\omega)\Big].
\end{split}$}
\end{equation}
In Appendix \ref{fluctuation_dissipation_theorem} we prove the GKB ansatz at equilibrium allowing then for a more intuitive physical interpretation based on the fluctuation-dissipation theorem, then in Appendix \ref{rigorous_derivation_GKBA} we prove the GKB ansatz also out-of-equilibrium provided that the phonon distribution function taking place in its definition is the Wigner distribution function \eqref{wigner_distr_function}. In what is commonly referred to as the quasiparticle approximation, the spectral function is represented as a Dirac delta function,
\begin{equation} \label{quasiparticle_approximation_main}
\mathrm{d}_{\nu}(\pm\omega)\to\delta(\omega\mp\omega_{\nu}),
\end{equation}
translating into the so-called quasiparticle ansatz \cite{lipavsky1986generalized}:
\begin{equation} \label{quasiparticle_ansatz}
\resizebox{0.52\textwidth}{!}{$
\begin{split}
G^{<}_{\nu}(\omega,\boldsymbol{R},t)&=-2\pi i\Big[n_{\nu}(\omega,\boldsymbol{R},t)\delta(\omega-\omega_{\nu})+\big[n_{\nu}(-\omega,\boldsymbol{R},-t)+1\big]\delta(\omega+\omega_{\nu})\Big],\\
G^{>}_{\nu}(\omega,\boldsymbol{R},t)&=-2\pi i\Big[n_{\nu}(-\omega,\boldsymbol{R},-t)\delta(\omega+\omega_{\nu})+\big[n_{\nu}(\omega,\boldsymbol{R},t)+1\big]\delta(\omega-\omega_{\nu})\Big].
\end{split}$}
\end{equation}
This formulation arises from the fact that in the quasiparticle picture, the system's excitations are treated as sharp, well-defined energy states, with negligible broadening and frequency lineshifts. Consequently, the spectral weight is concentrated entirely at the quasiparticle energy, leading to the Dirac delta function representation as prescribed by the Fermi golden rule. This approximation is valid when the quasiparticle lifetime is sufficiently long and translates into requiring exact energy conservation of individual phonon collisions. In this regime, it is justified to neglect frequency lineshifts at the standard BTE level because the leading diagrams obtained through perturbation theory of the self-energy are of the same order, namely the “loop” and “bubble” diagrams \cite{maradudin1962scattering,cowley1963lattice}. Moreover, the loop contributes only to the real part of the self-energy (lineshifts), while the bubble contributes to both the real and imaginary parts (lineshifts and linewidths) \cite{maradudin1962scattering}. If the effect of frequency lineshifts is neglected, the loop diagram can be entirely omitted, leaving only the contribution to the linewidths (and thus the lifetimes) from the bubble diagram. This is exactly what happens in the standard BTE approach, where the ansatz hierarchy assumes that, since broadening cannot be included (the spectral function is a Dirac delta function), frequency lineshifts should not occur within the Dirac delta either (see Section \ref{scf_broadening_Section_main}). Already at the level of the quasiparticle approximation \eqref{quasiparticle_approximation_main}, we clearly see that also energy renormalization effects are being neglected. These effects are quantified by the pole renormalization (or quasiparticle weight, or wavefunction renormalization),
\begin{equation} \label{z_pole_renormalization}
z_{\nu}(\omega)=\frac{1}{1-\frac{\partial\Delta_{\nu}(\omega)}{\partial\omega}},
\end{equation}
and are evidently related to the real part of the self-energy. This means that when the contributions to thermal transport coming from the real part of the self-energy are neglected (as in the standard BTE approach), then it is legitimate to use either the quasiparticle ansatz \eqref{quasiparticle_ansatz} or the GKB ansatz itself \eqref{GKB_ansatz_equation} with a more or less precise description of the broadening. We will show that a typical starting point is to assume a Lorentzian spectral function and, consequently, a broadening parameter, $\gamma_{\nu}(\omega)\equiv\gamma_{\nu}$, which does not depend on frequency. In this study, we focus specifically on how employing a GKB ansatz with a Lorentzian spectral function affects thermal transport. Delving into the more complex non-Lorentzian case—where the broadening parameter varies with frequency—lies beyond our present scope, as accurately resolving frequency-dependent broadening requires access to additional spectral features, such as satellite peaks. \\
However, if energy renormalization effects are to be included (see i.g. Ref. \cite{pavlyukh2016vertex} for electrons) then the GKB ansatz should be modified as follows
\begin{equation} \label{GKB_ansatz_equation_with_energy_renormalization_bosonic}
\begin{split}
&g^{<}_{\nu}(\omega,\boldsymbol{R},t)=-2\pi iz_{\nu}(\omega)n_{\nu}(\omega,\boldsymbol{R},t)\mathrm{d}_{\nu}(\omega),\\
&g^{>}_{\nu}(\omega,\boldsymbol{R},t)=-2\pi iz_{\nu}(\omega)\big[n_{\nu}(\omega,\boldsymbol{R},t)+1\big]\mathrm{d}_{\nu}(\omega),
\end{split}
\end{equation}
or
\begin{equation} \label{GKB_ansatz_equation_with_energy_renormalization}
\resizebox{0.55\textwidth}{!}{$
\begin{split}
G^{<}_{\nu}(\omega,\boldsymbol{R},t)&=-2\pi i\Big[z_{\nu}(\omega)n_{\nu}(\omega,\boldsymbol{R},t)\mathrm{d}_{\nu}(\omega)+z_{\nu}(-\omega)\big[n_{\nu}(-\omega,\boldsymbol{R},-t)+1\big]\mathrm{d}_{\nu}(-\omega)\Big],\\
G^{>}_{\nu}(\omega,\boldsymbol{R},t)&=-2\pi i\Big[z_{\nu}(-\omega)n_{\nu}(-\omega,\boldsymbol{R},-t)\mathrm{d}_{\nu}(-\omega)+z_{\nu}(\omega)\big[n_{\nu}(\omega,\boldsymbol{R},t)+1\big]\mathrm{d}_{\nu}(\omega)\Big].
\end{split}$}
\end{equation}
This means that when pole renormalization is considered, the spectral function \eqref{spectral_function_d_bosonic_main} must account for both frequency shifts and linewidths, the latter being modified to $z_{\nu}\gamma_{\nu}$ \cite{pavlyukh2016vertex,nery2018quasiparticles}. Furthermore, in the case of a quasiparticle ansatz, the action of the Dirac delta function ensures that pole renormalizations reduce to $z_{\nu}(\omega)\to z_{\nu}=\big(1-\big.\frac{\partial\Delta_{\nu}(\omega)}{\partial\omega}\big|_{\omega=\omega_{\nu}}\big)^{-1}$. In this work, we do not explore the impact of energy renormalization on phonon transport, as it falls beyond our current scope and will be addressed in a separate study.

\subsection{Driving term}

At this stage we are fully equipped to derive the extended phonon trasport equation. Starting from the Kadanoff--Baym equation in the Wigner mixed representation \eqref{KBE_to_use_for_derivation}, we derive the extended phonon transport equation by applying the gradient approximation to the driving term. This derivation employs the GKB ansatz \eqref{GKB_ansatz_equation} for the lesser Green’s function and exploits time-reversal invariance to combine positive- and negative-frequency contributions. The detailed steps are provided in Appendix \ref{dervation_of_driving_term_section} and lead to the final expression for the driving term:
\begin{equation} \label{DT}
\begin{split}
&\,\text{DT}_{\nu}(\omega,\boldsymbol{R},t)=\\
=&\,2\pi\Bigg[\frac{\omega}{\omega_{\nu}}\mathrm{d}_{\nu}(\omega)\frac{\partial n_{\nu}(\omega,\boldsymbol{R},t)}{\partial t}-\frac{\omega}{\omega_{\nu}}\mathrm{d}_{\nu}(-\omega)\frac{\partial n_{\nu}(-\omega,\boldsymbol{R},t)}{\partial t}+\\
&\hspace{0.5cm}+\left(\frac{\omega^{2}}{\omega_{\nu}^{2}}+1\right)\mathrm{d}_{\nu}(\omega)\frac{\boldsymbol{v}_{\nu}}{2}\cdot\frac{\partial n_{\nu}(\omega,\boldsymbol{R},t)}{\partial\boldsymbol{R}}+\\
&\hspace{0.5cm}+\left(\frac{\omega^{2}}{\omega_{\nu}^{2}}+1\right)\mathrm{d}_{\nu}(-\omega)\frac{\boldsymbol{v}_{\nu}}{2}\cdot\frac{\partial n_{\nu}(-\omega,\boldsymbol{R},t)}{\partial\boldsymbol{R}}\Bigg].
\end{split}
\end{equation}

\subsubsection{Quasiparticle ansatz}

Finally, we can obtain the quasiparticle approximation \eqref{quasiparticle_approximation_main} of the driving term \eqref{DT} by simply integrating over all frequencies:
\begin{equation} \label{DT_qp}
\begin{split}
\text{DT}_{\nu}&=4\pi\Bigg[\frac{\partial n_{\nu}(\boldsymbol{R},t)}{\partial t}
+\boldsymbol{v}_{\nu}\frac{\partial n_{\nu}(\boldsymbol{R},t)}{\partial\boldsymbol{R}}\Bigg].
\end{split}
\end{equation}
By having the Dirac delta functions act on the driving term, the general dependence on $\omega$ is reduced to a dependence on the wave vector $\boldsymbol{q}$ and band index $s$ through phonon frequencies $\omega_{\nu}$. Additionally, the phonon distribution $n$ in the driving term corresponds directly to the Wigner distribution function $\tilde{n}$. This demonstrates clearly that, when applying the quasiparticle ansatz \eqref{quasiparticle_ansatz} within the gradient approximation scheme, the driving term of the KBE reduces to that of the BTE.

Finally, inserting Eq. \eqref{DT_qp} into Eq. \eqref{KBE_to_use_for_derivation}, we can write the structure of the transport equation, including the driving term within the quasiparticle approximation and the scattering term to be evaluated in the following Sections:
\begin{equation} \label{skeleton_transport_equation}
\begin{split}
&\,\frac{\partial n_{\nu}(\boldsymbol{R},t)}{\partial t}
+\boldsymbol{v}_{\nu}\frac{\partial n_{\nu}(\boldsymbol{R},t)}{\partial\boldsymbol{R}}=\\
=&\,\frac{1}{2\pi}\sum_{\mu}\int d\omega\Big[\Sigma_{\mu}^{>}(\omega,\boldsymbol{R},t)G_{\nu}^{<}(\omega,\boldsymbol{R},t)-\\
&\hspace{2cm}-\Sigma_{\mu}^{<}(\omega,\boldsymbol{R},t)G_{\nu}^{>}(\omega,\boldsymbol{R},t)\Big]=\\
=&\,\text{CT}_{\nu}(\omega,\boldsymbol{R},t),
\end{split}
\end{equation}
where $\mu$ are simply the general wave vector and band index associated with the general frequency $\omega$.

\subsection{Formal derivation of space-time--dependent collision term}

For the drift term, no additional derivations are needed, as it is already expressed through commutators (see the LHS of Eq. \eqref{KBE_skeleton}), preserving its structure across different representations. However, the collision term involves anticommutators of products between $G$ and the self-energy $\Sigma$ (see the right-hand side (RHS) of Eq. \eqref{KBE_skeleton}), requiring the additional step of evaluating the functional form of  $\Sigma$ terms of $G$. Specifically, since we aim to use Wigner’s mixed representation to achieve a coherent framework that accommodates both the drift and collision dynamics accurately, we need to rewrite the functional expression of $\Sigma$ in terms of $G$ within this representation (see Appendix \ref{self_energy_wigner}).

In this work, we choose to consider the self-energy up to the first leading order in the diagrammatic expansion, derived rigorously using, e.g., the functional method of Martin and Schwinger \cite{wehner1967phonon,wehner1966infra,martin1959theory}, which was also employed by Kadanoff \cite{kadanoff2018quantum}. As mentioned previously, by neglecting frequency lineshifts (real part of the self-energy) the leading loop diagram with a bare fourth-order interaction vertex can be safely neglacted as it only contributes to frequency lineshifts \cite{maradudin1962scattering}. Therefore, this choice translates into having a self-energy given by the bubble diagram only:
\begin{equation} \label{sigma_eq_full}
\begin{split}
\Sigma_{\nu}^{\lessgtr}(\omega)=&\,4i\pi\hbar\sum_{\nu'\nu''}|\mathcal{F}_{\nu-\nu'-\nu''}|^{2}\int\frac{d\omega_{1}}{2\pi}\int\frac{d\omega_{2}}{2\pi}\cdot\\
&\cdot\delta(\omega-\omega_{1}-\omega_{2})G_{\nu'}(\omega_{1})G_{\nu''}(\omega_{2})\Gamma_{3,\nu\nu'\nu''}(\omega,\omega_{1},\omega_{2}),
\end{split}
\end{equation}
where the matrix element $\mathcal{F}$ is defined in as \cite{ziman2001electrons}
\begin{equation}
\mathcal{F}_{\nu-\nu'-\nu''}=\delta_{\boldsymbol{q}-\boldsymbol{q}'-\boldsymbol{q}'',\boldsymbol{K}}|\uppsi_{\nu-\nu'-\nu''}|,
\end{equation}
$\boldsymbol{K}$ being a reciprocal lattice vector. Note that the delta function $\delta(\omega-\omega_{1}-\omega_{2})$ appears due to the time dependence being only in the form of time difference $t_{1}-t_{2}$. In this formulation, the interaction is purely local in both time and space: the three phonons involved do not change their positions during the scattering event, which is assumed to occur instantaneously. This corresponds to the so-called completed-collision limit \cite{rossi2002theory,iotti2005quantum,rossi2011theory}. Extensions beyond this approximation—allowing for the formation of a transient “molecular scattering state” that persists over a finite collision time, during which quasiparticles can readjust their positions \cite{di2026nonlocalselfenergy}—have been discussed in the context of heavy-ion reactions \cite{morawetz2001retarded,morawetz2013nonequilibrium,morawetz2017nonequilibrium,lipavsky2001kinetic}. The third-order vertex function entering the bubble diagram, $\Gamma_{3}$, is diagrammatically defined as going beyond the simple bare three-phonon interaction $\Gamma_{3}^{0}\propto\uppsi_{3}$ \cite{klein1969derivation,wehner1967phonon}. When vertex corrections are considered, these are obtained from a self-consistent equation satisfied by the vertex function $\Gamma_{3}$ in terms of so-called ladder diagrams \cite{sham1967equilibrium,sham1967temperature,kadanoff2018quantum,danielewicz1984quantum1,danielewicz1984quantum2,botermans1990quantum}. Such equation reads \cite{klein1969derivation}
\begin{equation} \label{gamma_3_integral_eq}
\begin{split}
&\Gamma_{3}(1,2,3)=\\
=&\,\Gamma_{3}^{0}(1,2,3)-\\
&-i\,12\,\uppsi_{4}(1,2,4,5)G(4,4')G(5,5')\Gamma_{3}(4',5',4)\\
&+\Gamma_{3}^{0}(1,4,5)G(4,4')G(5,5')G(6,6')\Gamma_{3}(4',6,2)\Gamma_{3}(5',6',3)+\cdots,
\end{split}
\end{equation}
which represents an integral equation involving integrations over all internal variables, including frequencies and wave vectors. Its self-consistent character follows from the appearance of the fully dressed vertex $\Gamma_{3}$ on both sides of the equation. A similar Bethe--Salpeter-type equation can also be formulated for the fourth-order vertex $\Gamma_{4}$ \cite{wehner1967phonon,klein1969derivation,klein1968linear}. Within the compact notation adopted in Eq. \eqref{gamma_3_integral_eq}, all integrations are implicitly encoded through numerical labels. We recall that the connection between the equations governing vertex parts or ladder diagrams and transport equations has been extensively discussed by Abrikosov, Gorkov, and Dzyaloshinskii in their seminal book \cite{abrikosov1965quantum}. It is evident that repeated iteration of the third term on the RHS of Eq. \eqref{gamma_3_integral_eq} generates ladder diagrams in which the vertex corrections associated with the second and third terms on the RHS involve four- and three-phonon processes, respectively. Ladder diagrams, or vertex corrections, can be relevant in modifying several crystal properties such as sound absorption \cite{klein1967ultrasonic}, infrared radiation absorption \cite{wehner1966infra}, and scattering processes \cite{cochran1967phonons}. Moreover, they play a central role in the derivation of a BTE-like equation (understood as a partial differential equation describing the time evolution of a distribution function) for the vertex function $\Gamma_{3}$ \cite{sham1967equilibrium}. As shown by Sham \cite{sham1967equilibrium,sham1967temperature}, starting from the nonlinear integral equation \eqref{gamma_3_integral_eq} one can indeed obtain the Bethe--Salpeter equation for $\Gamma_{3}$. Using this equation as a starting point, it is possible to derive an equation formally analogous to the BTE, in which the vertex function replaces the phonon distribution and the RHS assumes the structure of Peierls' collision operator \cite{klein1969derivation}. Finally, Eq. \eqref{gamma_3_integral_eq} admits further simplifications in specific situations. In particular, vertex corrections (or ladder diagrams) can become relevant for a phonon whose frequency is comparable to the linewidth of the phonons with which it interacts, as already pointed out by Klein and Sham \cite{klein1969derivation,sham1967equilibrium}. Consequently, $\Gamma_{3}$ on the RHS of Eq. \eqref{gamma_3_integral_eq} needs to be evaluated only when long-wavelength acoustic phonons enter and exit the self-energy diagram. The internal phonons of the bubble diagram are thermal phonons with energies of order $k_{\mathrm{B}}T$. Hence, the relevant situation corresponds to one of the three phonons in $\Gamma_{3}(1,2,3)$ having a frequency much smaller than $k_{\mathrm{B}}T$, while the remaining two are thermal phonons with wave vectors of nearly equal magnitude but opposite direction. Accounting for this circumstance, and for the dependence on different wave vectors of the various $\Gamma_{3}$ terms in Eq. \eqref{gamma_3_integral_eq}, the second $\Gamma_{3}$ on the RHS of Eq. \eqref{gamma_3_integral_eq} can be replaced by its lowest-order contribution, namely $\Gamma_{3}\equiv\Gamma_{3}^{0}\propto\uppsi_{3}$, leading to
\begin{equation} \label{gamma_3_integral_eq_bis}
\begin{split}
&\,\Gamma_{3}(1,2,3)=\\
=&\,\Gamma_{3}^{0}(1,2,3)-\\
&-\Big[i12\uppsi_{4}(1,2,5,6)-\Gamma_{3}^{0}(1,4,5)\Gamma_{3}^{0}(2,4',6)G(4,4')\Big]\cdot\\
&\hspace{0.5cm}\cdot G(5,5')G(6,6')\Gamma_{3}(5',6',3),
\end{split}
\end{equation}
which makes it possible to explicitly account for vertex corrections since, upon iteration, it simply yields the sum of a geometric series \cite{di2026vertex}:
\begin{equation} 
\Gamma_{3}^{{\rm{N}}}\simeq\Gamma_{3}^{0}+\frac{\Gamma_{3}^{0}}{1+\big(i12\uppsi_{4}-\Gamma_{3}^{0}\Gamma_{3}^{0}G\big)GG}.
\end{equation}
The phonon self-energy may be improved following two distinct strategies. The first consists in retaining higher-order terms in the perturbative expansion in Feynman's diagrams, which would require the introduction of fourth-order and higher-order vertices. In this context, one should also note that another leading tad-pole diagram \cite{SSCHA_0,monacelli2024simulating,
calandra2007anharmonic,maradudin1962scattering,paulatto2015first} should be considered. This contribution provides information about the atomic motion due to anharmonicity; at zero temperature, this is attributed to zero-point motion. This diagram is not relevant for a high-symmetry crystal because it only provides corrections at the Gamma point in the Brillouin zone \cite{lazzeri2003anharmonic}. The loop diagram depends on $\uppsi_{4}$ once, while the bubble depends on $\uppsi_{3}$ and $\Gamma_{3}$. If the vertex $\Gamma_{3}$ is not dressed, then the bubble depends on $\uppsi_{3}$ twice. Although the tad-pole also depends on $\uppsi_{3}$ twice, it is diagrammatically distinct from the bubble at zero-order in the vertex corrections. In the self-energy perturbative treatment discussed above, we have assumed that the atoms oscillate around their harmonic positions, and under this assumption, the tad-pole is absent, as is the case in high-symmetry systems where the atomic positions are fixed by the symmetry group \cite{lazzeri2003anharmonic}.

The second strategy concerns a more accurate treatment of the vertex part itself, namely going beyond the approximation $\Gamma_{3}\equiv\Gamma_{3}^{0}$ by including the additional terms appearing in its functional derivative definition. Nevertheless, a systematic treatment of vertex corrections beyond the bare three-phonon interaction lies outside the scope of this work.

In summary, here we adopt the bubble approximation together with the zeroth-order approximation for the vertex function $\Gamma_{3}$. This corresponds to retaining Eq. \eqref{sigma_eq_full} and only the first term in Eq. \eqref{gamma_3_integral_eq}. This choice is consistent with neglecting the second term, proportional to $\uppsi_{4}$, since $\uppsi_{4}$ contributions are also omitted in the definition of the vertex function $\Gamma_{3}$.\\
By setting $\Gamma_{3}\equiv\Gamma_{3}^{0}\propto\uppsi_{3}$ into Eq. \eqref{sigma_eq_full}, one recovers the bubble self-energy in the form commonly employed for the evaluation of several anharmonic effects \cite{xu2008nonequilibrium,volz2020quantum,maradudin1962scattering,cowley1966anharmonic,semwal1972thermal}:
\begin{equation} \label{sigma_eq}
\begin{split}
\Sigma_{\nu}^{\lessgtr}(\omega)=&\,4i\pi\hbar\sum_{\nu'\nu''}|\mathcal{F}_{\nu-\nu'-\nu''}|^{2}\int\frac{d\omega_{1}}{2\pi}\int\frac{d\omega_{2}}{2\pi}\cdot\\
&\cdot\delta(\omega-\omega_{1}-\omega_{2})G_{\nu'}^{\lessgtr}(\omega_{1})G_{\nu''}^{\lessgtr}(\omega_{2}).
\end{split}
\end{equation}
In order to go from the KBE to a BTE-like equation using Wigner’s mixed representation, we need to ensure that the functional form of the self-energy with respect to the Green's function remains unchanged, even when adopting the Wigner formalism. The procedure for showing this is detailed in Appendix \ref{self_energy_wigner}; here, we present the final result (we reuse the tilde symbol here to emphasize that the self-energy has been converted to Wigner's mixed representation):
\begin{equation} \label{final_selfenergy_wigner_main} 
\begin{split}
\tilde{\Sigma}_{\nu}^{\lessgtr}(\omega;\boldsymbol{R},t)=&\,4i\pi\hbar\sum_{\nu'\nu''}|\mathcal{F}_{\nu-\nu'-\nu''}|^{2}\int\frac{d\omega_{1}}{2\pi}\int\frac{d\omega_{2}}{2\pi}\cdot\\
&\cdot\delta(\omega-\omega_{1}-\omega_{2})\tilde{G}_{\nu'}^{\lessgtr}(\omega_{1};\boldsymbol{R},t)\tilde{G}_{\nu''}^{\lessgtr}(\omega_{2};\boldsymbol{R},t).
\end{split}
\end{equation}

\subsubsection{Quasiparticle ansatz}
\label{CT_qp}

When we use the quasiparticle ansatz \eqref{quasiparticle_ansatz} to write the Green's function on the RHS of Eq. \eqref{skeleton_transport_equation} and in the self-energy expression \eqref{final_selfenergy_wigner_main}, we obtain the standard CT of the BTE (the detailed derivation of this result is presented in the SM \cite{supplementary}):
\begin{widetext}
\begin{equation} \label{final_BTE_scattering_qp_non_homo_main}
\resizebox{\textwidth}{!}{$
\begin{split}
{\rm{CT}}_{\nu}(\boldsymbol{R},t)=8\pi\hbar\sum_{\nu'\nu''}\Bigg\lbrace&\frac{1}{2}|\mathcal{F}_{\nu-\nu'-\nu''}|^{2}\Bigg[\big(n_{\nu}(\boldsymbol{R},t)+1\big)n_{\nu'}(\boldsymbol{R},t)n_{\nu''}(\boldsymbol{R},t)-n_{\nu}(\boldsymbol{R},t)\big(n_{\nu'}(\boldsymbol{R},t)+1\big)\big(n_{\nu''}(\boldsymbol{R},t)+1\big)\Bigg]\delta(\omega_{\nu}-\omega_{\nu'}-\omega_{\nu''})+\\
&+|\mathcal{F}_{\nu\nu'-\nu''}|^{2}\Bigg[\big(n_{\nu}(\boldsymbol{R},t)+1\big)\big(n_{\nu'}(\boldsymbol{R},t)+1\big)n_{\nu''}(\boldsymbol{R},t)-n_{\nu}(\boldsymbol{R},t)n_{\nu'}(\boldsymbol{R},t)\big(n_{\nu''}(\boldsymbol{R},t)+1\big)\Bigg]\delta(\omega_{\nu}+\omega_{\nu'}-\omega_{\nu''})\Bigg\rbrace,
\end{split}$}
\end{equation}
\end{widetext}
where the three-phonon scattering terms, including phonon coalescences and decays (with related factor of $1/2$ for avoiding double counting), can already be distinguished. It is important to note that the result \eqref{final_BTE_scattering_qp_non_homo_main} was obtained by integrating over all frequencies $\omega$, simultaneously with the driving term, as done for Eq. \eqref{DT_qp}. This approach allowed for the selection of $\omega=\omega_{\nu}$ even within the collision term. Eq. \eqref{final_BTE_scattering_qp_non_homo_main} represents the first key results of this work. While the derivation of the BTE from the KBE for electrons has been well-established in previous studies \cite{lipavsky1986generalized,ponce2020first}, only partial results for phonons had been obtained, based on inconsistencies in the choice of the bosonic ansatz within a vibronic notation expression of the transport equation \cite{kwok1966unified,niklasson1968theory}. Most importantly, to the best of our knowledge, this result allows to overcome the long-standing challenge \cite{spohn2006phonon} of deriving the BTE collision term, rigorously accounting for non-homogeneity and time-dependence—often obtained from empirical approximations \cite{vasko2006quantum}. This shows that the time and spatial dependence of the BTE naturally emerge from applying Wigner's mixed representation to the KBE, thus clarifying the most appropriate approach for dealing with time-dependent and non-homogeneous BTEs, usually solved employing Monte Carlo methods \cite{carrete2017almabte,raya2022bte}.

Using Eq. \eqref{final_BTE_scattering_qp_non_homo_main} for the scattering operator, the BTE remains a highly complex integro-differential equation. As is commonly the case in most situations of interest, phonons are fairly close to equilibrium \cite{ziman2001electrons,fugallo2013ab}. To simplify the problem, we therefore assume small temperature perturbations, allowing us to define a local temperature $T(\boldsymbol{R},t)$ and to describe the out-of-equilibrium distribution $n_{\nu}(\boldsymbol{R},t)$ as a small deviation from the local thermal equilibrium given by the Bose-Einstein distribution $\bar{n}_{\nu}(\boldsymbol{R},t)$. Under these conditions, it is unnecessary to evaluate the full scattering integral; instead, we can linearize it for small deviations $n_{\nu}-\bar{n}_{\nu}$. This linearization introduces the deviation from equilibrium function $h_{\nu}$, defined as ($\boldsymbol{R},t$ dependence is omitted):
\begin{equation} \label{expansion_n_bar_n_delta_n_main}
n_{\nu}=\bar{n}_{\nu}+\Delta n_{\nu}=\bar{n}_{\nu}+\bar{n}_{\nu}[\bar{n}_{\nu}+1]h_{\nu},
\end{equation}
where $h_{\nu}=\dfrac{\hbar\omega_{\nu}}{k_{{\rm B}}T^{2}}\,\nabla T\cdot\boldsymbol{{\rm f}}_{\nu}$, 
and $\boldsymbol{{\rm f}}_{\nu}$ is the solution of the LBTE, recast in matrix form as a linear algebra problem. By inserting Eq. \eqref{expansion_n_bar_n_delta_n_main} into Eq. \eqref{final_BTE_scattering_qp_non_homo_main} we get \cite{fugallo2013ab} (see SI \cite{supplementary} for details)
\begin{equation} \label{final_scattering_linerized_qp_matrix_main}
\begin{split}
{\rm{CT}}_{\nu}=\sum_{\nu'}A_{\nu\nu'}h_{\nu'}=\underbrace{-A^{\text{out}}_{\nu\nu}h_{\nu}}_{\text{diagonal}}+\underbrace{\sum_{\nu'\ne\nu}A^{\text{in}}_{\nu\nu'}h_{\nu'}}_{\text{non-diagonal}},
\end{split}
\end{equation}
where
\begin{equation} \label{final_A_in_A_out_qp_main}
\begin{split}
&A^{\text{out}}_{\nu\nu}=8\pi\hbar\sum_{\nu'\nu''}\Big(\mathcal{L}_{\nu\nu'}^{\nu''}{\rm{R}}_{\nu\nu'}^{\nu''}+\frac{1}{2}\mathcal{L}_{\nu}^{\nu'\nu''}{\rm{R}}_{\nu}^{\nu'\nu''}\Big),\\
&A^{\text{in}}_{\nu\nu'}=8\pi\hbar\sum_{\nu''}\Big(\mathcal{L}_{\nu\nu''}^{\nu'}{\rm{R}}_{\nu\nu''\nu'}-\mathcal{L}_{\nu\nu'}^{\nu''}{\rm{R}}_{\nu\nu'}^{\nu''}+\mathcal{L}_{\nu}^{\nu'\nu''}{\rm{R}}_{\nu}^{\nu'\nu''}\Big).
\end{split}
\end{equation}
The condensed notation for the collision term is as follows:
\begin{equation} \label{pieces_in_collision_term_qp}
\begin{split}
&\mathcal{L}_{\nu}^{\nu'\nu''}=\delta(\omega_{\nu}-\omega_{\nu'}-\omega_{\nu''})|\mathcal{F}_{\nu-\nu'-\nu''}|^{2},\\
&\mathcal{L}_{\nu\nu'}^{\nu''}=\delta(\omega_{\nu}+\omega_{\nu'}-\omega_{\nu''})|\mathcal{F}_{\nu\nu'-\nu''}|^{2},\\
&{\rm{R}}_{\nu\nu'}^{\nu''}=(\bar{n}_{\nu}+1)(\bar{n}_{\nu'}+1)\bar{n}_{\nu''}\\
&{\rm{R}}_{\nu}^{\nu'\nu''}=(\bar{n}_{\nu}+1)\bar{n}_{\nu'}\bar{n}_{\nu''}.
\end{split}
\end{equation}
The derivation of the expressions for the $R$ terms are also derived in the SM \cite{supplementary}. The decomposition in Eq. \eqref{final_A_in_A_out_qp_main} is particularly useful as it clearly highlights that the total scattering consists of two components: $A^{\text{out}}_{\nu\nu}$, representing the depopulation rate of the phonon $\nu$ due to scattering with other phonons, and $A^{\text{in}}_{\nu\nu'}$, describing the repopulation of the mode $\nu$ as a result of scattering involving incoming (already scattered) phonons. In this way, we can once again observe that the BTE is indeed a transport equation that can be understood as a specific case of the KBE. In fact, the latter also features a collision term that can generally be rewritten as ${\rm{CT_{in}}-CT_{out}}$. In the case of the quasiparticle ansatz \eqref{quasiparticle_ansatz}, due to the exact energy conservation that holds for each individual scattering event, it can be demonstrated that \cite{simoncelli2022wigner,spohn2006phonon} $A^{\text{in}}_{\nu\nu'}$ and $A^{\text{out}}_{\nu\nu}$ are specifically related by the following exact relation:
\begin{equation} \label{global_scattering_energy_cons}
A^{\text{out}}_{\nu\nu}=\sum_{\nu'\ne\nu}A^{\text{in}}_{\nu\nu'}\frac{\omega_{\nu'}}{\omega_{\nu}}.
\end{equation}
This result naturally arises from the energy conservation of the entire scattering matrix, as expected for a bulk isolated system (see Appendix \ref{energy_cons_demonstrations}). \\
Moreover, it can be demonstrated \cite{fugallo2013ab} that the matrix $A^{\text{out}}_{\nu\nu}$ and the linewidths $\Gamma_{\nu}$ (which are the inverse of the phonon lifetimes $\tau_{\nu}=\frac{1}{2\Gamma_{\nu}}$) are related by the following expression \cite{supplementary,fugallo2013ab}
\begin{equation} \label{A_out_as_Gamma}
A^{\text{out}}_{\nu\nu}=\bar{n}_{\nu}(\bar{n}_{\nu}+1)\Gamma_{\nu}.
\end{equation}
Such relation naturally emerges from the derivation of the scattering matrix $A^{\text{out}}_{\nu\nu}$ and is further validated by the NEGF approach when using Eq. \eqref{gamma_def_NEGF} (see the SM \cite{supplementary}). Moreover, since Eq. \eqref{global_scattering_energy_cons} is exact within the quasiparticle ansatz, it can be shown that, as expected by construction, calculating the linewidths using Eq. \eqref{global_scattering_energy_cons} yields the exact same result (see the SM \cite{supplementary} for details).

As is commonly done to facilitate a straightforward comparison with experimental data, the spectral function of the system is calculated as given in Eq. \eqref{spectral_function_d_bosonic_main} and it is related to the phonon self-energy as
\begin{equation} \label{self_energy_general_def}
\Sigma_{\nu}(\omega)=\Delta_{\nu}(\omega)+i\Gamma_{\nu}(\omega).
\end{equation}
It is important to note that in other works based on self-consistent phonon (SCPH) theories \cite{werthamer1970self}, this expression may differ, incorporating additional renormalization effects. Approaches based on SCPH theories, such as self-consistent ab initio lattice dynamics (SCAILD) \cite{souvatzis2008entropy} and the stochastic self-consistent harmonic approximation (SSCHA) \cite{errea2014anharmonic}, aim to include anharmonic effects within the phonon frequencies. The SCPH approach relies on the Dyson's equation as starting point and derives a self-consistent equation that must be solved to obtain renormalized frequencies. These frequencies are typically renormalized through the fourth-order interaction term (the loop diagram), which does not require external frequencies as it is free of vertices to dress. This allows for a self-consistent solution that captures anharmonic effects \cite{tadano2015self}. Additionally, these methods often include the effect of temperature on phonon bands, typically through ab initio molecular dynamics (AIMD) simulations or SSCHA itself \cite{SSCHA_0,SSCHA_1,SSCHA_2,SSCHA_3}. Moreover, Tadano et al. have extended this approach further by incorporating lineshift effects due to the bubble diagram in perturbation theory \cite{tadano2022first}. The discussion of frequency renormalization due to anharmonic effects and temperature contributions is not explicitly covered in this work. However, we aim to emphasize that the LGBTE presented here provides a foundational framework based on the the spectral function to be used in the GKB ansatz. In fact, for general renormalized and temperature-dependent spectral functions, the GKB ansatz remains valid. Therefore, spectral functions from SCPH or SSCHA approaches can be integrated with our framework to account for temperature and renormalization effects. 

Finally, as anticipated above (see the discussion before Eq. \eqref{KB_equation_ongoing}), writing the retarded Green’s function explicitly in terms of the self-energy \eqref{self_energy_general_def} shows that, already at the lowest level of the hierarchy, that is, the quasiparticle ansatz \eqref{quasiparticle_ansatz}, the system exhibits a Lorentzian spectral function (bearing in mind that frequency line shifts are neglected): 
\begin{equation} \label{lorentzian_spectral_function}
\begin{split}
\mathrm{d}_{\nu}(\omega)=\frac{1}{\pi}\frac{\gamma_{\nu}}{(\omega-\omega_{\nu})^{2}+\gamma_{\nu}^{2}},
\end{split}
\end{equation}
with the broadening given by $\gamma_{\nu}$. It is important to note that the spectral function is Lorentzian because the linewidths are evaluated at the phonon frequency $\omega_{\nu}$, causing the frequency dependence to disappear. This would no longer hold if one were to consider the effect of satellites on thermal transport, in which case frequency resolution would indeed be required. In summary, we construct the transport equation starting from an ansatz with a Dirac delta spectral function, which in turn results in a Lorentzian spectral function of the system. Then it becomes natural to iterate the procedure with the GKB ansatz (instead of the quasiparticle one) incorporating the Lorentzian spectral function \eqref{lorentzian_spectral_function} in it. As we will see, this procedure leads to an ansätze hierarchy, each yielding a distinct transport equation. Furthermore, this hierarchy promotes the self-consistent calculation of phonon broadenings/linewidths. For example, the Lorentzian broadening will be obtained by iterating the procedure of substituting the system's spectral function into the ansatz at the next order. In this way, the broadening can be defined as a true collisional broadening with physical meaning, independent of computational schemes.

\subsubsection{GKB ansatz with Lorentzian spectral function}

Following this path, we insert the GKB ansatz \eqref{GKB_ansatz_equation} with a Lorentzian spectral function into the collision term of Eq. \eqref{skeleton_transport_equation} (we omit the dependence $\boldsymbol{R},t$ for simplicity in the notation), yielding:
\begin{widetext}
\begin{equation} \label{scattering_with_Lorentzian_broadening_main}
\begin{split}
{\rm{CT}}=8\hbar\sum_{\nu'\nu''}\upgamma_{\nu\nu'\nu''}\Bigg[&\frac{1}{2}\frac{|\mathcal{F}_{\nu\nu'\nu''}|^{2}\left[(n_{\nu}+1)(n_{\nu'}+1)(n_{\nu''}+1)-n_{\nu}n_{\nu'}n_{\nu''}\right]}{(\omega_{\nu}+\omega_{\nu'}+\omega_{\nu''})^{2}+\upgamma_{\nu\nu'\nu''}^{2}}+\\
&+\frac{1}{2}\frac{|\mathcal{F}_{\nu-\nu'-\nu''}|^{2}\left[(n_{\nu}+1)n_{\nu'}n_{\nu''}-n_{\nu}(n_{\nu'}+1)(n_{\nu''}+1)\right]}{(\omega_{\nu}-\omega_{\nu'}-\omega_{\nu''})^{2}+\upgamma_{\nu\nu'\nu''}^{2}}+\\
&+\frac{|\mathcal{F}_{\nu\nu'-\nu''}|^{2}\left[(n_{\nu}+1)(n_{\nu'}+1)n_{\nu''}-n_{\nu}n_{\nu'}(n_{\nu''}+1)\right]}{(\omega_{\nu}+\omega_{\nu'}-\omega_{\nu''})^{2}+\upgamma_{\nu\nu'\nu''}^{2}}\Bigg],
\end{split}
\end{equation}
\end{widetext}
where
\begin{equation} \label{gamma_as_convolution}
\upgamma_{\nu\nu'\nu''}=\gamma_{\nu}+\gamma_{\nu'}+\gamma_{\nu''}
\end{equation}
comes from a sequence of convolution integrals of Lorentzian distributions with $\gamma_{\nu}$, $\gamma_{\nu'}$ and $\gamma_{\nu''}$ as broadening (see the SM \cite{supplementary} for the detailed derivation). This new collision term shares both similarities and differences with the one obtained through the quasiparticle ansatz. As can be seen, the second and third terms on the RHS are the same two terms found on the RHS of the standard BTE collision term in quasiparticle ansatz, simply extended via a Lorentzian broadening to relax the strict energy conservation for individual scattering events. In the SM \cite{supplementary}, we provide the full algebraic derivation of the collision term for both the quasiparticle ansatz and the Lorentzian GKBA. In contrast, the first term on the RHS is absent at the quasiparticle anstaz level. This is because it involves three phonons whose frequencies all share the same sign, making such a process entirely forbidden if strict energy conservation is imposed for each individual phonon scattering event. In this case, since we relax this condition, the process becomes allowed and can be taken into account. In summary, in addition to the standard terms representing the creation of 2 phonons and annihilation of 1 phonon, and the creation of 1 phonon and annihilation of 2 phonons, we also have the term that accounts for the possibility of creating or annihilating 3 phonons. In Appendix \ref{section_small_broadening_annihilation_creation_three_phonons}, we provide an argument showing that, in the small-broadening limit, this contribution is negligible compared with the decay and coalescence processes. For this reason, we do not examine it further. At this point, it might seem natural to object that the collision term as written is mathematically correct but physically incorrect because it does not conserve energy overall. In the following, we will address this aspect in detail and show how global energy conservation could be restored, ensuring the definition of a local temperature and thermal conductivity \cite{allen2018temperature}.

As done previously, we write the scattering in the linear response framework (details can be found in the SM \cite{supplementary}):
\begin{equation} \label{final_scattering_linerized_broadening_matrix}
{\rm{CT}}=\sum_{\nu'}A_{\nu\nu'}h_{\nu'}=\underbrace{-A^{\text{out}}_{\nu\nu}h_{\nu}}_{\text{diagonal}}+\underbrace{\sum_{\nu'\ne\nu}A^{\text{in}}_{\nu\nu'}h_{\nu'}}_{\text{non-diagonal}},
\end{equation}
where
\begin{equation} \label{final_A_in_A_out_broadening_not_conserving}
\resizebox{0.5\textwidth}{!}{$
\begin{split}
&A^{\text{out}}_{\nu\nu}=8\pi\hbar\sum_{\nu'\nu''}\Bigg[\frac{1}{2}\Big(\tilde{\mathcal{L}}_{\nu}^{\nu'\nu''}-\tilde{\mathcal{L}}_{\nu\nu'\nu''}\Big){\rm{\Lambda}}_{\nu\nu'\nu''}-\tilde{\mathcal{L}}_{\nu\nu'}^{\nu''}{\rm{\Theta}}_{\nu\nu'\nu''}\Bigg],\\
&A^{\text{in}}_{\nu\nu'}=8\pi\hbar\sum_{\nu''}\Bigg[\Big(\tilde{\mathcal{L}}_{\nu\nu'\nu''}+\tilde{\mathcal{L}}_{\nu\nu''}^{\nu'}\Big){\rm{\Lambda}}_{\nu'\nu\nu''}+\Big(\tilde{\mathcal{L}}_{\nu}^{\nu'\nu''}+\tilde{\mathcal{L}}_{\nu\nu'}^{\nu''}\Big){\rm{\Theta}}_{\nu'\nu\nu''}\Bigg].
\end{split}$}
\end{equation}
Here, the condensed notation for the collision term is as follows:
\begin{equation} \label{pieces_in_collision_term_broadening}
\begin{split}
&\tilde{\mathcal{L}}_{\nu\nu'\nu''}=\frac{1}{\pi}\frac{\upgamma_{\nu\nu'\nu''}}{(\omega_{\nu}+\omega_{\nu'}+\omega_{\nu''})^{2}+\upgamma_{\nu\nu'\nu''}^{2}}|\mathcal{F}_{\nu\nu'\nu''}|^{2},\\
&\tilde{\mathcal{L}}_{\nu\nu'}^{\nu''}=\frac{1}{\pi}\frac{\upgamma_{\nu\nu'\nu''}}{(\omega_{\nu}+\omega_{\nu'}-\omega_{\nu''})^{2}+\upgamma_{\nu\nu'\nu''}^{2}}|\mathcal{F}_{\nu\nu'-\nu''}|^{2},\\
&\tilde{\mathcal{L}}_{\nu}^{\nu'\nu''}=\frac{1}{\pi}\frac{\upgamma_{\nu\nu'\nu''}}{(\omega_{\nu}-\omega_{\nu'}-\omega_{\nu''})^{2}+\upgamma_{\nu\nu'\nu''}^{2}}|\mathcal{F}_{\nu-\nu'-\nu''}|^{2},\\
&{\rm{\Lambda}}_{ijk}=\bar{n}_{i}(\bar{n}_{i}+1)(\bar{n}_{j}+\bar{n}_{k}+1),\text{ where }i,j,k\in\{\nu,\nu',\nu''\},\\
&{\rm{\Theta}}_{i,j,k}=\bar{n}_{i}(\bar{n}_{i}+1)(\bar{n}_{k}-\bar{n}_{j}),\text{ where }i,j,k\in\{\nu,\nu',\nu''\}.\\
\end{split}
\end{equation}
Note that in this case, the simplification of the ${\rm{\Lambda}}$s and ${\rm{\Theta}}$s expressions using strict energy conservation (see the SM \cite{supplementary}) is not possible. The repumping and depumping terms, as expressed in Eq. \eqref{final_A_in_A_out_broadening_not_conserving}, do not inherently ensure global energy conservation (this is also analytically shown in the SM \cite{supplementary}), as expected for a bulk isolated system. As is well known from the literature, approximations such as the relaxation time approximation (RTA), where the collision term is reduced solely to scattering-out (which defines phonon linewidths)—i.e., only its diagonal part—do not conserve energy by construction, leading to non-physical results for thermal conductivity \cite{fugallo2014thermal}, especially at low temperatures. As we already mentioned, the concept of energy conservation is closely tied to the definition of a local temperature in the system, which, in turn, is needed to define thermal conductivity. As summarized in Ref. \cite{allen2018temperature}, there are at least three distinct strategies to define a local temperature. These three versions are derived from thermal susceptibility, thermal conductivity, and energy conservation, respectively \cite{allen2018temperature}, and are simultaneously satisfied only by the exact solutions of the LBTE, where “exact” means taking into account the full collision term (both scattering-in and scattering-out). In the present work, we refer to the version based on energy conservation expressed by the continuity equation \cite{allen2018temperature,simoncelli2020generalization}:
\begin{equation}
\frac{\partial E(\boldsymbol{r},t)}{\partial t}+\nabla\cdot\boldsymbol{j}(\boldsymbol{r},t)=0,
\end{equation}
describing the rate of local energy increase (in absence of external heating) as balanced by energy current, $\boldsymbol{j}$, flowing in. The exact solution of the LBTE satisfies local energy conservation, or, in other words, local energy conservation allows for the definition of a local temperature that is consistent with the exact solution of the LBTE.

In the quasiparticle ansatz, integrating the product of phonon frequencies and the collision term over all phonon wave vectors confirms that the energy conservation condition (see Appendix \ref{energy_cons_demonstrations})
\begin{equation} \label{energy_cons_condition}
\sum_{\nu}\omega_{\nu}{\rm{CT}}_{\nu}=0
\end{equation}
is always fulfilled. However, once collisional broadening is introduced, the exact energy conservation condition at the level of an individual scattering event is relaxed. As a consequence, the global energy conservation condition \eqref{energy_cons_condition} is no longer guaranteed to hold. We show this explicitly through an analytical argument in the SM \cite{supplementary}. In fact, when directly evaluating $\sum_{\nu}\omega_{\nu}\mathrm{CT}_{\nu}$ using the scattering matrix in Eq. \eqref{final_A_in_A_out_broadening_not_conserving}—that is, when the phonon linewidths are obtained from the scattering-out matrix (see Eq. \eqref{A_out_as_Gamma})—we cannot verify the global energy conservation condition \eqref{energy_cons_condition}. To address this limitation, we first recall that the relation linking the scattering-in and scattering-out terms \cite{simoncelli2022wigner}, given in Eq. \eqref{global_scattering_energy_cons}, is satisfied by construction—independently of whether collisional broadening is included (see Eq. \eqref{LBTE_with_n_T})—provided that total energy is conserved and the full scattering matrix can be symmetrized under the exchange $\nu \leftrightarrow \nu'$. In particular, when total energy conservation is enforced by construction—as in the BTE within the quasiparticle ansatz—Eq. \eqref{global_scattering_energy_cons} yields exactly the same phonon linewidths as those obtained from the scattering-out matrix in Eq. \eqref{final_A_in_A_out_qp_main} (see the SM \cite{supplementary} for details). For this reason, we fix $A^{\text{in}}_{\nu\nu'}$ to the form obtained in Eq. \eqref{final_A_in_A_out_broadening_not_conserving} and then derive $A^{\text{out}}_{\nu\nu}$ from such $A^{\text{in}}_{\nu\nu'}$ using Eq. \eqref{global_scattering_energy_cons}, thereby enforcing total energy conservation a posteriori. This procedure has also been used and total energy conservation has been verified numerically in the calculations presented in Section \ref{result_section}.
In the broadening case we obtain 
\begin{equation} \label{almost_final_A_in_A_out_broadening_still_not_conserving}
\begin{split}
&A^{\text{out}}_{\nu\nu}=8\pi\hbar\sum_{\nu'\nu''}\frac{\omega_{\nu'}}{\omega_{\nu}}\Bigg[\Big(\tilde{\mathcal{L}}_{\nu\nu'\nu''}+\tilde{\mathcal{L}}_{\nu\nu''}^{\nu'}\Big){\rm{\Lambda}}_{\nu'\nu\nu''}+\\
&\hspace{3cm}+\Big(\tilde{\mathcal{L}}_{\nu\nu'}^{\nu''}+\tilde{\mathcal{L}}_{\nu}^{\nu'\nu''}\Big){\rm{\Theta}}_{\nu'\nu\nu''}\Bigg],\\
&A^{\text{in}}_{\nu\nu'}=8\pi\hbar\sum_{\nu''}\Bigg[\Big(\tilde{\mathcal{L}}_{\nu\nu'\nu''}+\tilde{\mathcal{L}}_{\nu\nu''}^{\nu'}\Big){\rm{\Lambda}}_{\nu'\nu\nu''}+\\
&\hspace{3cm}+\Big(\tilde{\mathcal{L}}_{\nu}^{\nu'\nu''}+\tilde{\mathcal{L}}_{\nu\nu'}^{\nu''}\Big){\rm{\Theta}}_{\nu'\nu\nu''}\Bigg].
\end{split}
\end{equation}
As anticipated, the validity of Eq. \eqref{global_scattering_energy_cons} relies on the fundamental requirement that the total scattering matrix be symmetrizable under the exchange $\nu \leftrightarrow \nu'$. If this condition is not satisfied, the Bose–Einstein eigenvector cannot be properly defined \cite{simoncelli2020generalization} (see Appendix \ref{bose_einstein_Section}):
\begin{equation}
\big|\theta_{\nu}^{0}\big\rangle=\frac{\sqrt{\bar{n}_{\nu}[\bar{n}_{\nu}+1]}}{k_{\text{B}}\bar{T}^{2}}\hbar\omega_{\nu}.
\end{equation}
In fact, Eq. \eqref{global_scattering_energy_cons} follows directly from the fact that the Bose–Einstein eigenvector is a right eigenvector of the scattering matrix with zero eigenvalue (see Eq. \eqref{LBTE_with_n_T} in Appendix \ref{bose_einstein_Section}). In contrast, the energy conservation condition in Eq. \eqref{energy_cons_condition} is satisfied only if the Bose–Einstein eigenvector is a left eigenvector of the total scattering matrix with zero eigenvalue, since $\big|\theta_{\nu}^{0}\big\rangle\propto\omega$), as discussed in Appendix \ref{energy_cons_demonstrations} and Ref. \cite{allen2018temperature}. These two conditions can only be simultaneously satisfied if the scattering matrix can be written in symmetric form, as in this case the right and left eigenvalues coincide:
\begin{equation} \label{left_right_eigenvalue}
\sum_{\nu'}\tilde{\Omega}_{\nu\nu'}\big|\theta_{\nu'}^{0}\big\rangle=\sum_{\nu'}\big\langle\theta_{\nu'}^{0}\big|\tilde{\Omega}_{\nu\nu'}.
\end{equation}
This requirement is automatically satisfied within the quasiparticle ansatz, but it fails in the presence of collisional broadening when three-phonon annihilation and creation processes are included. As detailed in the SM \cite{supplementary}, this breakdown originates from the impossibility of constructing a symmetric scattering matrix via the standard transformation \cite{hardy1970phonon,chaput2013direct,cepellotti2016thermal}
\begin{equation} \label{symmetric_matrix_main}
\tilde{{\rm{\Omega}}}_{\nu\nu'}=\frac{1}{\sqrt{\bar{n}_{\nu'}(\bar{n}_{\nu'}+1)}}A_{\nu\nu'}\frac{1}{\sqrt{\bar{n}_{\nu}(\bar{n}_{\nu}+1)}}.
\end{equation}
On the other hand, if three-phonon annihilation and creation processes are neglected—which, as anticipated above, is a safe approximation in the small-broadening limit (see Appendix \ref{section_small_broadening_annihilation_creation_three_phonons})—condition \eqref{left_right_eigenvalue} is restored and energy conservation follows from Eq. \eqref{global_scattering_energy_cons}.

It is worth mentioning that the effect of collisional broadening has been previously addressed in the context of electron transport, as discussed in Refs. \cite{reggiani1987quantum, aksamija2009energy, kim1987inclusion}. In particular, Reggiani et al. \cite{reggiani1987quantum} highlighted that the accumulation of broadening during simulations can lead to the introduction of unphysical energy into the electron population (see also Ref. \cite{ferrari2006introducing}). To mitigate this issue, the concept of non-accumulated broadening (NAB) was introduced in Ref. \cite{register2000improved}. This approach prevents quasiparticles from undergoing multiple broadening collisions, allowing them to only experience the effects of a single broadening event at a time. While this method successfully eliminates the unphysical drift toward high energies and resolves the runaway broadening problem, it also restricts the quasiparticles from fully exploring the impact of collisional broadening on their states. In contrast, Aksamija et al. \cite{aksamija2009energy} proposed dividing the Lorentzian distribution by the electronic density of states (DOS) to obtain an energy distribution that conserves energy “on average” across multiple scattering events. The Lorentzian distribution itself is symmetric around the energy-conserving point and would, in the absence of a DOS slope, uphold energy conservation on average. To neutralize the effect of the DOS, Aksamija et al. simply divided the Lorentzian distribution by the DOS curve, using the resulting distribution to select the broadened final states. Notably, a recent study \cite{lihm2024self} introduces a self-consistent framework for computing electron linewidths, effectively eliminating piezoelectric divergence and influencing scattering from polar optical phonons in both piezoelectric and nonpiezoelectric materials. This approach bridges the gap between theory and experimental angle-resolved photoemission measurements in InSe \cite{lihm2024non}. In the present work, we adopt a similar approach by utilizing a scattering matrix that conserves energy “on average”, as prescribed by Eq. \eqref{global_scattering_energy_cons}. 

\subsection{Thermal conductivity}

Within this framework, thermal conductivity is ultimately obtained by solving the linearized steady-state BTE and considering solutions that are linear in the temperature gradient. In this case, the time derivative in the driving term of Eq. \eqref{skeleton_transport_equation} vanishes, and the spatial derivative reduces to $\boldsymbol{v}_{\nu}\cdot\nabla T\,\partial\bar{n}_{\nu}(\boldsymbol{R},t)/\partial T$. Accordingly, to leading (harmonic) order, the heat flux reads \cite{hardy1963energy,simoncelli2022wigner}
\begin{equation}
Q=\frac{1}{\mathcal{V}}\sum_{\nu}\hbar\omega_{\nu}v_{\nu}n_{\nu}=\frac{1}{\mathcal{V}k_{{\rm{B}}}T^{2}}\sum_{\nu}\bar{n}_{\nu}(\bar{n}_{\nu}+1)\hbar\omega_{\nu}v_{\nu}f_{\nu}\nabla T.
\end{equation}
This equation can be directly compared with the definition of thermal conductivity $Q=-\kappa\nabla T$, obtaining
\begin{equation} \label{thermal_conductivity_definition_main}
\kappa=-\frac{1}{\mathcal{V}k_{{\rm{B}}}T^{2}}\sum_{\nu}\bar{n}_{\nu}(\bar{n}_{\nu}+1)\hbar\omega_{\nu}v_{\nu}f_{\nu}.
\end{equation}

\subsection{Self-consistent collisional broadening} \label{scf_broadening_Section_main}

The next step is to determine the appropriate broadening parameter $\gamma_{\nu}$ to be used in the Lorentzian distributions and how to compute it. We identify three possible approaches. The natural starting point is the quasiparticle ansatz, where the starting spectral function is represented by a Dirac delta function, meaning no broadening is included. This yields the standard LBTE, which is widely used in computational implementations \cite{paulatto2013anharmonic,phono3py,cepellotti2022phoebe,ShengBTE_2014}, where, in practice, the Dirac delta functions are commonly approximated with Gaussian smearing. However, the LBTE solution obtained using this approach is often affected by convergence issues of the thermal conductivity with respect to the smearing parameter, particularly in highly conducting materials. Moreover, in the small-smearing (Dirac delta-like) limit, it leads to artificially overdamped phonon lifetimes in two-dimensional crystals. 

\begin{figure}[!htb]
\centering
\includegraphics[width=0.5\textwidth]{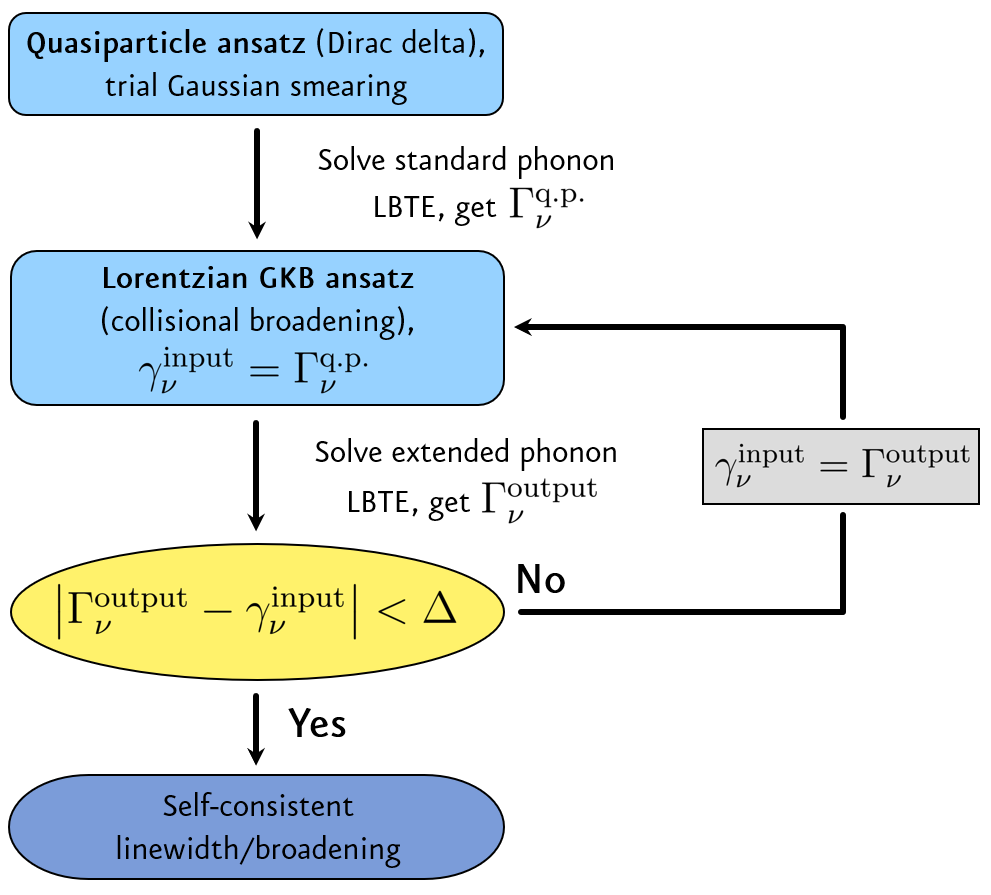}
\caption{Computational protocol for evaluating the self-consistent collisional broadening or phonon linewidths. $\Gamma_{\nu}$s are the computed linewidths, $\gamma_{\nu}$s are collisional broadening of Lorentzian spectral functions, while $\Delta$ represents convergence threshold for each phonon $\nu=\boldsymbol{q}s$.}
\label{fig:scf_broadening}
\end{figure}

In this work, we instead propose to compute the Lorentzian collisional broadening $\gamma_{\nu}$ through a physically rigorous approach based on its self-consistent determination. In this way, the broadening originates solely from physical scattering processes rather than numerical artifacts. The importance of such a self-consistent treatment of collisional broadening has been emphasized in previous studies \cite{gu2019revisiting,turney2009predicting}. Self-consistency is dictated by the structure of the spectral function. As anticipated above (see the discussion preceding Eqs. \eqref{KB_equation_ongoing} and \eqref{lorentzian_spectral_function}), the spectral function derived within the quasiparticle ansatz (Eq. \eqref{quasiparticle_ansatz}) already exhibits a Lorentzian shape. Moreover, this functional form is preserved when starting from a Lorentzian GKB ansatz (Eq. \eqref{GKB_ansatz_equation} with Lorentzian $d_{\nu}$). Therefore, it is natural to iterate the procedure by progressively updating the linewidths obtained from the solution of the transport equation and reintroducing them as the broadening parameters of the spectral function in the initial ansatz.
It can indeed be shown that $\Gamma_{\nu}=\gamma_{\nu}$ (see the SM \cite{supplementary}), which defines the first step of a self-consistent loop. This iterative scheme guarantees convergence of the collisional broadening parameter and is schematically illustrated in Fig. \ref{fig:scf_broadening}. 
The goal is to demonstrate that, although an initial numerical smearing must always be chosen to implement the Dirac delta functions of the quasiparticle ansatz, the iterative procedure outlined in Fig. \ref{fig:scf_broadening} ensures the self-consistency of broadening/linewidths. In other words, by the final iteration, the results no longer depend on the specific (numerical) Gaussian or Lorentzian smearing initially selected. 

Finally, it is clear that going beyond the quasiparticle ansatz corresponds to advancing to a higher level in the hierarchical scheme. The present framework can be naturally extended to include frequency lineshifts. For consistency, however, once lineshifts are incorporated, the spectral function cannot be retained at the lowest hierarchical level (the Dirac-delta form); at least the next-order Lorentzian approximation must be adopted. Indeed, both broadening and lineshift represent higher-order corrections in the hierarchy, since $\Delta_{\nu}^{\text{q.p.}}$ (the frequency shift) and $\Gamma_{\nu}^{\text{q.p.}}$ (the quasiparticle broadening) are obtained by solving the standard LBTE starting from the quasiparticle ansatz. Moreover, as discussed by Bonitz for electrons (see Eq. 7.12 of Ref. \cite{bonitz2016quantum}), broadening and lineshifts jointly account for the influence of the surrounding quasiparticle medium on a given phonon and should therefore be treated consistently within the same hierarchical level.

\subsubsection{Comparison with adaptive smearing}

At this point, it is natural to ask under which conditions the renormalization of phonon lifetimes induced by collisional broadening becomes significant, and whether it can exceed the corrections already captured by established approaches such as the adaptive smearing scheme introduced in, e.g., Refs. \cite{ShengBTE_2014,li2012thermal}. To clarify this issue, we define, for simplicity, $\epsilon_{\nu'\nu''}=\pm\omega_{\nu'}+\omega_{\nu''}$. The corresponding energy mismatch in the phonon dispersion arising from a small variation of the wavevector can then be written as \cite{li2012thermal}:
\begin{equation} \label{energy_mismatch}
\Delta\epsilon_{\nu'\nu''}=\left|\frac{\partial\Omega}{\partial \boldsymbol{Q}}\right|\Delta\boldsymbol{Q}=|\boldsymbol{v}_{\nu'}-\boldsymbol{v}_{\nu''}|\,\Delta\boldsymbol{Q},
\end{equation}
where $\Delta\boldsymbol{Q}$ is simply the spacing of the sampling $\boldsymbol{q}$ points in the Brillouin zone. Eq. \eqref{energy_mismatch} thus scales linearly with the phonon group velocity, and the adaptive smearing is chosen to be of the same order as this dispersion-induced energy mismatch \cite{ShengBTE_2014,li2012thermal}. To compare such energy mismatch with the collisional broadening, we introduce an integrated descriptor that captures the average velocity of the phonons contributing to a given linewidth $\Gamma_{\nu}$. We refer to this quantity as the phonon dispersion energy scale, defined as
\begin{equation} \label{dispersion_energy_scale}
W_{\nu}=\frac{1}{\mathcal{N}_{\nu}}\sum_{\nu'\nu''}|\boldsymbol{v}_{\nu'}-\boldsymbol{v}_{\nu''}|\,\Delta\boldsymbol{Q},
\end{equation}
where $\mathcal{N}_{\nu}$ is the number of $(\nu',\nu'')$ pairs included for the mode $\nu$, i.e. the cardinality of the set over which the sum is performed (in practice, the number of allowed triplets/pairs contributing to $\Gamma_{\nu}$). An analogous definition applies to contributions from other types of scattering channels. When the dispersion energy scale \eqref{dispersion_energy_scale} becomes large and comparable to the broadening scale adaptive smearing has been proved to be useful \cite{ShengBTE_2014} (see e.g. also the supplementary information of Ref. \cite{cheng2020experimental}). 

In detail, there are several reasons why the self-consistent collisional broadening surpasses the adaptive smearing technique. First, it does not rely on computational parameters, as it can be determined self-consistently and is therefore physically driven, as dictated by the hierarchy of ansätze. Second, adaptive smearing is constructed from purely harmonic quantities, namely phonon frequencies and group velocities, whereas collisional broadening is iteratively renormalized by anharmonic phonon linewidths. Moreover, collisional broadening is mode-resolved, so that no single global parameter is applied uniformly to all phonon modes, as is instead the case for purely numerical Gaussian smearing schemes commonly implemented in phonon BTE solvers available in the literature. This last aspect has an additional and crucial consequence: in contrast to adaptive smearing, collisional broadening does not break the symmetry of the scattering matrix. In fact, adaptive smearing depends explicitly on the phonon wave vectors $\nu'$ and $\nu''$. By contrast, the self-consistent collisional broadening is always defined as a convolution over the three phonon modes involved in the scattering process, $\upgamma_{\nu\nu'\nu''}=\gamma_{\nu}+\gamma_{\nu'}+\gamma_{\nu''}$ (see Eq. \eqref{gamma_as_convolution}). This distinction has important implications for the symmetrization of the scattering matrix, as discussed in detail in Appendix \ref{symmetrization_with_adaptive_smearing}, and, consequently, for global energy conservation. In particular, one can show that the transformation in Eq. \eqref{symmetric_matrix_main} successfully symmetrizes the scattering matrix when collisional broadening is employed, precisely because the total broadening satisfies $\upgamma_{\nu\nu'\nu''}=\upgamma_{\nu'\nu\nu''}$ (this holds true also for constant numerical smearing schemes that do not depend on phonon wave vectors, although, as discussed above, such approaches introduce other significant drawbacks). In contrast, when adaptive smearing is used, the extension of the detailed-balance relations for phonon populations and frequencies (see Eq. \eqref{extended_detailed_balance_relations_1_adaptive}) breaks this symmetry. Specifically, in the evaluation of the matrix element $\tilde{\Omega}_{\nu'\nu}$, adaptive smearing relaxes the strict energy conservation enforced by the Dirac delta function and replaces it with a smeared representation that depends on combinations of phonon group velocities $\boldsymbol{v}_{\nu'}$ and $\boldsymbol{v}_{\nu''}$. These combinations are not invariant under the exchange of $\nu$ and $\nu'$, and therefore the transformation in Eq. \eqref{symmetric_matrix_main} no longer guarantees symmetrization. As a result, adaptive smearing fails to preserve the symmetry of the scattering matrix. Consequently, a well-defined Bose-Einstein eigenvector no longer exists, total energy conservation is not guaranteed, and the temperature and thermal conductivity become ill defined.

\subsection{Analytical solution to overdamped phonon lifetimes in 2D systems via collisional broadening} \label{overdamped_phonon_section_main}

As shown in Ref. \cite{bonini2012acoustic}, both numerical and analytical analyses indicate that, in the long-wavelength limit of two-dimensional systems, absorption processes are significantly weaker than decay processes. In particular, the absorption contribution decreases linearly with the crystal momentum $\boldsymbol{q}$ and vanishes as $\boldsymbol{q}\to0$. By contrast, the long-wavelength scaling of decay processes involving quadratic flexural out-of-plane (ZA) modes remains finite in the $\boldsymbol{q}\to0$ limit, leading to overdamped phonon lifetimes. In the following, we therefore focus on the dominant three-phonon Normal decay channel in which an in-plane longitudinal acoustic (LA) or transverse acoustic (TA) mode (initial state, $i$) decays into two lower-energy flexural acoustic (ZA) modes (final state, $f$), i.e., LA (TA) $\to$ ZA + ZA, in a generic two-dimensional crystal. The extended scattering rate of an in-plane acoustic mode can be written as
\begin{equation} \label{scattering_rate_main}
\tau_{i}(\boldsymbol{q}_{i})^{-1}\propto\sum_{\boldsymbol{q}_{f}}F(\boldsymbol{q}_{i},\boldsymbol{q}_{f})\,\mathrm{d}\!\left(\Delta\omega(\boldsymbol{q}_{i},\boldsymbol{q}_{f})\right),
\end{equation}
where
\begin{equation}
\Delta\omega(\boldsymbol{q}_{i},\boldsymbol{q}_{f})=\omega_{i}(\boldsymbol{q}_{i})-\omega_{f}(\boldsymbol{q}_{f})-\omega_{f}(\boldsymbol{q}_{i}+\boldsymbol{q}_{f}),
\end{equation}
and
\begin{equation}
F(\boldsymbol{q}_{i},\boldsymbol{q}_{f})=|\uppsi_{\boldsymbol{q}_{i}\boldsymbol{q}_{f}}|^{2}\Big[n\!\big(\omega_{f}(\boldsymbol{q}_{f})\big)+n\!\big(\omega_{f}(\boldsymbol{q}_{i}+\boldsymbol{q}_{f})\big)+1\Big].
\end{equation}
In the long-wavelength limit of a 2D system, the phonon dispersions behave as $\omega_{\text{LA/TA}} \sim A q$ and $\omega_{\text{ZA}} \sim B q^{2}$. Imposing exact energy conservation within the quasiparticle ansatz, i.e. $\mathrm{d}(\Delta\omega(\boldsymbol{q}_{i},\boldsymbol{q}_{f}))\equiv\delta(\Delta\omega(\boldsymbol{q}_{i},\boldsymbol{q}_{f}))$ and $\Delta\omega(q_i,q_f,\theta)=0$, one obtains the following scaling for the scattering rate \cite{bonini2012acoustic} (see also the SM \cite{supplementary}):
\begin{equation} \label{bonini_res}
\tau_i(\boldsymbol q_i)^{-1}\sim\text{const.}\qquad q_i\to0.
\end{equation}
Therefore, the in-plane acoustic linewidth remains finite
in the long-wavelength limit, while $\omega_{i}(\boldsymbol q_i)\sim q_i\to0$. The Landau criterion \cite{landau1959theory,landau1957theory} for having a well-defined quasiparticle, $\tau_{\nu}^{-1}<\omega_{\nu}$, is violated, and the acoustic modes become overdamped. A detailed analytical derivation of the asymptotic behavior of the individual contributions entering the scattering rate in Eq. \eqref{scattering_rate_main}, within the quasiparticle ansatz and in the long-wavelength limit, is given in Ref. \cite{bonini2012acoustic} and is revised in the SM \cite{supplementary}.

Going beyond the quasiparticle ansatz, we replace the Dirac delta with a Lorentzian, which in the small-broadening limit can be written as
\begin{equation}
L(\Delta\omega)\simeq\frac{1}{\nabla_{q_f}\Delta\omega\big|_{q_{f,0}}}\frac{1}{\pi}\frac{\frac{\gamma}{\nabla_{q_f}\Delta\omega\big|_{q_{f,0}}}}{(q_f-q_{f,0})^{2}+\Big(\frac{\gamma}{\nabla_{q_f}\Delta\omega\big|_{q_{f,0}}}\Big)^{2}}.
\end{equation}
Performing the long-wavelength scaling analysis in the joint limit $q_i,q_f\to0$ with $q_i<q_f$, we find
\begin{equation}
\tau_i(\boldsymbol q_i)^{-1}\sim2\pi\sqrt{q_i}\int dq_f\,\frac{1}{q_f}L(q_f).
\end{equation}
Since the Lorentzian is peaked at $q_f\sim\sqrt{q_i}$,
the integral scales logarithmically,
\begin{equation}
\tau_i(\boldsymbol q_i)^{-1}\sim2\frac{q_{i}}{\gamma}\ln\left(\frac{\gamma}{q_{i}}\right).
\end{equation}
The constant nature of $\gamma$ at this stage is a direct consequence of the quasiparticle ansatz. Indeed, this expression already shows that, if one uses the linewidth obtained from the previous iteration (i.e., iteration zero corresponding to the quasiparticle ansatz) the resulting phonon lifetime for the scattering process under consideration is not overdamped:
\begin{equation} \label{lorentzian_lifetime_vanishing_limit}
\tau_i(\boldsymbol q_i)^{-1}\to0,\qquad q_i\to0.
\end{equation}
A detailed analytical derivation is presented in Appendix \ref{solution_overdamped_section}, where we further show that the vanishing limit in Eq. \eqref{lorentzian_lifetime_vanishing_limit} is systematically recovered within the fully self-consistent collisional broadening scheme. In particular, $\tau_i(\boldsymbol q_i)^{-1}$ converges exactly to zero in the long-wavelength limit at self-consistency, even in cases where the Lorentzian broadening at the first iteration may not be numerically very sharp and approaches a Dirac delta-like form.

\section{Results} \label{result_section}

To demonstrate the validity and robustness of the transport theory developed in this work—from the full quantum KBE to the resulting extended BTE beyond the quasiparticle regime, including self-consistent collisional broadening—we apply the method to two markedly different systems: the two-dimensional monolayer $\alpha$-GeSe, a thermal insulator, and three-dimensional bulk diamond, a prototypical thermal conductor.

Monolayer $\alpha$-GeSe belongs to the family of orthorhombic group IV–VI compounds (GeS, GeSe, SnS, SnSe), which in bulk form possess a puckered (hinge-like) layered structure similar to black phosphorus \cite{zhao2014ultralow,zhao2016ultrahigh,parenteau1990influence,guo2015first,carrete2014low,shi2015quasiparticle}. These materials are particularly attractive due to their earth-abundance, low toxicity, chemical stability, and environmental compatibility, making them promising candidates for large-scale applications in photovoltaics and thermoelectrics \cite{zhao2014ultralow,zhao2016ultrahigh,chen2014thermoelectric,tan2014thermoelectrics,zhu2014band}. The isolation of two-dimensional materials following the discovery of graphene has stimulated intense interest in 2D analogues of these compounds. In particular, monolayer SnSe has been experimentally synthesized \cite{li2013single,antunez2011tin} and predicted to exhibit promising optoelectronic, piezoelectric, and thermoelectric properties \cite{fei2015giant,wang2015thermoelectric,zhang2016tinselenidene,qin2016diverse}. 

\begin{figure}[!htb]
\centering
\includegraphics[width=0.48\textwidth]{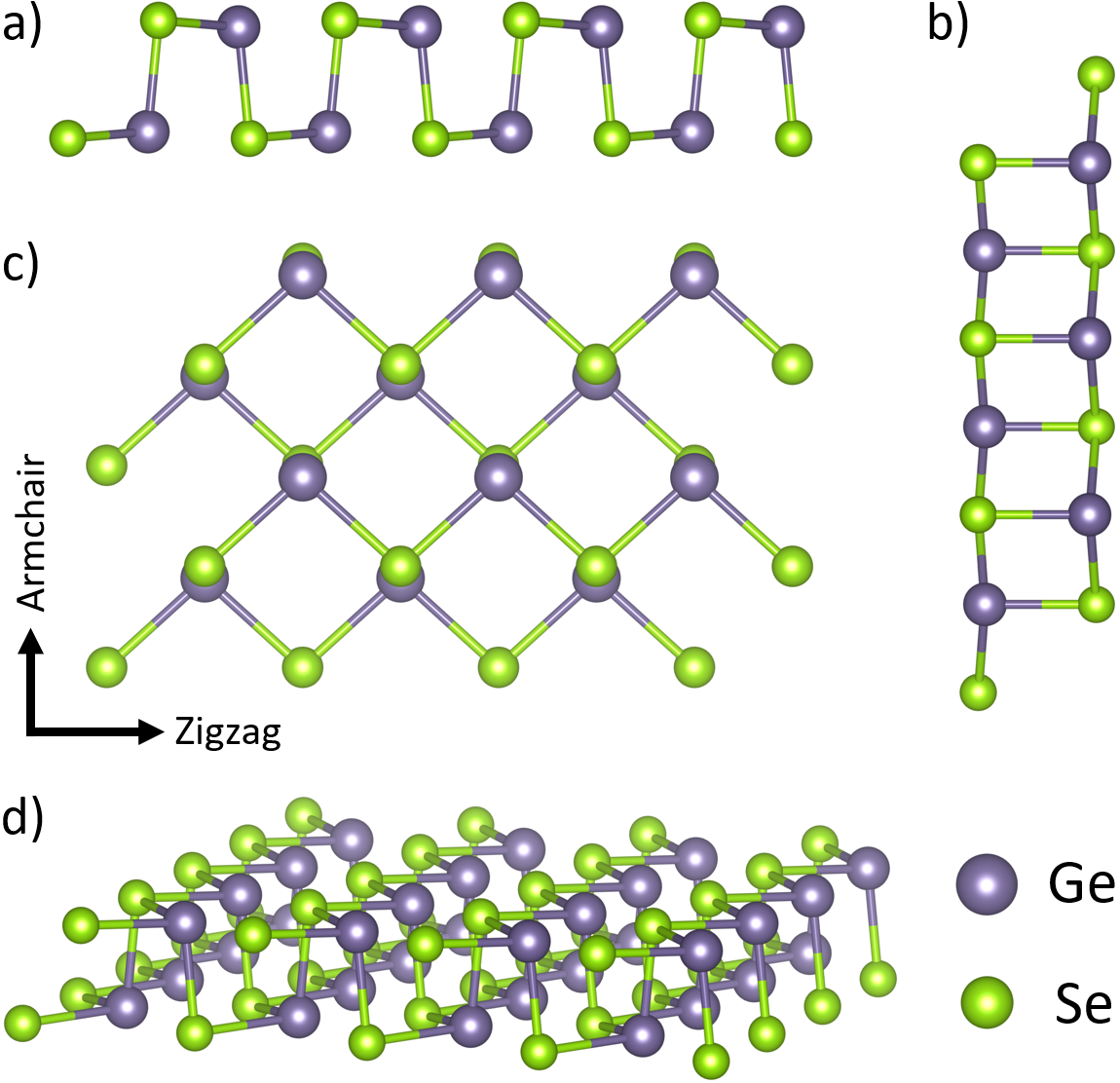}
\caption{Crystal structure of orthorhombic monolayer $\alpha$-GeSe. (a,b) Side views highlighting the puckered (hinge-like) geometry characteristic of group IV–VI compounds. (c) Top view with the in-plane zigzag and armchair directions indicated; the lengths of the arrows are drawn proportional to the relative anisotropy of the lattice thermal conductivity along the two directions. (d) Perspective view of monolayer $\alpha$-GeSe. Ge atoms are shown in lavender and Se atoms in lime green.}
\label{fig:crystal_structure_GeSe}
\end{figure}

From the thermal transport viewpoint, monolayer $\alpha$-GeSe is particularly compelling because it combines intrinsically low lattice thermal conductivity with pronounced in-plane anisotropy \cite{qin2016diverse}. In addition, it belongs to a class of two-dimensional materials that is fundamentally different from hexagonal lattice systems such as graphene or most 2D transition-metal dichalcogenides (TMDs), which crystallize in honeycomb-derived or trigonal prismatic/1T structures. In contrast, $\alpha$-GeSe stabilizes in a puckered (buckled) orthorhombic structure (see Fig. \ref{fig:crystal_structure_GeSe}).
This structural difference has important consequences for phonon transport. In graphene, reflection symmetry with respect to the basal plane imposes strict selection rules on three-phonon processes involving flexural (ZA) modes. In particular, scattering events containing an odd number of out-of-plane acoustic phonons are symmetry-forbidden. This suppression of large classes of three-phonon channels leads to unusually weak intrinsic scattering of flexural modes and makes higher-order processes, such as four-phonon scattering, comparatively more relevant. Monolayer $\alpha$-GeSe, by contrast, lacks this mirror symmetry due to its puckered structure. As a consequence, no analogous symmetry-protected suppression of three-phonon scattering channels occurs, and flexural modes are allowed to couple efficiently with in-plane modes already at the three-phonon level. Moreover, in the orthorhombic $\alpha$-phase considered here, the phonon dispersion does not exhibit a significant phonon band gap separating acoustic and optical branches. The absence of such a gap enhances the available three-phonon phase space and prevents the reduction of scattering channels that would otherwise arise from energy-conservation constraints \cite{feng2016quantum,feng2017four,feng2018four,feng2018phonon,yang2019stronger}. For these reasons, higher-order phonon scattering is not expected to play a dominant role in thermal transport for this phase \cite{feng2016quantum}. In contrast, in hexagonal 2D materials \cite{sun2023four} and in several 2D transition-metal dichalcogenides \cite{chaudhuri2024understanding,zhang2022four,guo2024four,zhang2015phonon,tang2023strong}, four-phonon scattering has been shown to provide a non-negligible, and sometimes essential, correction to thermal conductivity due to the reduced third-order anharmonic phase space. Further confirmation comes from comparative studies of different structural phases of GeSe.
\begin{figure}[!htb]
\centering
\includegraphics[width=0.48\textwidth]{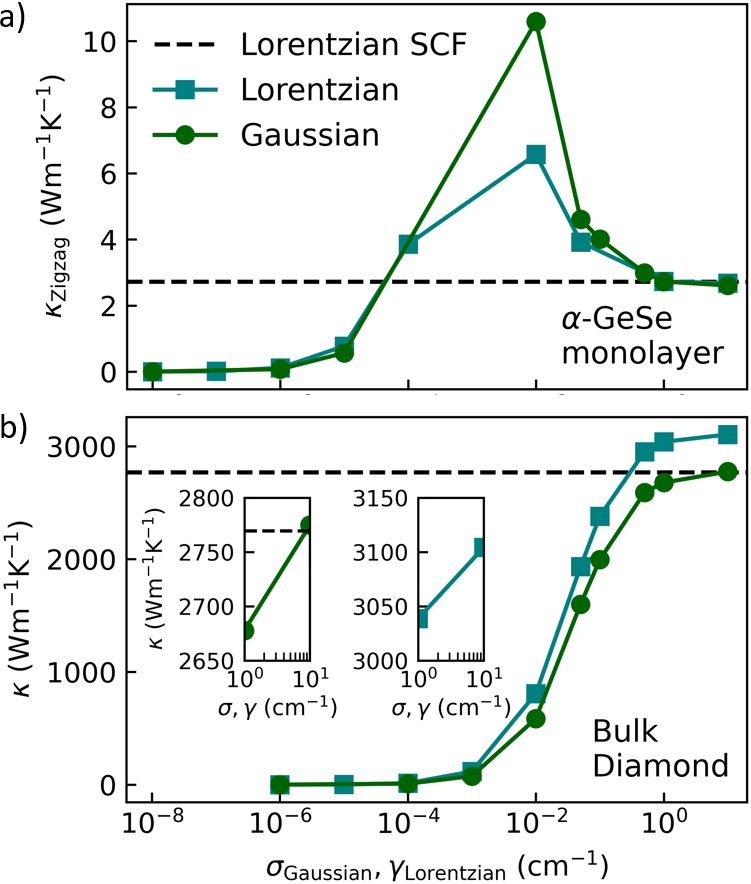}
\caption{Room-temperature lattice thermal conductivity as a function of the smearing parameter for Gaussian (green) and Lorentzian (teal) numerical broadening schemes. (a) In-plane lattice thermal conductivity along the zigzag direction of $\alpha$-GeSe monolayer. Very small smearing values lead to strongly overdamped phonon lifetimes and, consequently, to unphysically low thermal conductivity. Even within the commonly used smearing range of $10^{-1}$--$10^{0}$ cm$^{-1}$, a variation of approximately 32\% is observed. (b) Lattice thermal conductivity of bulk diamond. Both the overall trend and the zoomed inset, focusing on the smearing range $10^{0}$--$10^{1}$ cm$^{-1}$, demonstrate that convergence with respect to the smearing parameter is never achieved. The result obtained with the present Lorentzian self-consistent collisional broadening (SCF) approach is shown as a black dashed line and is independent of the smearing.}
\label{fig:kappa_vs_smearing}
\end{figure}
A significant contribution from four-phonon scattering has been reported only for the $\gamma$-phase monolayer of GeSe, which—differently from the orthorhombic $\alpha$-phase investigated here and the $\beta$-phase—exhibits a phonon band gap of approximately 25 cm$^{-1}$ \cite{wang2022first,liu2018first}. The presence of this gap reduces the efficiency of certain three-phonon channels, thereby enhancing the relative importance of four-phonon processes. In the $\alpha$-phase considered in the present work, however, no phonon band gap is present, and four-phonon scattering has been reported to be essentially negligible \cite{wang2022first}. Moreover, since the primitive unit cell contains only four atoms, the phonon band structure consists of just twelve branches (see the SM \cite{supplementary}). As a consequence, heat transport is entirely particle-like, and the wave-like (coherent) contribution to thermal conduction within the unified Wigner formalism \cite{simoncelli2019unified,simoncelli2021thermal,simoncelli2022wigner,di2023crossover} is fully negligible (as we have verified numerically, not shown). This makes monolayer $\alpha$-GeSe particularly interesting: it is an excellent thermal insulator, yet its low thermal conductivity does not originate from strong phonon wave-tunneling or coherence effects, which are commonly found in many other low-thermal-conductivity materials \cite{simoncelli2022wigner,di2023crossover}.

In Fig. \ref{fig:kappa_vs_smearing} we show the room-temperature lattice thermal conductivity evolution as a function of the numerical smearing parameter, both when the Dirac delta function enforcing exact energy conservation is approximated by a Lorentzian and when it is approximated by a Gaussian. Panel (a) reports the lattice thermal conductivity along the zigzag direction of monolayer $\alpha$-GeSe, while panel (b) shows the lattice thermal conductivity of bulk diamond. In both cases the conductivity does not converge with respect to the numerical smearing employed, and this lack of convergence is particularly evident and significant for diamond. Indeed, even within the commonly used smearing range from $10^{0}$ cm$^{-1}$ to $10^{1}$ cm$^{-1}$, the Lorentzian scheme yields values of 3038.67 W\,m$^{-1}$K$^{-1}$ and 3104.28 W\,m$^{-1}$K$^{-1}$, respectively, while the Gaussian scheme gives 2677.68 W\,m$^{-1}$K$^{-1}$ and 2774.88 W\,m$^{-1}$K$^{-1}$, as shown in the insets of Fig. \ref{fig:kappa_vs_smearing}(b). The maximum phonon frequency of monolayer $\alpha$-GeSe is approximately 240 cm$^{-1}$ (see Fig. \ref{fig:phonons_GeSe}), while in bulk diamond it reaches about 1400 cm$^{-1}$. Therefore, smearing values larger than 10 cm$^{-1}$ (the largest value considered in Fig. \ref{fig:kappa_vs_smearing}) could, in principle, represent reasonable choices to further test numerical convergence. However, as shown in Fig. \ref{fig:kappa_vs_smearing_diamond}, even when increasing the smearing up to 100 cm$^{-1}$ no convergence of the lattice thermal conductivity is achieved. Instead, after reaching a maximum around 10 cm$^{-1}$, the conductivity starts to decrease, highlighting the intrinsic limitations of the standard smearing-based approach (see also Ref. \cite{fugallo2013ab}, where the same trend is clearly observed). For both materials we also report the black dashed horizontal line corresponding to our Lorentzian self-consistent collisional broadening (SCF) result, which is independent of the smearing parameter and resolves the convergence issues associated with the standard approach (see also Fig. \ref{fig:kappa_convergence}). Within this method we obtain $\kappa_{\rm{Zigzag}} = 2.73$ W\,m$^{-1}$K$^{-1}$ and $\kappa_{\rm{Armchair}} = 2.22$ W\,m$^{-1}$K$^{-1}$ for monolayer $\alpha$-GeSe, and $\kappa = 2769.523$ W\,m$^{-1}$K$^{-1}$ for diamond at room temperature. As evident, monolayer $\alpha$-GeSe exhibits in-plane anisotropy in the lattice thermal conductivity, although reduced compared to results obtained from models solving the standard BTE \cite{qin2016diverse}. We note that in Ref. \cite{qin2016diverse} the smearing value used in the simulations performed with the \texttt{ShengBTE} code is not specified, however, the values reported therein for monolayer $\alpha$-GeSe are $\kappa_{\rm{Zigzag}} = 5.89$ W\,m$^{-1}$K$^{-1}$ and $\kappa_{\rm{Armchair}} = 4.57$ W\,m$^{-1}$K$^{-1}$, which are consistent with those we obtain within the smearing range $0.01$--$1$ cm$^{-1}$. 

\begin{figure}[!htb]
\centering
\includegraphics[width=0.48\textwidth]{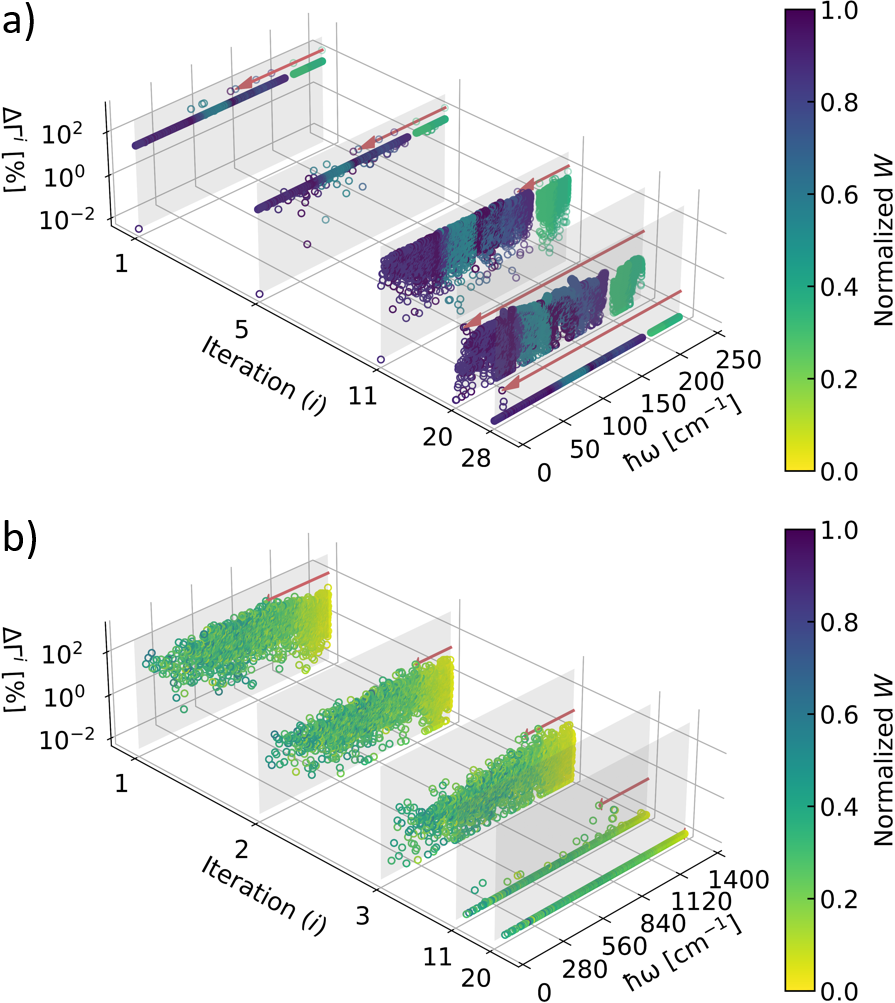}
\caption{Difference between consecutive iterations of the room-temperature Lorentzian phonon broadening (linewidths) for monolayer $\alpha$-GeSe (a) and bulk diamond (b). The results are obtained following the self-consistent procedure outlined in Fig. \ref{fig:scf_broadening}, starting from an initial Lorentzian numerical smearing of $10^{-8}$ cm$^{-1}$ for monolayer $\alpha$-GeSe and $8$ cm$^{-1}$ for bulk diamond. Owing to these markedly different initial smearings, the linewidths of bulk diamond clearly exhibit a faster convergence behavior. The relative percentage difference in linewidths between consecutive iterations, defined as $\Delta\Gamma_{\nu}^{i}$ = $\left|\Gamma_{\nu}^{i}-\Gamma_{\nu}^{i-1}\right|/\Gamma_{\nu}^{i-1}$, is shown for selected iteration pairs: 1--0 (first), 5--4, 11--10, 20--19, and 28--27 (last) for monolayer $\alpha$-GeSe, and 1--0 (first), 2--1, 3--2, 11--10, and 20--19 (last) for bulk diamond, clearly demonstrating the achievement of self-consistency. Importantly, the largest impact of collisional broadening (the largest value of $\Delta\Gamma_{\nu}^{i}$, highlighted by the red arrows) does not systematically occur in the acoustic phonon branches, despite their higher group velocities. This demonstrates that there is no strict correlation with the phonon energy scale $W_{\nu}$ used in the color bar. Instead, the self-consistent collisional broadening captures anharmonic effects beyond those described such adaptive smearing schemes. For visualization, $W_{\nu}$ is linearly normalized to the interval $[0,1]$ as $(W_{\nu}-W_{\nu}^{\mathrm{min}})/(W_{\nu}^{\mathrm{max}}-W_{\nu}^{\mathrm{min}})$. As self-consistency is approached, $\Delta\Gamma_{\nu}^{i}$$\to$ 0, reflecting the convergence of the phonon linewidths and, correspondingly, of the lattice thermal conductivity.}
\label{fig:3D_percent_gamma_diff}
\end{figure}

\begin{figure}[!htb]
\centering
\includegraphics[width=0.48\textwidth]{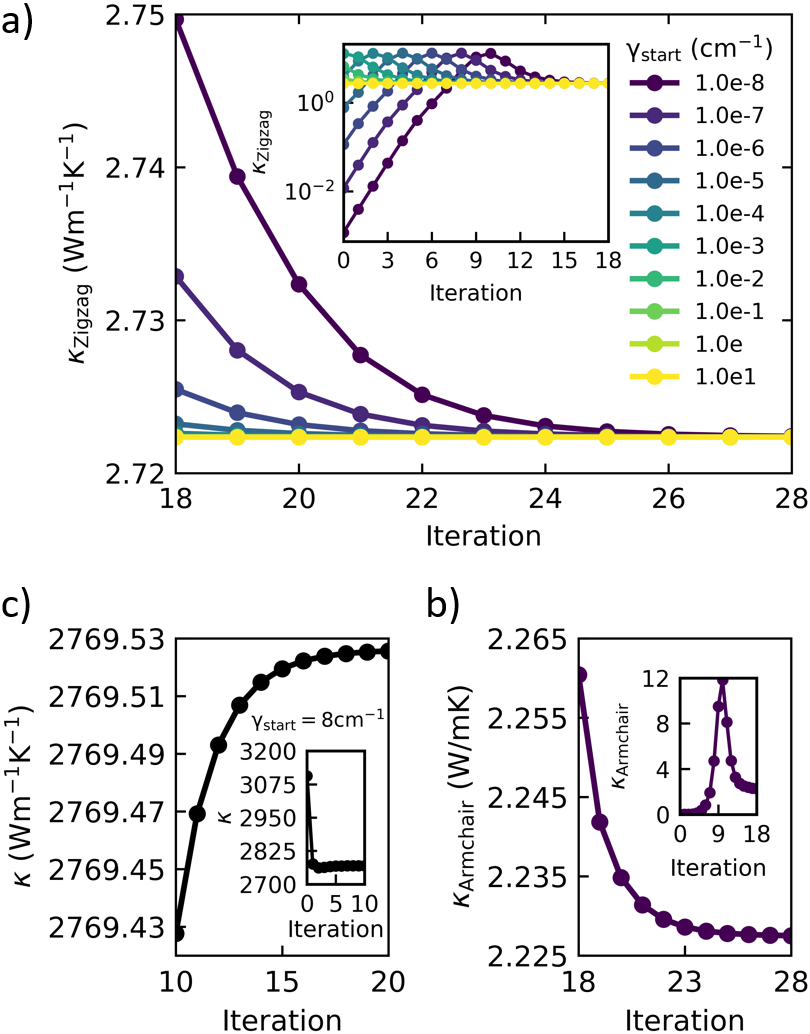}
\caption{Self-consistent iterations of the room-temperature lattice thermal conductivity of monolayer $\alpha$-GeSe (zigzag and armchair directions in panels a and b, respectively) and bulk diamond (c). For the zigzag lattice thermal conductivity of monolayer $\alpha$-GeSe, we report the self-consistent evolution starting from different initial smearing values, namely the same ones sampled in Fig. \ref{fig:kappa_vs_smearing}a, in order to explicitly demonstrate that the final self-consistent conductivity is independent of the initial smearing and therefore free from any computational smearing bias. The armchair direction of monolayer $\alpha$-GeSe exhibits an analogous behavior; therefore, we only show the representative case corresponding to $\gamma_{\rm{start}} = 10^{-8}$ cm$^{-1}$. For bulk diamond, we adopt $\gamma_{\rm{start}} = 8$ cm$^{-1}$, leading to a faster convergence because the initial value is physically more realistic. The convergence of the lattice thermal conductivity clearly mirrors the convergence behavior of the linewidths shown in Fig. \ref{fig:3D_percent_gamma_diff}.}
\label{fig:kappa_convergence}
\end{figure}

In Fig. \ref{fig:3D_percent_gamma_diff}a and Fig. \ref{fig:3D_percent_gamma_diff}b we show the variation of the phonon linewidths/broadening across selected consecutive iterations of the self-consistent cycle reported in Fig. \ref{fig:scf_broadening}, for monolayer $\alpha$-GeSe and bulk diamond, respectively. The initial Lorentzian numerical smearing used to illustrate the self-consistency is set to $10^{-8}$ cm$^{-1}$ in order to emphasize that, independently of how far the starting point is from the final solution, self-consistency is always achieved (see also Fig. \ref{fig:kappa_convergence}).
Indeed, the progressive reduction of the difference between successive iterations demonstrates the convergence of the linewidths to a self-consistent solution, thereby removing any residual dependence on computational parameters. Consequently, the resulting thermal conductivity is fully parameter-free and consistently incorporates anharmonic collisional broadening effects. Moreover, isotopic scattering has been included by accounting for the natural isotopic abundances and implementing the corresponding scattering channel within the Lorentzian broadening framework, thus ensuring full methodological consistency.

Fig. \ref{fig:kappa_convergence} shows the evolution of the room-temperature lattice thermal conductivity during the self-consistent (SCF) cycle. For monolayer $\alpha$-GeSe (zigzag direction), different initial Lorentzian smearings are considered to explicitly demonstrate that the iterative procedure converges to a unique, smearing-independent solution. Regardless of how far the starting value is from the final result, the conductivity systematically approaches the same fixed point, confirming that the SCF scheme removes any residual dependence on the numerical broadening.
The armchair direction exhibits an analogous behavior, and a representative case is shown. For bulk diamond, starting from a physically more realistic initial smearing leads to a faster convergence rate, consistent with the smaller deviations from the final solution in the early iterations. In all cases, the convergence of the thermal conductivity closely follows that of the phonon linewidths (Fig. \ref{fig:3D_percent_gamma_diff}), demonstrating the internal consistency of the approach and the robustness of the resulting parameter-free conductivity. 

\begin{figure}[!htb]
\centering
\includegraphics[width=0.48\textwidth]{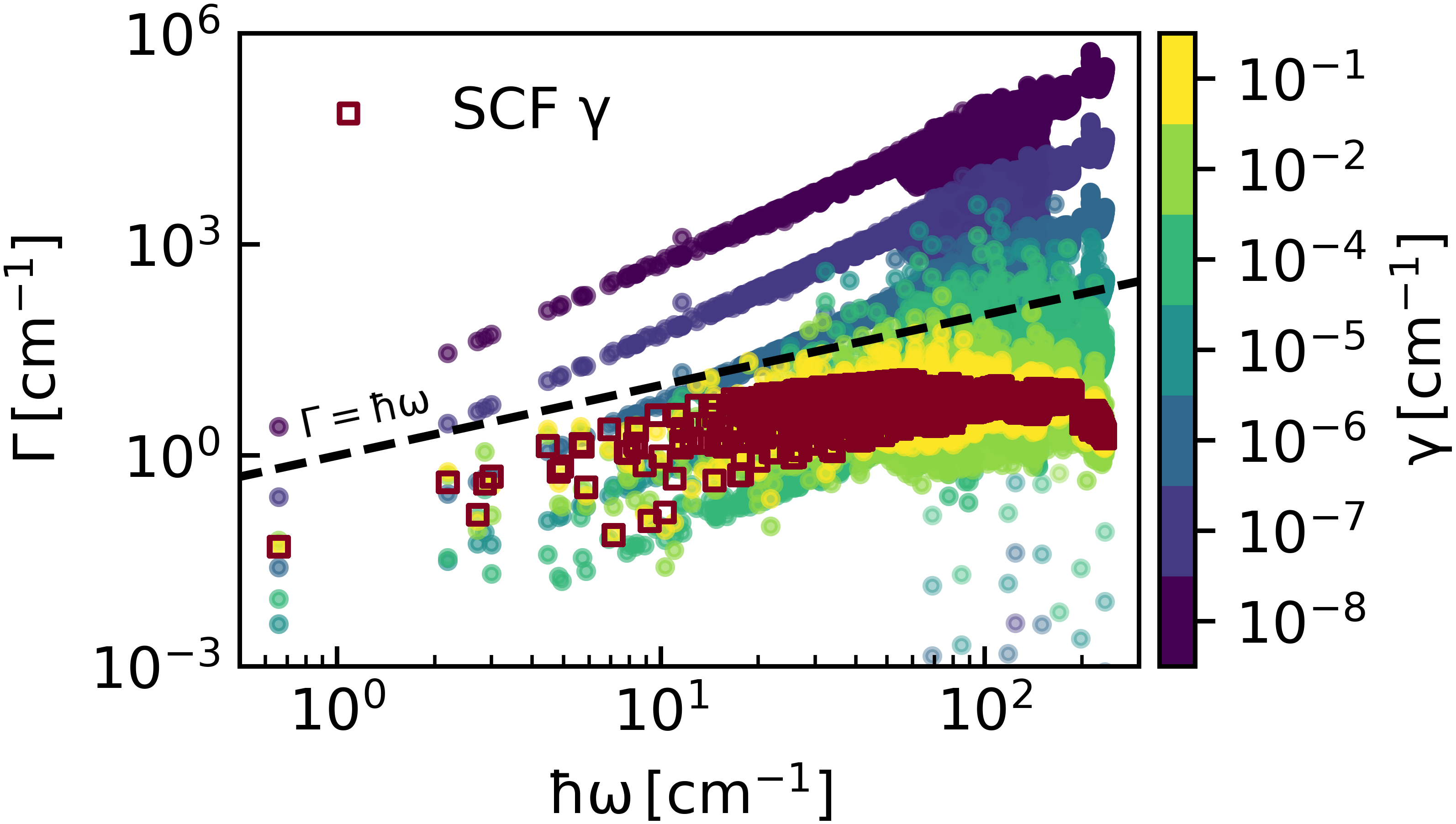}
\caption{Comparison between phonon linewidths of monolayer $\alpha$-GeSe calculated using a numerical Lorentzian smearing $\gamma$ (full circles) and those obtained within the self-consistent Lorentzian collisional broadening scheme (SCF $\gamma$, red open squares). The dashed black line indicates the Landau quasiparticle criterion, $\Gamma = \hbar\omega$, above which phonon modes become overdamped. As the numerical smearing is progressively reduced—thereby improving the approximation of the Dirac delta function enforcing energy conservation—the calculated linewidths increase and eventually cross the threshold set by the Landau criterion, leading to unphysically overdamped phonon lifetimes. In contrast, the self-consistent Lorentzian collisional broadening remains systematically below the dashed black line, ensuring that phonons are never overdamped, in full agreement with the analytical result discussed in Section \ref{overdamped_phonon_section_main}.}
\label{fig:Gamma_vs_omega_lorentzian}
\end{figure}

Finally, the microscopic origin of the smearing dependence observed in Fig. \ref{fig:kappa_vs_smearing} becomes evident when analyzing the phonon linewidths. Fig. \ref{fig:Gamma_vs_omega_lorentzian} directly compares the linewidths obtained with a standard numerical Lorentzian smearing and those resulting from the self-consistent Lorentzian collisional broadening scheme. As the numerical smearing is reduced in an attempt to better approximate the Dirac delta function enforcing exact energy conservation, the computed linewidths increase substantially. In two-dimensional systems such as monolayer $\alpha$-GeSe, this effect is amplified by the presence of flexural acoustic branches, which are especially sensitive to the sharp spectral constraint imposed by the delta function. As a consequence, several modes cross the threshold set by the Landau criterion, $\Gamma=\hbar\omega$, entering an overdamped regime that is physically inconsistent with a well-defined phonon excitation picture. This overdamping directly explains the anomalous suppression of the lattice thermal conductivity at small smearings reported in Fig. \ref{fig:kappa_vs_smearing}. The artificial broadening induced by the Dirac delta function approximation enhances scattering rates and leads to unphysically short lifetimes, thereby reducing the conductivity. In contrast, the self-consistent collisional broadening scheme systematically yields linewidths that remain below the Landau quasiparticle criterion (open red squares in Fig. \ref{fig:Gamma_vs_omega_lorentzian}), thereby preventing overdamping and restoring a physically meaningful description of phonon transport. Consistently, as shown in Fig. \ref{fig:kappa_convergence}, the corresponding lattice thermal conductivity converges toward a unique fixed point that is independent of the initial smearing. Although certain empirically chosen smearing values may occasionally yield numerically reasonable conductivities, such agreement is accidental and the results remain strongly dependent on this computational parameter. The present approach instead provides a computationally inexpensive and physically grounded strategy that removes this arbitrariness, ensuring that the predicted transport properties are free from ad hoc numerical parametrizations and dictated solely by the intrinsic anharmonic physics of the system.
Within a Dirac delta (quasiparticle) description of three-phonon processes involving flexural ZA modes, the linewidth of in-plane acoustic phonons approaches a finite value as $q \to 0$, thereby violating the Landau quasiparticle criterion and producing overdamped excitations \cite{bonini2012acoustic} (see also the SM \cite{supplementary}). This result is strictly asymptotic—an essential requirement for obtaining a controlled analytical demonstration—and pertains to specific ZA-mediated channels in the $q \to 0$ limit. The behavior displayed in Fig. \ref{fig:Gamma_vs_omega_lorentzian} provides direct numerical support for this mechanism: as the smearing is reduced toward the Dirac delta limit, the linewidths systematically increase and cross the threshold set by the Landau quasiparticle criterion. While the overdamping observed at higher frequencies can arise from enhanced phase space and matrix elements at finite wavevectors, it reflects the same underlying mechanism—namely, that enforcing increasingly sharp $\delta$-like energy conservation artificially amplifies scattering rates and can drive modes beyond the Landau quasiparticle criterion. The analytical result thus identifies the fundamental source of the instability, whereas the numerical evidence provides a sufficient confirmation of the mechanism, showing that its impact is not confined to the asymptotic regime but propagates across a broad portion of the phonon spectrum. The self-consistent collisional broadening scheme eliminates this pathology at all frequencies, restoring $\Gamma < \hbar\omega$ and a physically well-defined quasiparticle picture. Taken together, Figs. \ref{fig:3D_percent_gamma_diff}, \ref{fig:kappa_convergence}, and \ref{fig:Gamma_vs_omega_lorentzian} provide a coherent picture: the self-consistent procedure removes the artificial smearing dependence at the level of the phonon linewidths, avoids unphysical overdamping of flexural modes, and yields a parameter-free thermal conductivity that consistently incorporates anharmonic collisional broadening.

\section{Conclusions}

We have introduced a formalism that bridges the quantum (Kadanoff-Baym equation) and semiclassical (Boltzmann transport equation) descriptions of phonon-mediated thermal transport, enabling a systematic transition between them while rigorously tracking the approximations performed at different levels of theory. This allows us to derive from first principles a linearized generalized Boltzmann transport equation (LGBTE) that accounts for collisional broadening, overcoming the universal failure of the celebrated linearized Boltzmann transport equation in two-dimensional materials. Our findings also address longstanding convergence issues associated with the strong sensitivity to numerical parameters in extreme three-dimensional conductors.
Notably, our formalism naturally incorporates frequency lineshifts and pole renormalizations through the real part of the self-energy, which can be especially important in 2D systems, where out-of-plane flexural modes with quadratic dispersion can produce overdamped phonon lifetimes. We also clarify the treatment of lineshifts, showing that they should be included in the broadened spectral function rather than in the Dirac delta approximation at the base of the ansatz hierarchy, where neither broadening nor lineshifts are present. More generally, this formulation retains pole spectral resolution, thus extending its range of applicability toward the overdamped regime.
We provide the first formal and rigorous derivation of the time-dependent, spatially non-homogeneous phonon BTE\cite{spohn2006phonon}, starting from the Wigner mixed representation of the quasiparticle distribution function. This paves the way for studying systems with spatial inhomogeneities.
Moreover, by exploiting the relation between repumping and depumping scattering, we avoid resummation issues and preserve energy conservation over macroscopic timescales, an essential condition for defining a local temperature and, consequently, thermal conductivity.

We demonstrated the practical impact of the formalism through its application to monolayer $\alpha$-GeSe and bulk diamond. In both systems, we show that conventional numerical smearing schemes fail to provide converged thermal conductivity values. For monolayer $\alpha$-GeSe, reducing the smearing parameter to better approximate the Dirac delta function leads to unphysically large linewidths and overdamped phonon modes, particularly in the low-energy spectrum. For bulk diamond, even within commonly adopted smearing ranges, significant variations in the computed conductivity persist, indicating the absence of clear numerical convergence. In contrast, the present self-consistent broadening scheme yields a unique, smearing-independent solution for the phonon linewidths and, correspondingly, for the lattice thermal conductivity. The iterative procedure systematically converges to a fixed point regardless of the initial smearing choice, removing any dependence on computational parameters. For monolayer $\alpha$-GeSe, the method captures its intrinsically low and anisotropic in-plane thermal conductivity.
% and clarifies the origin of the discrepancies reported in previous theoretical studies. 
For bulk diamond, the approach yields a stable and physically consistent conductivity value for a system known to be highly sensitive to numerical broadening, in quantitative agreement with experimental measurements. In both cases, the convergence of the lattice thermal conductivity directly mirrors that of the phonon linewidths, confirming the internal consistency of the theoretical framework. Moreover, we have shown analytically and verified numerically that self-consistent collisional broadening eliminates the unphysical overdamping that the Fermi golden rule universally yields for quadratic phonon modes interacting with linear acoustic modes.
%This result may have implications beyond phononic systems. Indeed, a quadratic dispersion at small momentum is not exclusive to flexural modes in 2D materials, but also arises in the magnon spectrum of ferromagnets \cite{dyson1956general,holstein1940field,bloch1930theorie,sato2016magnon,vsmejkal2023chiral,jenni2023magnon}. Consequently, analogous self-consistent broadening mechanisms may address the failure of the semiclassical Boltzmann transport equation in magnetic systems featuring quadratic bands.

Finally, we discussed how the combination of energy-non-conserving scattering processes and collisional broadening could also have a profound impact on frequency-dependent thermal conductivity.
Moreover, the developed formalism can be directly generalized to include vertex corrections and is inherently spectral-function based, offering broad flexibility in incorporating renormalizations and temperature effects, for example through Matsubara temperature-dependent Green's functions. This ensures compatibility with self-consistent phonon theories (SCP) of spectral functions, such as those calculated using the stochastic self-consistent harmonic approximation (SSCHA)\cite{SSCHA_0}. In particular, the framework presented here accounts for the term representing off-resonant propagation associated with the satellite structures described by SCP. This versatility opens pathways for exploring complex systems with advanced spectral features, ensuring that thermal transport calculations remain consistent with the underlying physics.

\section{Acknowledgments}

This research was supported by the Swiss National Science Foundation (SNSF) through Grant No. CRSII5\_189924 (“Hydronics” project). N.M. acknowledges NCCR MARVEL, a National Centre of Competence in Research, funded by the Swiss National Science Foundation (Grant No. 205602)

\onecolumngrid
\appendix

\section{Formal derivation of the relation between vibronic and bosonic Green's functions} \label{relation_g_bosonic_and_G_vibronic}

Here we show the relation between the Green's functions in the vibronic and bosonic representations (see also Ref. \cite{caldarelli2022many}) introduced in Sections \ref{vibronic_g_Section} and \ref{bosonic_g_Section}. We start by considering the phonon propagator in the vibronic representation:
\begin{equation} \label{passage}
\begin{split}
G_{\nu_{1}\nu_{2}}(t_{1},t_{2})&=-i\hat{T}\langle A_{\nu_{1}}(t_{1}) A^{\dagger}_{\nu_{2}}(t_{2})\rangle=-i\hat{T}\left\langle\Big( a_{\nu_{1}}(t_{1})+ a^{\dagger}_{-\nu_{1}}(t_{1})\Big)\Big( a_{\nu_{2}}^{\dagger}(t_{2})+ a_{-\nu_{2}}(t_{2})\Big)\right\rangle=\\
&=-i\hat{T}\left\langle a_{\nu_{1}}(t_{1}) a_{\nu_{2}}^{\dagger}(t_{2})+ a^{\dagger}_{-\nu_{1}}(t_{1}) a_{-\nu_{2}}(t_{2})\right\rangle=-i\hat{T}\left\langle a_{\nu_{1}}(t_{1}) a_{\nu_{2}}^{\dagger}(t_{2})\right\rangle-i\hat{T}\left\langle a^{\dagger}_{-\nu_{1}}(t_{1}) a_{-\nu_{2}}(t_{2})\right\rangle,
\end{split}
\end{equation}
where terms $aa$ and $a^{\dagger}a^{\dagger}$ do not contribute \cite{mahan2000many}. The first term on the RHS of Eq. \eqref{passage} is simply the Green's function in the bosonic representation:
\begin{equation}
\begin{split}
g_{\nu_{1}\nu_{2}}(t_{1},t_{2})=-i\hat{T}\left\langle a_{\nu_{1}}(t_{1}) a_{\nu_{2}}^{\dagger}(t_{2})\right\rangle.
\end{split}
\end{equation}
We know that the Green's function $g$ in wave vector space is the Fourier's transform of that in position space. The bra-ket of phonon operators defining the Green's function $g$ is in turn a real function, therefore in $\boldsymbol{q}$ space it needs to follow the properties:
\begin{equation}
\begin{split}
&\langle a_{\nu_{1}} a_{\nu_{2}}^{\dagger}\rangle^{*}=\langle a_{-\nu_{1}} a_{-\nu_{2}}^{\dagger}\rangle,\\
&\langle a_{\nu_{1}}^{\dagger} a_{\nu_{2}}\rangle^{*}=\langle a_{-\nu_{1}}^{\dagger} a_{-\nu_{2}}\rangle
\end{split}
\end{equation}
where the operator $^{*}$ stands for the complex conjugate. So we can write the second term on the RHS of Eq. \eqref{passage} as
\begin{equation} \label{property_of_g_bosonic}
\begin{split}
-i\hat{T}\left\langle a^{\dagger}_{-\nu_{1}}(t_{1}) a_{-\nu_{2}}(t_{2})\right\rangle&=-i\hat{T}\left\langle a^{\dagger}_{\nu_{1}}(t_{1}) a_{\nu_{2}}(t_{2})\right\rangle^{*}=-i\hat{T}\left\langle e^{-i\omega_{\nu}t_{1}} a^{\dagger}_{\nu_{1}}e^{-i\omega_{\nu}t_{2}} a_{\nu_{2}}\right\rangle^{*}=-i\hat{T}\left\langle e^{i\omega_{\nu}t_{2}} a^{\dagger}_{\nu_{2}}e^{i\omega_{\nu}t_{1}} a_{\nu_{1}}\right\rangle=\\
&=-i\hat{T}\left\langle  a_{\nu_{2}}^{\dagger}(-t_{2}) a_{\nu_{1}}(-t_{1})\right\rangle.
\end{split}
\end{equation}
Using this property in Eq. \eqref{passage} we have
\begin{equation}
\begin{split}
G_{\nu_{1}\nu_{2}}(t_{1},t_{2})&=-i\hat{T}\left\langle a_{\nu_{1}}(t_{1}) a_{\nu_{2}}^{\dagger}(t_{2})\right\rangle-i\hat{T}\left\langle a^{\dagger}_{-\nu_{1}}(t_{1}) a_{-\nu_{2}}(t_{2})\right\rangle=-i\hat{T}\left\langle a_{\nu_{1}}(t_{1}) a_{\nu_{2}}^{\dagger}(t_{2})\right\rangle-i\hat{T}\left\langle  a_{\nu_{2}}^{\dagger}(-t_{2}) a_{\nu_{1}}(-t_{1})\right\rangle=\\
&=g_{\nu_{1}\nu_{2}}(t_{1},t_{2})+g_{\nu_{1}\nu_{2}}(-t_{1},-t_{2}).
\end{split}
\end{equation}
We can now translate this result in the notation of lesser and greater Green's functions. We finally have \cite{ryndyk2016theory,caldarelli2022many}
\begin{equation} \label{relation_G_g}
\begin{split}
&G^{<}_{\nu_{1}\nu_{2}}(t_{1},t_{2})=g^{<}_{\nu_{1}\nu_{2}}(t_{1},t_{2})+g^{>}_{\nu_{1}\nu_{2}}(-t_{1},-t_{2}),\\
&G^{>}_{\nu_{1}\nu_{2}}(t_{1},t_{2})=g^{>}_{\nu_{1}\nu_{2}}(t_{1},t_{2})+g^{<}_{\nu_{1}\nu_{2}}(-t_{1},-t_{2}).
\end{split}
\end{equation}

\section{Wigner distribution function} \label{wigner_distr_section}

In this Section, we show how the definition of the Wigner distribution function naturally emerges from the Wigner mixed representation of the Green’s functions. By using the relation \eqref{relation_G_g_main} to rewrite Eq. \eqref{Wigner's_mixed_representation}, we obtain
\begin{equation} \label{G_wigner_on_going}
\begin{split}
G_{\nu}^{<}(\omega;\boldsymbol{R},t)=&\iint d\tau d\boldsymbol{q}'\,e^{i\omega\tau+i\boldsymbol{q}'\cdot\boldsymbol{R}}G_{s}^{<}\left(\boldsymbol{q}+\frac{\boldsymbol{q}'}{2},t+\frac{\tau}{2};\boldsymbol{q}-\frac{\boldsymbol{q}'}{2},t-\frac{\tau}{2}\right)=\\
=&\iint d\tau d\boldsymbol{q}'\,e^{i\omega\tau+i\boldsymbol{q}'\cdot\boldsymbol{R}}g_{s}^{<}\left(\boldsymbol{q}+\frac{\boldsymbol{q}'}{2},t+\frac{\tau}{2};\boldsymbol{q}-\frac{\boldsymbol{q}'}{2},t-\frac{\tau}{2}\right)+\\
&+\iint d\tau d\boldsymbol{q}'\,e^{i\omega\tau+i\boldsymbol{q}'\cdot\boldsymbol{R}}g_{s}^{>}\left(\boldsymbol{q}+\frac{\boldsymbol{q}'}{2},-t-\frac{\tau}{2};\boldsymbol{q}-\frac{\boldsymbol{q}'}{2},-t+\frac{\tau}{2}\right).
\end{split}
\end{equation}
The second term of Eq. \eqref{G_wigner_on_going} can be rewritten using the change of variable $\tau\to-\tau'$, so that 
\begin{equation}
\begin{split}
\iint d\tau d\boldsymbol{q}'\,e^{i\omega\tau+i\boldsymbol{q}'\cdot\boldsymbol{R}}g_{s}^{>}\left(\boldsymbol{q}+\frac{\boldsymbol{q}'}{2},-t-\frac{\tau}{2};\boldsymbol{q}-\frac{\boldsymbol{q}'}{2},-t+\frac{\tau}{2}\right)=\iint d\tau'd\boldsymbol{q}'\,e^{-i\omega\tau'+i\boldsymbol{q}'\cdot\boldsymbol{R}}g_{s}^{>}\left(\boldsymbol{q}+\frac{\boldsymbol{q}'}{2},-t+\frac{\tau'}{2};\boldsymbol{q}-\frac{\boldsymbol{q}'}{2},-t-\frac{\tau'}{2}\right).
\end{split}
\end{equation}
In this way Eq. \eqref{G_wigner_on_going} becomes
\begin{equation}
\begin{split}
G_{\nu}^{<}(\omega;\boldsymbol{R},t)=&\iint d\tau d\boldsymbol{q}'\,e^{i\omega\tau+i\boldsymbol{q}'\cdot\boldsymbol{R}}g_{s}^{<}\left(\boldsymbol{q}+\frac{\boldsymbol{q}'}{2},t+\frac{\tau}{2};\boldsymbol{q}-\frac{\boldsymbol{q}'}{2},t-\frac{\tau}{2}\right)+\\
&+\iint d\tau'd\boldsymbol{q}'\,e^{-i\omega\tau'+i\boldsymbol{q}'\cdot\boldsymbol{R}}g_{s}^{>}\left(\boldsymbol{q}+\frac{\boldsymbol{q}'}{2},-t+\frac{\tau'}{2};\boldsymbol{q}-\frac{\boldsymbol{q}'}{2},-t-\frac{\tau'}{2}\right).
\end{split}
\end{equation}
Now we use the Wigner's mixed representation \eqref{Wigner's_mixed_representation} also for the phonon Green's functions in the bosonic representation and obtain
\begin{equation} \label{relation_G_g_in_wigner_space}
\begin{split}
G_{\nu}^{<}(\omega;\boldsymbol{R},t)=g_{\nu}^{<}(\omega;\boldsymbol{R},t)+g_{\nu}^{>}(-\omega;\boldsymbol{R},-t).
\end{split}
\end{equation}
By evaluating the Wigner's mixed representation of the bosonic Green's function in the limit of the fast relative time scale going to zero, $\tau\to0$ (see i.g. page 94 of Ref. \cite{haug2008quantum}), we get
\begin{equation}
\begin{split}
g^{<}_{\nu}(\boldsymbol{R},t)&=-i\int d\boldsymbol{q}'\,e^{i\boldsymbol{q}'\cdot\boldsymbol{R}}\left\langle a_{s}^{\dagger}\left(\boldsymbol{q}-\frac{\boldsymbol{q}'}{2},t\right)a_{s}\left(\boldsymbol{q}+\frac{\boldsymbol{q}'}{2},t\right)\right\rangle,
\end{split}
\end{equation}
and, by recalling the bosonic Green's function definition \cite{ryndyk2016theory,caldarelli2022many},
\begin{equation}
\begin{split}
g_{s}^{<}\left(\boldsymbol{q}+\frac{\boldsymbol{q}'}{2},t;\boldsymbol{q}-\frac{\boldsymbol{q}'}{2},t\right)=-i\left\langle a_{s}^{\dagger}\left(\boldsymbol{q}-\frac{\boldsymbol{q}'}{2},t\right)a_{s}\left(\boldsymbol{q}+\frac{\boldsymbol{q}'}{2},t\right)\right\rangle,
\end{split}
\end{equation}
we write 
\begin{equation} \label{wigner_distr_function}
\begin{split}
\tilde{n}_{s}(\boldsymbol{q},\boldsymbol{R},t)=\frac{1}{2\pi}\int d\boldsymbol{q}'\,e^{i\boldsymbol{q}'\cdot\boldsymbol{R}}\left\langle a_{s}^{\dagger}\left(\boldsymbol{q}-\frac{\boldsymbol{q}'}{2},t\right) a_{s}\left(\boldsymbol{q}+\frac{\boldsymbol{q}'}{2},t\right)\right\rangle,
\end{split}
\end{equation}
where we have defined the Wigner distribution function $\tilde{n}_{s}(\boldsymbol{q},\boldsymbol{R},t)$ given in Eq. \eqref{wigner_distr_function_main}. In this way we finally have 
\begin{equation} \label{G_to_wigner_distr_function_bis_bis}
\begin{split}
g^{<}_{\nu}(\boldsymbol{R},t)=-2\pi i\,\tilde{n}_{s}(\boldsymbol{q},\boldsymbol{R},t)
\end{split}
\end{equation}
and analogously
\begin{equation} \label{G_to_wigner_distr_function_bis_bis_bis}
\begin{split}
g^{>}_{\nu}(\boldsymbol{R},t)=-2\pi i\Big[\tilde{n}_{s}(\boldsymbol{q},\boldsymbol{R},t)+1\Big].
\end{split}
\end{equation}
With this definition we are implicitly assuming that it is legitimate to neglect the rapid oscillations of $g^{<}$ and $G^{<}$, since the characteristic time scale we are interested in is given by time $t$ (see Eq. \eqref{center-of-mass_difference_variables}).

\section{Proof of the GKB ansatz}

In this Section, we demonstrate the GKB ansatz used in the main text, first at thermal equilibrium, where a more intuitive physical interpretation can be provided, and then out of equilibrium, assuming that the out-of-equilibrium phonon distribution is described by the Wigner distribution function. For this we need to introduce the generalized Kadanoff-Baym equation (GKBE).

\subsection{Generalized Kadanoff-Baym equation}

It is convenient to rewrite the KBE in its integral form. Applying Eq. \eqref{Langreth_theorem_2} to Eq. \eqref{dyson_eq} yields
\begin{equation}
G^{<}=G^{0,<}+G^{0,R}\Sigma^{R}G^{<}+G^{0,R}\Sigma^{<}G^{A}+G^{0,<}\Sigma^{A}G^{A}.
\end{equation}
By iterating with respect to $G^{<}$ and regrouping we obtain
\begin{equation}
\begin{split}
G^{<}=&\,\,(1+G^{0,R}\Sigma^{R})G^{0,<}(1+\Sigma^{A}G^{A})+\\
&+(G^{0,R}+G^{0,R}\Sigma^{R}G^{0,R})\Sigma^{<}G^{A}+\\
&+G^{0,R}\Sigma^{R}G^{0,R}\Sigma^{R}G^{<}.
\end{split}
\end{equation}
If we keep iterating the last equation to infinite order we get to the so-called Keldysh equation \cite{keldysh1964diagram,kadanoff2018quantum,lipavsky1986generalized}
\begin{equation} \label{keldysh_eq}
G^{<}=(1+G^{R}\Sigma^{R})G^{0,<}(1+\Sigma^{A}G^{A})+G^{R}\Sigma^{<}G^{A}.
\end{equation}
The first term on the RHS of Eq. \eqref{keldysh_eq} represents an initial condition that applies to a time infinitely far in the past and can be substituted with a more physically relevant boundary condition indicating that $G^{<}$ should possess a specific value just before a collision event occurs \cite{lipavsky1986generalized}. Adopting this notation one can deal with the much simpler equation
\begin{equation} \label{GKB_equation}
G^{<}=G^{R}\Sigma^{<}G^{A},
\end{equation}
which goes under the name of generalized Kadanoff-Baym equation (GKBE) \cite{lipavsky1986generalized}.

\subsection{Equilibrium: fluctuation-dissipation theorem} \label{fluctuation_dissipation_theorem}

Here we seek the version of the fluctuation-dissipation theorem that links the spectral function $\mathrm{a}_{\nu}(\omega)$ to the phonon propagator $g_{\nu}^{<}(\omega)$ at equilibrium. This relation holds significance as it sets up an initial condition in equilibrium for the non-equilibrium lesser function. Furthermore, it presents a practical assumption and interpretation of the ansatz on the unknown out-of-equilibrium phonon distribution. The proof is a straightforward calculation using the complete eigenstates of the Hamiltonian, $H$, in the Lehmann representation \cite{lehmann1954eigenschaften}. In the following manipulations the wavevector and band index $\nu=\boldsymbol{q}s$ are not important and we suppress them. The concept behind this approach is to make use of the precise eigenstates of many-particle systems, even if we may not have explicit knowledge of them. Inserting the set of complete states, and choosing $t_{1}=t$ and $t_{2}=0$ in Eq. \eqref{G<_corr_function_boson} we get
\begin{equation}
\begin{split}
g^{<}(\omega)=\int dt e^{i\omega t}g^{<}(t;0)=-i\int dt e^{i\omega t}\langle a^{\dagger}(0) a(t)\rangle=-i\int dt e^{i\omega t}\sum_{n,m}\rho\langle n| a^{\dagger}(0)|m\rangle\langle m|e^{iHt} a(0)e^{-iHt}|n\rangle,
\end{split}
\end{equation}
where $\rho$ is the density matrix. We use the partition function for a grand-canonical ensemble of bosons
\begin{equation}
Z=e^{-\beta\omega}
\end{equation}
where we set $\hbar=1$ and the grand-canonical thermodynamic potential which in the case of phonons (where there is not a chemical potential since the number of phonons is not conserved upon scattering) reduces to the canonical thermodynamic potential given by
\begin{equation}
\omega_{\nu}=\frac{1}{\beta}\ln\left(1-e^{-\beta\hbar\omega_{\nu}}\right).
\end{equation}
In this way the density matrix is
\begin{equation}
\rho=\frac{e^{-\beta H}}{Z}.
\end{equation}
Being the states $|n\rangle$ and $|m\rangle$ eigenstates of $H$, we can proceed as
\begin{equation} \label{g<_eigenstates_hamiltonian}
\begin{split}
g^{<}(\omega)&=-\frac{i}{Z}\int dt e^{i\omega t}\sum_{n,m}e^{-\beta\omega_{n}}e^{i(\omega_{m}-\omega_{n})t}\langle n| a^{\dagger}(0)|m\rangle\langle m| a(0)|n\rangle=\\
&=-\frac{2\pi i}{Z}\sum_{n,m}\delta(\omega+\omega_{m}-\omega_{n})e^{-\beta\omega_{n}}\langle n| a^{\dagger}(0)|m\rangle\langle m| a(0)|n\rangle.
\end{split}
\end{equation}
Similarly
\begin{equation}
\begin{split}
g^{>}(\omega)=-\frac{2\pi i}{Z}\sum_{n,m}\delta(\omega+\omega_{n}-\omega_{m})e^{-\beta\omega_{n}}\langle n| a(0)|m\rangle\langle m| a^{\dagger}(0)|n\rangle.
\end{split}
\end{equation}
By taking into account that $\omega_{n}=\omega_{m}+\omega$ because of the Dirac delta function in Eq. \eqref{g<_eigenstates_hamiltonian}, we get
\begin{equation}
\begin{split}
g^{<}(\omega)=-\frac{2\pi i}{Z}\sum_{n,m}\delta(\omega+\omega_{m}-\omega_{n})e^{-\beta\omega_{m}}e^{-\beta\omega}\langle n| a^{\dagger}(0)|m\rangle\langle m| a(0)|n\rangle.
\end{split}
\end{equation}
Now, interchanging $n$ and $m$ we can finally write
\begin{equation} \label{g_>_g_<_rel_bis}
\begin{split}
g^{<}(\omega)=-e^{-\beta\omega}\frac{2\pi i}{Z}\sum_{n,m}\delta(\omega+\omega_{n}-\omega_{m})e^{-\beta\omega_{n}}\langle m| a^{\dagger}(0)|n\rangle\langle n| a(0)|m\rangle=e^{-\beta\omega}g^{>}(\omega),
\end{split}
\end{equation}
or 
\begin{equation} \label{g_>_g_<_rel}
\begin{split}
g^{>}(\omega)=e^{\beta\omega}g^{<}(\omega).
\end{split}
\end{equation}
Using Eq. \eqref{spectral_function_def} for the phonon spectral function in the bosonic representation and Eq. \eqref{g_>_g_<_rel}, we can express the spectral function as
\begin{equation}
\mathrm{a}_{\nu}(\omega)=2\pi\,\mathrm{d}_{\nu}(\omega)=\left[g_{\nu}^{>}(\omega)-g_{\nu}^{<}(\omega)\right]=i\left[e^{\beta\omega}g^{<}(\omega)-g_{\nu}^{<}(\omega)\right]=i\left[e^{\beta\omega}-1\right]g_{\nu}^{<}(\omega),
\end{equation}
which gives
\begin{equation} \label{fluctuation_dissipation}
g_{\nu}^{<}(\omega)=-2\pi i\frac{1}{e^{\beta\omega}-1}\mathrm{d}_{\nu}(\omega)=-2\pi i\,\mathrm{d}_{\nu}(\omega)\bar{n}_{\omega}.
\end{equation}
We see how Eq. \eqref{fluctuation_dissipation} is related to the “fluctuation-dissipation” theorem: the correlation function $g^{<}$ (which also carries information about fluctuations) is proportional to the dissipative part $\mathrm{d}_{\nu}$; the proportionality factor being the Bose-Einstein distribution. From Eqs. \eqref{g_>_g_<_rel_bis} and \eqref{spectral_function_def} we also get 
\begin{equation}
\mathrm{a}_{\nu}(\omega)=2\pi\mathrm{d}_{\nu}(\omega)=i\left[g_{\nu}^{>}(\omega)-g_{\nu}^{<}(\omega)\right]=i\left[g_{\nu}^{>}(\omega)-e^{-\beta\omega}g_{\nu}^{>}(\omega)\right]=i\left[1-\frac{1}{e^{\beta\omega}}\right]g_{\nu}^{>}(\omega)=i\frac{e^{\beta\omega}-1}{e^{\beta\omega}}g_{\nu}^{>}(\omega),
\end{equation}
and so 
\begin{equation}
g_{\nu}^{>}(\omega)=-2\pi i\,\mathrm{d}_{\nu}(\omega)\frac{e^{\beta\omega}}{e^{\beta\omega}-1}=-2\pi i\,\mathrm{d}_{\nu}(\omega)\left[\frac{1}{e^{\beta\omega}-1}+1\right]=-2\pi i\,\mathrm{d}_{\nu}(\omega)[\bar{n}_{\omega}+1].
\end{equation}
In the semiclassical picture of phonon thermal transport, the spectral function is considered to be sharply peaked, so that $\mathrm{d}_{\nu}(\omega)=\delta(\omega-\omega_{\nu})$. In such cases one can directly replace the energy argument in the Bose-Einstein distribution by the single-particle energy $\omega_{\nu}$.

\subsection{Out-of-equilibrium: ansatz for the Wigner distribution function} \label{rigorous_derivation_GKBA}

To begin with the demonstration of the GKB ansatz out of thermal equilibrium, we define the following auxiliary functions \cite{haug2008quantum}:
\begin{equation} \label{auxiliary_G}
\begin{split}
&\mathsf{g}_{\nu}^{R,<}(t_{1},t_{2})=\theta(t_{1}- t_{2})g^{<}_{\nu}(t_{1},t_{2}),\\
&\mathsf{g}_{\nu}^{A,<}(t_{1},t_{2})=\theta(t_{2}- t_{1})g^{<}_{\nu}(t_{1},t_{2}).
\end{split}
\end{equation}
These definitions allows for the following relation, by construction:
\begin{equation} \label{auxiliary_G_sum}
\begin{split}
g^{<}_{\nu}(t_{1},t_{2})=\mathsf{g}_{\nu}^{R,<}(t_{1},t_{2})+\mathsf{g}_{\nu}^{A,<}(t_{1},t_{2})
\end{split}
\end{equation}
We now concentrate on $\mathsf{g}_{\nu}^{R,<}(t_{1},t_{2})$ exclusively; the result we obtain can be easily extended to $\mathsf{g}_{\nu}^{A,<}(t_{1},t_{2})$. We operate $g_{\nu}^{R^{-1}}=\omega-\omega_{\nu}-\sigma_{\nu}^{R}$ from the left on the first of Eqs. \eqref{auxiliary_G}:
\begin{equation} \label{auxiliary_G_acting}
\begin{split}
g_{\nu}^{R^{-1}}\mathsf{g}_{\nu}^{R,<}(t_{1},t_{2})&=\int dt'\left\lbrace \left[i\frac{\partial}{\partial t_{1}}-\omega_{\nu}\right]\delta(t_{1}-t')-\sigma^{R}_{\nu}(t_{1},t')\right\rbrace\mathsf{g}_{\nu}^{R,<}(t',t_{2})=\\
&=\int dt'\left\lbrace i\frac{\partial}{\partial t_{1}}\delta(t_{1}-t')\mathsf{g}_{\nu}^{R,<}(t',t_{2})-\omega_{\nu}\delta(t_{1}-t')\mathsf{g}_{\nu}^{R,<}(t',t_{2})-\sigma_{\nu}^{R}(t_{1},t')\mathsf{g}_{\nu}^{R,<}(t',t_{2})\right\rbrace,
\end{split}
\end{equation}
where we used the quantum mechanical definition of the energy operator $\omega\leftrightarrow i\frac{\partial}{\partial t}$ and $\sigma$ is the self-energy in the bosonic representation. Thanks to Eq. \eqref{auxiliary_G}, the first term of Eq. \eqref{auxiliary_G_acting} reads
\begin{equation}  \label{auxiliary_G_acting_term1}
\int dt'i\frac{\partial}{\partial t_{1}}\delta(t_{1}-t')\mathsf{g}_{\nu}^{R,<}(t',t_{2})=\int dt'i\frac{\partial}{\partial t_{1}}\delta(t_{1}-t')\theta(t'-t_{2})g_{\nu}^{<}(t',t_{2})=i\delta(t_{1}-t_{2})g_{\nu}^{<}(t_{1},t_{2})+i\theta(t_{1}-t_{2})\frac{\partial g_{\nu}^{<}(t_{1},t_{2})}{\partial t_{1}}.
\end{equation}
The second term of Eq. \eqref{auxiliary_G_acting} is
\begin{equation}  \label{auxiliary_G_acting_term2}
-\omega_{\nu}\int dt'\delta(t_{1}-t')\mathsf{g}_{\nu}^{R,<}(t',t_{2})=-\omega_{\nu}\int dt'\delta(t_{1}-t')\theta(t'-t_{2})g_{\nu}^{<}(t',t_{2})=-\omega_{\nu}\theta(t_{1}-t_{2})g_{\nu}^{<}(t_{1},t_{2}).
\end{equation}
Finally, the third term of Eq. \eqref{auxiliary_G_acting} is
\begin{equation}  \label{auxiliary_G_acting_term3}
-\int dt'\sigma_{\nu}^{R}(t_{1},t')\mathsf{g}_{\nu}^{R,<}(t',t_{2})=-\int dt'\sigma_{\nu}^{R}(t_{1},t')\theta(t'-t_{2})g_{\nu}^{<}(t',t_{2})=-\theta(t_{1}-t_{2})\int_{t_{2}}^{t_{1}} dt'\sigma_{\nu}^{R}(t_{1},t')g_{\nu}^{<}(t',t_{2})
\end{equation}
where we exploited the fact that $\theta(t'-t_{2})\ne 0$ only for $t'>t_{2}$ and so we can factorize it out of the integral as $\theta(t_{1}-t_{2})\ne 0$, leaving the integral with $t_{2}$ as the lower bound. Note that $\theta(t_{1}-t_{2})$ also allows to truncate the integral with $t_{1}$ as upper bound. Putting together Eqs. \eqref{auxiliary_G_acting_term1}, \eqref{auxiliary_G_acting_term2} and \eqref{auxiliary_G_acting_term3}, Eq. \eqref{auxiliary_G_acting} becomes
\begin{equation} \label{last_auxiliary_G_acting}
\begin{split}
g_{\nu}^{R^{-1}}\mathsf{g}_{\nu}^{R,<}(t_{1},t_{2})=i\delta(t_{1}-t_{2})g_{\nu}^{<}(t_{1},t_{2})+\theta(t_{1}-t_{2})\left\lbrace\left[i\frac{\partial}{\partial t_{1}}-\omega_{\nu}\right]g_{\nu}^{<}(t_{1},t_{2})-\int_{t_{2}}^{t_{1}} dt'\sigma_{\nu}^{R}(t_{1},t')g_{\nu}^{<}(t',t_{2})\right\rbrace.
\end{split}
\end{equation}
Focusing on the term in curly brackets, from Eq. \eqref{GKB_equation} it is possible to show that the GKBE holds for the Green's functions in the bosonic representation \cite{haug2008quantum}, such that
\begin{equation} \label{GKB_equation_mod}
g_{\nu}^{R^{-1}}g_{\nu}^{<}=\sigma_{\nu}^{<}g_{\nu}^{A}.
\end{equation}
Writing it explicitly we get 
\begin{equation} \label{GR_acting_G<}
g_{\nu}^{R^{-1}}g_{\nu}^{<}(t_{1},t_{2})=\left[i\frac{\partial}{\partial t_{1}}-\omega_{\nu}\right]g_{\nu}^{<}(t_{1},t_{2})-\int_{-\infty}^{t_{1}}dt'\sigma_{\nu}^{R}(t_{1},t')g_{\nu}^{<}(t',t_{2})=\int_{-\infty}^{t_{2}}dt'\sigma_{\nu}^{<}(t_{1},t')g_{\nu}^{A}(t',t_{2})=\sigma_{\nu}^{<}g_{\nu}^{A}(t_{1},t_{2}).
\end{equation}
By substituting Eq. \eqref{GR_acting_G<} into the curly brackets of Eq. \eqref{last_auxiliary_G_acting} we get
\begin{equation}
\begin{split}
g_{\nu}^{R^{-1}}\mathsf{g}_{\nu}^{R,<}(t_{1},t_{2})&=i\delta(t_{1}-t_{2})g_{\nu}^{<}(t_{1},t_{2})+\theta(t_{1}-t_{2})\Bigg\lbrace\int_{-\infty}^{t_{1}}dt'\sigma_{\nu}^{R}(t_{1},t')g_{\nu}^{<}(t',t_{2})+\int_{-\infty}^{t_{2}}dt'\sigma_{\nu}^{<}(t_{1},t')g_{\nu}^{A}(t',t_{2})-\\
&\hspace{5.75cm}-\int_{t_{2}}^{t_{1}} dt'\sigma_{\nu}^{R}(t_{1},t')g_{\nu}^{<}(t',t_{2})\Bigg\rbrace=\\
&=i\delta(t_{1}-t_{2})g_{\nu}^{<}(t_{1},t_{2})+\theta(t_{1}-t_{2})\left\lbrace\int_{-\infty}^{t_{2}}dt'\sigma_{\nu}^{R}(t_{1},t')g_{\nu}^{<}(t',t_{2})+\sigma_{\nu}^{<}(t_{1},t')g_{\nu}^{A}(t',t_{2})\right\rbrace.
\end{split}
\end{equation}
Finally, operating with $g_{\nu}^{R}$ from the left,
\begin{equation}
\begin{split}
\int_{t_{2}}^{t_{1}} dt'' g_{\nu}^{R}(t_{1},t'') g_{\nu}^{R^{-1}}(t_{1},t_{2})\mathsf{g}_{\nu}^{R,<}(t_{1},t_{2})=&\,\,i\int_{t_{2}}^{t_{1}} dt'' g_{\nu}^{R}(t_{1},t'')\delta(t_{1}-t_{2})g_{\nu}^{<}(t_{1},t_{2})+\\
&\,\,+\int_{t_{2}}^{t_{1}} dt'' g_{\nu}^{R}(t_{1},t'')\theta(t_{1}-t_{2})\left\lbrace\int_{-\infty}^{t_{2}}dt'\sigma_{\nu}^{R}(t_{1},t')g_{\nu}^{<}(t',t_{2})+\sigma_{\nu}^{<}(t_{1},t')g_{\nu}^{A}(t',t_{2})\right\rbrace,
\end{split}
\end{equation}
we get to
\begin{equation} \label{G^{R}_{<}_equation}
\begin{split}
\mathsf{g}_{\nu}^{R,<}(t_{1},t_{2})=ig_{\nu}^{R}(t_{1},t_{2})g_{\nu}^{<}(t_{2},t_{2})+\int_{t_{2}}^{t_{1}} dt''\int_{-\infty}^{t_{2}}dt'g_{\nu}^{R}(t_{1},t'')\left[\sigma_{\nu}^{R}(t'',t')g_{\nu}^{<}(t',t_{2})+\sigma_{\nu}^{<}(t'',t')g_{\nu}^{A}(t',t_{2})\right].
\end{split}
\end{equation}
Similarly, one can derive the auxiliary function $\mathsf{g}_{\nu}^{A,<}(t_{1},t_{2})$:
\begin{equation} \label{G^{A}_{<}_equation}
\begin{split}
\mathsf{g}_{\nu}^{A,<}(t_{1},t_{2})=-ig_{\nu}^{<}(t_{1},t_{1})g_{\nu}^{A}(t_{1},t_{2})+\int_{t_{1}}^{t_{2}} dt''\int_{-\infty}^{t_{1}}dt'\left[g_{\nu}^{<}(t_{1},t')\sigma_{\nu}^{A}(t',t'')+g_{\nu}^{R}(t_{1},t')\sigma_{\nu}^{<}(t',t'')\right]g_{\nu}^{A}(t'',t_{2}).
\end{split}
\end{equation}
Eq. \eqref{auxiliary_G_sum}, together with Eqs. \eqref{G^{R}_{<}_equation} and \eqref{G^{A}_{<}_equation}, allow for an iterative construction of $g_{\nu}^{<}(t_{1},t_{2})$ from its time-diagonal component which contains the Wigner function $\tilde{n}$. The first terms of Eqs. \eqref{G^{R}_{<}_equation} and \eqref{G^{A}_{<}_equation} give the GKB ansatz, and we will keep only these terms in the following. A detailed analysis of the iterative corrections coming from the second terms of Eqs. \eqref{G^{R}_{<}_equation} and \eqref{G^{A}_{<}_equation} is discussed in Refs. \cite{haug2008quantum,lipavsky1986generalized}. Using Eqs. \eqref{auxiliary_G_sum}, \eqref{G^{R}_{<}_equation} and \eqref{G^{A}_{<}_equation} we have
\begin{equation} \label{G^{<}_equation_final_1}
\begin{split}
g_{\nu}^{<}(t_{1},t_{2})=\mathsf{g}_{\nu}^{R,<}(t_{1},t_{2})+\mathsf{g}_{\nu}^{A,<}(t_{1},t_{2})\simeq i\Big[g_{\nu}^{R}(t_{1},t_{2})g_{\nu}^{<}(t_{2},t_{2})-g_{\nu}^{<}(t_{1},t_{1})g_{\nu}^{A}(t_{1},t_{2})\Big].
\end{split}
\end{equation}
At this stage, one should quickly note that, being the Green's functions $g_{\nu}^{<}(t_{2},t_{2})$ and $g_{\nu}^{<}(t_{1},t_{1})$ calculated at the same time, from Eq. \eqref{times_expressions} this can be seen as having $\tau=0$, which will lead to the definition of the Wigner distribution function \eqref{wigner_distr_function}. On the other hand, the retarded and advanced functions, $g_{\nu}^{R}(t_{1},t_{2})$ and $g_{\nu}^{A}(t_{1},t_{2})$, are readily transformed according to Wigner's representation \eqref{Wigner's_mixed_representation}. In Wigner's mixed representation we have
\begin{equation}
\begin{split}
&\,\,\,\,\,\,\,\,\iint d\tau d\boldsymbol{q}'\,e^{i\omega\tau+i\boldsymbol{q}'\cdot\boldsymbol{R}}g_{s}^{<}\left(\boldsymbol{q}+\frac{\boldsymbol{q}'}{2},t+\frac{\tau}{2};\boldsymbol{q}-\frac{\boldsymbol{q}'}{2},t-\frac{\tau}{2}\right)=\\
&=i\iint d\tau d\boldsymbol{q}'\,e^{i\omega\tau+i\boldsymbol{q}'\cdot\boldsymbol{R}}g_{s}^{R}\left(\boldsymbol{q}+\frac{\boldsymbol{q}'}{2},t+\frac{\tau}{2};\boldsymbol{q}-\frac{\boldsymbol{q}'}{2},t-\frac{\tau}{2}\right)\iint d\tau d\boldsymbol{q}'\,e^{i\omega\tau+i\boldsymbol{q}'\cdot\boldsymbol{R}}g_{s}^{<}\left(\boldsymbol{q}+\frac{\boldsymbol{q}'}{2},t-\frac{\tau}{2};\boldsymbol{q}-\frac{\boldsymbol{q}'}{2},t-\frac{\tau}{2}\right)-\\
&\,\,\,\,\,\,\,\,-i\iint d\tau d\boldsymbol{q}'\,e^{i\omega\tau+i\boldsymbol{q}'\cdot\boldsymbol{R}}g_{s}^{<}\left(\boldsymbol{q}+\frac{\boldsymbol{q}'}{2},t+\frac{\tau}{2};\boldsymbol{q}-\frac{\boldsymbol{q}'}{2},t+\frac{\tau}{2}\right)\iint d\tau d\boldsymbol{q}'\,e^{i\omega\tau+i\boldsymbol{q}'\cdot\boldsymbol{R}}g_{s}^{A}\left(\boldsymbol{q}+\frac{\boldsymbol{q}'}{2},t+\frac{\tau}{2};\boldsymbol{q}-\frac{\boldsymbol{q}'}{2},t-\frac{\tau}{2}\right).
\end{split}
\end{equation}
Writing the Wigner's representation implicitly and setting $\tau=0$ in the integrals of $g_{s}^{<}$ then leads to
\begin{equation} \label{proof_ansatz_piece1}
\begin{split}
g_{\nu}^{<}(\omega;\boldsymbol{R},t)&=i\left[g_{\nu}^{R}(\omega;\boldsymbol{R},t)\int d\boldsymbol{q}'\,e^{i\boldsymbol{q}'\cdot\boldsymbol{R}}g_{s}^{<}\left(\boldsymbol{q}+\frac{\boldsymbol{q}'}{2},t;\boldsymbol{q}-\frac{\boldsymbol{q}'}{2},t\right)-\int d\boldsymbol{q}'\,e^{i\boldsymbol{q}'\cdot\boldsymbol{R}}g_{s}^{<}\left(\boldsymbol{q}+\frac{\boldsymbol{q}'}{2},t;\boldsymbol{q}-\frac{\boldsymbol{q}'}{2},t\right)g_{\nu}^{A}(\omega;\boldsymbol{R},t)\right]=\\
&=i\int d\boldsymbol{q}'\,e^{i\boldsymbol{q}'\cdot\boldsymbol{R}}g_{s}^{<}\left(\boldsymbol{q}+\frac{\boldsymbol{q}'}{2},t;\boldsymbol{q}-\frac{\boldsymbol{q}'}{2},t\right)\left[g_{\nu}^{R}(\omega;\boldsymbol{R},t)-g_{\nu}^{A}(\omega;\boldsymbol{R},t)\right]=\\
&=-i\,\tilde{n}_{s}(\boldsymbol{q},\boldsymbol{R},t)\mathrm{a}_{\nu}(\omega)=-2\pi i\,\tilde{n}_{s}(\boldsymbol{q},\boldsymbol{R},t)\mathrm{d}_{\nu}(\omega).
\end{split}
\end{equation}
In the same way we can derive
\begin{equation} \label{proof_ansatz_piece2}
\begin{split}
g_{\nu}^{>}(\omega;\boldsymbol{R},t)=-2\pi i\big[\tilde{n}_{s}(\boldsymbol{q},\boldsymbol{R},t)+1\big]\mathrm{d}_{\nu}(\omega),
\end{split}
\end{equation}
and extend this result to the vibronic representation (using Eq. \eqref{relation_G_g_in_wigner_space}):
\begin{equation}
\begin{split}
&G^{<}_{\nu}(\omega,\boldsymbol{R},t)=g_{\nu}^{<}(\omega;\boldsymbol{R},t)+g_{\nu}^{>}(-\omega;\boldsymbol{R},-t)=-2\pi i\Big[n_{s}(\boldsymbol{q},\boldsymbol{R},t)\mathrm{d}_{\nu}(\omega)+\big[n_{s}(\boldsymbol{q},\boldsymbol{R},-t)+1\big]\mathrm{d}_{\nu}(-\omega)\Big],\\
&G^{>}_{\nu}(\omega,\boldsymbol{R},t)=g_{\nu}^{<}(-\omega;\boldsymbol{R},-t)+g_{\nu}^{>}(\omega;\boldsymbol{R},t)=-2\pi i\Big[n_{s}(\boldsymbol{q},\boldsymbol{R},-t)\mathrm{d}_{\nu}(-\omega)+\big[n_{s}(\boldsymbol{q},\boldsymbol{R},t)+1\big]\mathrm{d}_{\nu}(\omega)\Big].
\end{split}
\end{equation}

\section{Formal derivation of the driving term} \label{dervation_of_driving_term_section}

In order to derive the driving term, DT, in the form of Eq. \eqref{DT} we begin by applying the gradient approximation \eqref{gradient_expansion_2} to the LHS of Eq. \eqref{KBE_to_use_for_derivation}:
\begin{equation}
\begin{split}
&\text{DT}_{\nu}(\omega,\boldsymbol{R},t)=\\
=&\left[G^{0^{-1}}_{\nu}(\omega)\,,\,G^{<}_{\nu}(\omega,\boldsymbol{R},t)\right]=\\
=&-i\frac{\partial G^{0^{-1}}_{\nu}(\omega)}{\partial t}\frac{\partial G^{<}_{\nu}(\omega,\boldsymbol{R},t)}{\partial\omega}+i\frac{\partial G^{0^{-1}}_{\nu}(\omega)}{\partial\omega}\frac{\partial G^{<}_{\nu}(\omega,\boldsymbol{R},t)}{\partial t}+\\
&+i\frac{\partial G^{0^{-1}}_{\nu}(\omega)}{\partial\boldsymbol{R}}\cdot\frac{\partial G^{<}_{\nu}(\omega,\boldsymbol{R},t)}{\partial\boldsymbol{q}}-i\frac{\partial G^{0^{-1}}_{\nu}(\omega)}{\partial\boldsymbol{q}}\cdot\frac{\partial G^{<}_{\nu}(\omega,\boldsymbol{R},t)}{\partial\boldsymbol{R}}.
\end{split}
\end{equation}
Substituting the expression for $G^{0^{-1}}_{\nu}(\omega)=2\omega_{\nu}/(\omega^{2}-\omega_{\nu}^{2})$ (see SI \cite{supplementary}) and noting that $G^{0^{-1}}_{\nu}(\omega)$ is time- and position-independent (only the $-\frac{\partial A}{\partial\omega}\frac{\partial B}{\partial t}$ and $\frac{\partial A}{\partial\boldsymbol{q}}\cdot\frac{\partial B}{\partial\boldsymbol{R}}$ terms contribute) we get
\begin{equation}
\begin{split}
&\,\text{DT}_{\nu}(\omega,\boldsymbol{R},t)=\\
=&\,i\frac{\omega}{\omega_{\nu}}\frac{\partial G^{<}_{\nu}(\omega,\boldsymbol{R},t)}{\partial t}+i\frac{\boldsymbol{v}_{\nu}}{2}\left(\frac{\omega^{2}}{\omega_{\nu}^{2}}+1\right)\cdot\frac{\partial G^{<}_{\nu}(\omega,\boldsymbol{R},t)}{\partial\boldsymbol{R}},
\end{split}
\end{equation}
where $\boldsymbol{v}_{\nu}=\frac{\partial\omega_{\nu}}{\partial\boldsymbol{q}}$ is the phonon group velocity. Substituting the GKB ansatz \eqref{GKB_ansatz_equation} for the phonon Green's function we then get
\begin{equation} \label{DT_on_going}
\begin{split}
&\,\,\text{DT}_{\nu}(\omega,\boldsymbol{R},t)=\\
=&\,\,2\pi\frac{\omega}{\omega_{\nu}}\mathrm{d}_{\nu}(\omega)\frac{\partial n_{\nu}(\omega,\boldsymbol{R},t)}{\partial t}+\\
&+2\pi\frac{\omega}{\omega_{\nu}}\mathrm{d}_{\nu}(-\omega)\frac{\partial\big[n_{\nu}(-\omega,\boldsymbol{R},-t)+1\big]}{\partial t}+\\
&+2\pi\left(\frac{\omega^{2}}{\omega_{\nu}^{2}}+1\right)\mathrm{d}_{\nu}(\omega)\frac{\boldsymbol{v}_{\nu}}{2}\cdot\frac{\partial n_{\nu}(\omega,\boldsymbol{R},t)}{\partial\boldsymbol{R}}+\\
&+2\pi\left(\frac{\omega^{2}}{\omega_{\nu}^{2}}+1\right)\mathrm{d}_{\nu}(-\omega)\frac{\boldsymbol{v}_{\nu}}{2}\cdot\frac{\partial\big[n_{\nu}(-\omega,\boldsymbol{R},-t)+1\big]}{\partial\boldsymbol{R}}.
\end{split}
\end{equation}
The second time derivative in Eq. \eqref{DT_on_going} can be written as
\begin{equation}
\begin{split}
\frac{\partial\big[n_{\nu}(-\omega,\boldsymbol{R},-t)+1\big]}{\partial t}=-\frac{\partial n_{\nu}(-\omega,\boldsymbol{R},t)}{\partial t},
\end{split}
\end{equation}
while the second space derivative reads
\begin{equation} \label{time_reversal_invariance}
\frac{\partial\big[n_{\nu}(-\omega,\boldsymbol{R},-t)+1\big]}{\partial\boldsymbol{R}}=\frac{\partial n_{\nu}(-\omega,\boldsymbol{R},-t)}{\partial\boldsymbol{R}}=\frac{\partial n_{\nu}(-\omega,\boldsymbol{R},t)}{\partial\boldsymbol{R}};
\end{equation}
the latter equality stems from the time-reversal invariance of quantum kinetic equations in the NEGF formalism \cite{bonitz2016quantum}. Indeed, as the KBE can be directly derived from the equations of motion of the field operators in second quantization, which are time-reversal invariant, it can be shown that the KBE shares the same symmetry properties \cite{bonitz2016quantum,scharnke2017time}. So in the end the driving term reads
\begin{equation} 
\begin{split}
\text{DT}_{\nu}(\omega,\boldsymbol{R},t)
=2\pi\Bigg[&
\frac{\omega}{\omega_{\nu}}\mathrm{d}_{\nu}(\omega)\frac{\partial n_{\nu}(\omega,\boldsymbol{R},t)}{\partial t}
-\frac{\omega}{\omega_{\nu}}\mathrm{d}_{\nu}(-\omega)\frac{\partial n_{\nu}(-\omega,\boldsymbol{R},t)}{\partial t} \\
&+\left(\frac{\omega^{2}}{\omega_{\nu}^{2}}+1\right)\mathrm{d}_{\nu}(\omega)\frac{\boldsymbol{v}_{\nu}}{2}\cdot\frac{\partial n_{\nu}(\omega,\boldsymbol{R},t)}{\partial\boldsymbol{R}} \\
&+\left(\frac{\omega^{2}}{\omega_{\nu}^{2}}+1\right)\mathrm{d}_{\nu}(-\omega)\frac{\boldsymbol{v}_{\nu}}{2}\cdot\frac{\partial n_{\nu}(-\omega,\boldsymbol{R},t)}{\partial\boldsymbol{R}}
\Bigg],
\end{split}
\end{equation}
which is Eq. \eqref{DT} of the main text.

\section{Phonon bubble self-energy in Wigner's mixed representation} \label{self_energy_wigner}

In order to write the bubble self-energy in Wigner's mixed representation we apply a Fourier transform to the self-energy in $\boldsymbol{q}\omega$ space, allowing it to be expressed in $\boldsymbol{q}t$ space (note that Eq. \eqref{sigma_eq} is equivalent to, e.g., Eq. A23 in Ref. \cite{volz2020quantum} with $\omega_{2}=\omega-\omega_{1}$):
\begin{equation}
\begin{split}
\Sigma^{\lessgtr}_{s_{1}s_{2}}(\boldsymbol{q}_{1},t_{1};\boldsymbol{q}_{2},t_{2})=4i\pi\hbar\sum_{\substack{\nu_{1}'\nu_{2}'\\\nu_{1}''\nu_{2}''}}\mathcal{F}_{\nu_{1}-\nu_{1}'-\nu_{1}''}\mathcal{F}_{\nu_{2}-\nu_{2}'-\nu_{2}''}G^{\lessgtr}_{s_{1}'s_{2}'}(\boldsymbol{q}_{1}',t_{1};\boldsymbol{q}_{2}',t_{2})G^{\lessgtr}_{s_{1}''s_{2}''}(\boldsymbol{q}_{1}'',t_{1};\boldsymbol{q}_{2}'',t_{2}).
\end{split}
\end{equation}
Inverting the definition of Wigner's mixed representation \eqref{Wigner's_mixed_representation}, we can write
\begin{equation}
G^{\lessgtr}_{s_{1}^{i}s_{2}^{i}}(\boldsymbol{q}_{1}^{i},t_{1};\boldsymbol{q}_{2}^{i},t_{2})\equiv G_{s^{i}}^{\lessgtr}\left(\boldsymbol{q}^{i}+\frac{\boldsymbol{Q}^{i}}{2},t+\frac{\tau}{2};\boldsymbol{q}^{i}-\frac{\boldsymbol{Q}^{i}}{2},t-\frac{\tau}{2}\right)=\iint\frac{d\omega_{1} d\boldsymbol{R}^{i}}{(2\pi)^{4}}e^{-i\omega_{1}\tau-i\boldsymbol{Q}^{i}\cdot\boldsymbol{R}^{i}}\tilde{G}_{\nu^{i}}^{\lessgtr}(\omega_{1};\boldsymbol{R}^{i},t),
\end{equation}
where $i=$ $',''$.
In this way can write:
\begin{equation}
\begin{split}
\Sigma^{\lessgtr}_{s_{1}s_{2}}(\boldsymbol{q}_{1},t_{1};\boldsymbol{q}_{2},t_{2})&\equiv \Sigma^{\lessgtr}_{s}\left(\boldsymbol{q}+\frac{\boldsymbol{Q}}{2},t+\frac{\tau}{2};\boldsymbol{q}-\frac{\boldsymbol{Q}}{2},t-\frac{\tau}{2}\right)=\\
&=4i\pi\hbar\sum_{\nu'\nu''}|\mathcal{F}_{\nu-\nu'-\nu''}|^{2}\iint\frac{d\omega_{1} d\boldsymbol{R}'}{(2\pi)^{4}}e^{-i\omega_{1}\tau-i\boldsymbol{Q}'\cdot\boldsymbol{R}'}\tilde{G}_{\nu'}^{\lessgtr}(\omega_{1};\boldsymbol{R}',t)\cdot\\
&\hspace{3.6cm}\cdot\iint\frac{d\omega_{2} d\boldsymbol{R}''}{(2\pi)^{4}}e^{-i\omega_{2}\tau-i\boldsymbol{Q}''\cdot\boldsymbol{R}''}\tilde{G}_{\nu''}^{\lessgtr}(\omega_{2};\boldsymbol{R}'',t).
\end{split}
\end{equation}
So, finally, the self-energy in Wigner's mixed representation reads
\begin{equation}
\begin{split}
\tilde{\Sigma}_{\nu}^{\lessgtr}(\omega;\boldsymbol{R},t)&=\iint d\tau d\boldsymbol{Q}\,e^{i\omega\tau+i\boldsymbol{Q}\cdot\boldsymbol{R}}\Sigma_{s}^{\lessgtr}\left(\boldsymbol{q}+\frac{\boldsymbol{Q}}{2},t+\frac{\tau}{2};\boldsymbol{q}-\frac{\boldsymbol{Q}}{2},t-\frac{\tau}{2}\right)=\\
&=4i\pi\hbar\iint d\tau d\boldsymbol{Q}\,e^{i\omega\tau+i\boldsymbol{Q}\cdot\boldsymbol{R}}\sum_{\nu'\nu''}|\mathcal{F}_{\nu-\nu'-\nu''}|^{2}\iint\frac{d\omega_{1} d\boldsymbol{R}'}{(2\pi)^{4}}e^{-i\omega_{1}\tau-i\boldsymbol{Q}'\cdot\boldsymbol{R}'}\tilde{G}_{\nu'}^{\lessgtr}(\omega_{1};\boldsymbol{R}',t)\cdot\\
&\hspace{5.25cm}\cdot\iint\frac{d\omega_{2} d\boldsymbol{R}''}{(2\pi)^{4}}e^{-i\omega_{2}\tau-i\boldsymbol{Q}''\cdot\boldsymbol{R}''}\tilde{G}_{\nu''}^{\lessgtr}(\omega_{2};\boldsymbol{R}'',t).
\end{split}
\end{equation}
We recall center-of-mass and relative variables,
\begin{equation} \label{center-of-mass_difference_variables_bis}
\begin{split}
&\boldsymbol{Q}^{i}=\boldsymbol{q}_{1}^{i}-\boldsymbol{q}_{2}^{i},\hspace{1cm}\tau=t_{1}-t_{2},\\
&\boldsymbol{q}^{i}=\frac{1}{2}(\boldsymbol{q}_{1}^{i}+\boldsymbol{q}_{2}^{i}),\hspace{1cm}t=\frac{1}{2}(t_{1}+t_{2})
\end{split}
\end{equation}
and
\begin{equation} \label{times_expressions_bis}
\begin{split}
&\boldsymbol{q}_{1}^{i}=\boldsymbol{q}^{i}+\frac{\boldsymbol{Q}^{i}}{2},\hspace{1cm}t_{1}=t+\frac{\tau}{2},\\
&\boldsymbol{q}_{2}^{i}=\boldsymbol{q}^{i}-\frac{\boldsymbol{Q}^{i}}{2},\hspace{1cm}t_{2}=t-\frac{\tau}{2},
\end{split}
\end{equation}
along with their related momentum and energy conservation laws: 
\begin{equation}
\begin{split}
&\boldsymbol{q}_{1}=\boldsymbol{q}_{1}'+\boldsymbol{q}_{1}''\\
&\boldsymbol{q}_{2}=\boldsymbol{q}_{2}'+\boldsymbol{q}_{2}''\\
&\boldsymbol{Q}=\boldsymbol{q}_{1}-\boldsymbol{q}_{2}=\boldsymbol{q}_{1}'+\boldsymbol{q}_{1}''-(\boldsymbol{q}_{2}'+\boldsymbol{q}_{2}'')=\boldsymbol{Q}'+\boldsymbol{Q}''\\
&\omega=\omega_{1}+\omega_{2}.
\end{split}
\end{equation}
In this way the self-energy in Wigner's mixed representation reads
\begin{equation} \label{almost_done_selfenergy_wigner}
\begin{split}
\tilde{\Sigma}_{\nu}^{\lessgtr}(\omega;\boldsymbol{R},t)&=4i\pi\hbar\int d\boldsymbol{Q}\sum_{\nu'\nu''}|\mathcal{F}_{\nu-\nu'-\nu''}|^{2}\iint\frac{d\omega_{1} d\boldsymbol{R}'}{(2\pi)^{4}}\iint\frac{d\omega_{2} d\boldsymbol{R}''}{(2\pi)^{4}}2\pi\delta(\omega-\omega_{1}-\omega_{2})\cdot\\
&\hspace{1.1cm}\cdot e^{-i\boldsymbol{Q}\cdot(\boldsymbol{R}''-\boldsymbol{R})+i\boldsymbol{Q}'\cdot(\boldsymbol{R}''-\boldsymbol{R}')}\tilde{G}_{\nu'}^{\lessgtr}(\omega_{1};\boldsymbol{R}',t)\tilde{G}_{\nu''}^{\lessgtr}(\omega_{2};\boldsymbol{R}'',t)=\\
&=4i\pi\hbar\sum_{\nu'\nu''}|\mathcal{F}_{\nu-\nu'-\nu''}|^{2}\iint\frac{d\omega_{1} d\boldsymbol{R}'}{(2\pi)^{4}}\iint\frac{d\omega_{2} d\boldsymbol{R}''}{(2\pi)^{4}}2\pi\delta(\omega-\omega_{1}-\omega_{2})\cdot\\
&\hspace{3.6cm}\cdot\int d\boldsymbol{Q}e^{i\boldsymbol{Q}\cdot\boldsymbol{R}}e^{-i\boldsymbol{Q}'\cdot\boldsymbol{R}'}e^{-i\boldsymbol{Q}''\cdot\boldsymbol{R}''}\tilde{G}_{\nu'}^{\lessgtr}(\omega_{1};\boldsymbol{R}',t)\tilde{G}_{\nu''}^{\lessgtr}(\omega_{2};\boldsymbol{R}'',t).
\end{split}
\end{equation}
Imposing $\boldsymbol{Q}'=\boldsymbol{Q}-\boldsymbol{Q}''$ we get
\begin{equation}
\begin{split}
\tilde{\Sigma}_{\nu}^{\lessgtr}(\omega;\boldsymbol{R},t)&=4i\pi\hbar\sum_{\nu'\nu''}|\mathcal{F}_{\nu-\nu'-\nu''}|^{2}\iint\frac{d\omega_{1} d\boldsymbol{R}'}{(2\pi)^{4}}\iint\frac{d\omega_{2} d\boldsymbol{R}''}{(2\pi)^{4}}2\pi\delta(\omega-\omega_{1}-\omega_{2})\cdot\\
&\hspace{3.6cm}\cdot\int d\boldsymbol{Q}e^{i\boldsymbol{Q}\cdot(\boldsymbol{R}-\boldsymbol{R}')}e^{i\boldsymbol{Q}''\cdot(\boldsymbol{R}'-\boldsymbol{R}'')}\tilde{G}_{\nu'}^{\lessgtr}(\omega_{1};\boldsymbol{R}',t)\tilde{G}_{\nu''}^{\lessgtr}(\omega_{2};\boldsymbol{R}'',t).
\end{split}
\end{equation}
The term $(2\pi)^{-3}\int d\boldsymbol{Q}e^{i\boldsymbol{Q}\cdot(\boldsymbol{R}-\boldsymbol{R}')}$ is nothing but the exponential representation of the Dirac delta, so we get:
\begin{equation} \label{almost_done_selfenergy_wigner_1} 
\begin{split}
\tilde{\Sigma}_{\nu}^{\lessgtr}(\omega;\boldsymbol{R},t)&=4i\pi\hbar\sum_{\nu'\nu''}|\mathcal{F}_{\nu-\nu'-\nu''}|^{2}\iint\frac{d\omega_{1} d\boldsymbol{R}'}{(2\pi)^{4}}\iint\frac{d\omega_{2} d\boldsymbol{R}''}{(2\pi)^{4}}\frac{1}{(2\pi)^{2}}\delta(\omega-\omega_{1}-\omega_{2})\cdot\\
&\hspace{3.6cm}\cdot\delta(\boldsymbol{R}-\boldsymbol{R}')e^{i\boldsymbol{Q}''\cdot(\boldsymbol{R}'-\boldsymbol{R}'')}\tilde{G}_{\nu'}^{\lessgtr}(\omega_{1};\boldsymbol{R}',t)\tilde{G}_{\nu''}^{\lessgtr}(\omega_{2};\boldsymbol{R}'',t).
\end{split}
\end{equation}
Similarly, we can use $\boldsymbol{Q}''=\boldsymbol{Q}-\boldsymbol{Q}'$ in Eq. \eqref{almost_done_selfenergy_wigner} obtaining
\begin{equation} \label{almost_done_selfenergy_wigner_2} 
\begin{split}
\tilde{\Sigma}_{\nu}^{\lessgtr}(\omega;\boldsymbol{R},t)&=4i\pi\hbar\sum_{\nu'\nu''}|\mathcal{F}_{\nu-\nu'-\nu''}|^{2}\iint\frac{d\omega_{1} d\boldsymbol{R}'}{(2\pi)^{4}}\iint\frac{d\omega_{2} d\boldsymbol{R}''}{(2\pi)^{4}}\frac{1}{(2\pi)^{2}}\delta(\omega-\omega_{1}-\omega_{2})\cdot\\
&\hspace{3.6cm}\cdot\delta(\boldsymbol{R}-\boldsymbol{R}'')e^{-i\boldsymbol{Q}'\cdot(\boldsymbol{R}'-\boldsymbol{R}'')}\tilde{G}_{\nu'}^{\lessgtr}(\omega_{1};\boldsymbol{R}',t)\tilde{G}_{\nu''}^{\lessgtr}(\omega_{2};\boldsymbol{R}'',t),
\end{split}
\end{equation}
from which we find that Eq. \eqref{almost_done_selfenergy_wigner_1} equals Eq. \eqref{almost_done_selfenergy_wigner_2} only if we also have $\boldsymbol{R}'=\boldsymbol{R}''$. So, finally the self-energy in Wigner's mixed representation reads
\begin{equation} \label{final_selfenergy_wigner} 
\begin{split}
\tilde{\Sigma}_{\nu}^{\lessgtr}(\omega;\boldsymbol{R},t)&=4i\pi\hbar\sum_{\nu'\nu''}|\mathcal{F}_{\nu-\nu'-\nu''}|^{2}\int\frac{d\omega_{1}}{2\pi}\int\frac{d\omega_{2}}{2\pi}\delta(\omega-\omega_{1}-\omega_{2})\tilde{G}_{\nu'}^{\lessgtr}(\omega_{1};\boldsymbol{R},t)\tilde{G}_{\nu''}^{\lessgtr}(\omega_{2};\boldsymbol{R},t).
\end{split}
\end{equation}

\section{Small-broadening analysis of annihilation and creation of three phonons} \label{section_small_broadening_annihilation_creation_three_phonons}

We consider three-phonon scattering processes described by Lorentzian spectral functions of the following form:
\begin{equation}
L_\gamma(\Delta) \equiv \frac{1}{\pi}\frac{\gamma}{\Delta^2+\gamma^2},
\end{equation}
where $\Delta=\omega_\nu+\omega_{\nu'}+\omega_{\nu''}$ for the process of e.g. annihilation of three phonons, and $\Delta=\omega_\nu+\omega_{\nu'}-\omega_{\nu''}$ for the process of e.g. annihilation of phonon $\nu$ and $\nu'$ and creation of phonon $\nu''$. In the small-broadening limit it is useful to analyze the related Lorentzian distributions in terms of the dimensionless ratio $\Delta/\gamma$:
\begin{equation}
L_\gamma(\Delta)=\frac{1}{\pi}\frac{\gamma}{\Delta^2+\gamma^2}
=\frac{1}{\pi\gamma}\frac{1}{1+(\Delta/\gamma)^2}.
\end{equation}
In the case of annihilation (or creation) of three phonons, $\Delta=\omega_\nu+\omega_{\nu'}+\omega_{\nu''}$ is strictly positive and bounded from below by a finite value $\Delta_{\min}>0$ for physical phonon modes. Consequently,
\begin{equation}
\frac{\Delta}{\gamma}\xrightarrow{\gamma\to0}\infty,
\end{equation}
and the Lorentzian is always evaluated in its asymptotic tail,
\begin{equation}
L_\gamma(\Delta)
\sim \frac{1}{\pi}\frac{\gamma}{\Delta^2}
\qquad (\Delta/\gamma\gg1),
\end{equation}
which vanishes linearly with $\gamma$ in the small-broadening limit.\\
For phonon decay and coalescence processes, $\Delta=\omega_\nu+\omega_{\nu'}-\omega_{\nu''}$ is not sign-definite and can approach zero for energy-conserving mode triplets. In this case,
\begin{equation}
\frac{\Delta}{\gamma}=\mathcal{O}(1)
\end{equation}
and the Lorentzian probes its central peak,
\begin{equation}
L_\gamma(\Delta)
\sim \frac{1}{\pi\gamma}
\qquad (\Delta/\gamma\lesssim1),
\end{equation}
which is still finite for small $\gamma$.

\section{Energy conservation} \label{energy_cons_demonstrations}

As discussed in detail in the main text, having a collision term that conserves the total energy of the system is a necessary condition for defining a local temperature and, consequently, a thermal conductivity. The local energy at position $\boldsymbol{R}$ and time $t$ is defined as
\begin{equation}
E(\boldsymbol{R},t)=\frac{1}{\mathcal{V}}\sum_{\nu}\hbar\omega_{\nu}n_{\nu}(\boldsymbol{R},t).
\end{equation}
This satisfies the local conservation law (or continuity equation) \cite{allen2018temperature}
\begin{equation}
\frac{\partial E(\boldsymbol{R},t)}{\partial t}+\nabla\cdot\boldsymbol{j}_{E}(\boldsymbol{R},t)=0,
\end{equation}
where the energy current density, $\boldsymbol{j}_{E}$, is given by
\begin{equation}
\boldsymbol{j}_{E}(\boldsymbol{R},t)=\frac{1}{V}\sum_{\nu}\hbar\omega_{\nu}\boldsymbol{v}_{\nu}n_{\nu}(\boldsymbol{R},t).
\end{equation}
In this Section, we first demonstrate that in the standard LBTE framework, this holds true by construction, as each scattering event individually conserves energy \cite{spohn2006phonon}. We then show that the same analysis can be extended to the case with broadening, where an exact expression for energy conservation cannot be derived. However, it can be demonstrated that the Lorentzian collisional broadening represents a necessary condition to break energy conservation, but not a sufficient one.\\
We start from the standard LBTE:
\begin{equation}
\frac{\partial h_{\nu}(\boldsymbol{r},t)}{\partial t}+\boldsymbol{v}_{\nu}\cdot\frac{\partial h_{\nu}(\boldsymbol{r},t)}{\partial\boldsymbol{r}}=\frac{1}{\bar{n}_{\nu}[\bar{n}_{\nu}+1]}\sum_{\nu'}A_{\nu\nu'}h_{\nu'}.
\end{equation}
By studying the effect of small perturbations in the temperature, we can expand the deviation from equilibrium, $h_{\nu}$ (see Eq. \eqref{expansion_n_bar_n_delta_n_main}), in proximity of the local equilibrium:
\begin{equation}
h_{\nu}(\boldsymbol{r},t)=\frac{1}{\bar{n}_{\nu}[\bar{n}_{\nu}+1]}\left.\frac{\partial\bar{n}_{\nu}}{\partial T}\right|_{T=\bar{T}}\cdot\left[T(\boldsymbol{r},t)-\bar{T}\right]+h_{\nu}^{\delta}(\boldsymbol{r},t)=h_{\nu}^{T}(\boldsymbol{r},t)+h_{\nu}^{\delta}(\boldsymbol{r},t).
\end{equation}
Substituting this expansion in the LBTE, keeping only linear terms in the temperature gradient, we get
\begin{equation} \label{LBTE}
\frac{\partial\bar{n}_{\nu}}{\partial T}\frac{\partial T(\boldsymbol{r},t)}{\partial t}+\bar{n}_{\nu}[\bar{n}_{\nu}+1]\frac{\partial h_{\nu}^{\delta}(\boldsymbol{r},t)}{\partial t}+\frac{\partial\bar{n}_{\nu}}{\partial T}\boldsymbol{v}_{\nu}\cdot \frac{\partial T(\boldsymbol{r},t)}{\partial\boldsymbol{r}}=\sum_{\nu'}A_{\nu\nu'}\big[h_{\nu'}^{T}(\boldsymbol{r},t)+h_{\nu'}^{\delta}(\boldsymbol{r},t)\big],
\end{equation}
where derivatives with respect to temperature are computed at equilibrium and $\boldsymbol{v}_{\nu}\cdot\frac{\partial h_{\nu}^{\delta}(\boldsymbol{r},t)}{\partial\boldsymbol{r}}$ has been neglected being of second order in deviations from equilibrium \cite{simoncelli2020generalization}. Taking the steady-state limit we get
\begin{equation} \label{LBTE_steady_state}
\frac{\partial\bar{n}_{\nu}}{\partial T}\boldsymbol{v}_{\nu}\cdot \frac{\partial T(\boldsymbol{r},t)}{\partial\boldsymbol{r}}=\sum_{\nu'}A_{\nu\nu'}\big[h_{\nu'}^{T}(\boldsymbol{r},t)+h_{\nu'}^{\delta}(\boldsymbol{r},t)\big].
\end{equation}
Multiplying Eq. \eqref{LBTE_steady_state} for the phonon energy $\hbar\omega_{\nu}$ and summing over all wave vectors we get
\begin{equation} \label{LBTE_steady_state_short}
\sum_{\nu}\,\omega_{\nu}\frac{\partial\bar{n}_{\nu}}{\partial T}\boldsymbol{v}_{\nu}\cdot \frac{\partial T(\boldsymbol{r},t)}{\partial\boldsymbol{r}}=\sum_{\nu}\omega_{\nu}\left[\sum_{\nu'}A_{\nu\nu'}\big[h_{\nu'}^{T}(\boldsymbol{r},t)+h_{\nu'}^{\delta}(\boldsymbol{r},t)\big]\right].
\end{equation}
Eq. \eqref{LBTE_steady_state_short} basically translates into evaluating the total energy of the system throughout the transport phenomenon, and so addressing its global conservation. Then, knowing that $\omega_{\nu}$ is an even function of $\boldsymbol{q}$, $\boldsymbol{v}_{\nu}$ is an odd function of $\boldsymbol{q}$ and that
\begin{equation} \label{bose_distr_property}
\frac{\partial\bar{n}_{\nu}}{\partial T}=\frac{\partial}{\partial T}\left[\frac{1}{e^{\frac{\hbar\omega_{\nu}}{k_{{\rm{B}}}T}}-1}\right]=-\frac{-\frac{\hbar\omega_{\nu}}{k_{{\rm{B}}}T^{2}}e^{\frac{\hbar\omega_{\nu}}{k_{{\rm{B}}}T}}}{\left(e^{\frac{\hbar\omega_{\nu}}{k_{{\rm{B}}}T}}-1\right)^{2}}=\frac{\hbar\omega_{\nu}}{k_{{\rm{B}}}T^{2}}\bar{n}_{\nu}(\bar{n}_{\nu}+1)
\end{equation}
is clearly even in $\boldsymbol{q}$, the integral over $\boldsymbol{q}$ in the LHS of Eq. \eqref{LBTE_steady_state_short} is a symmetric integral of an odd function and so it vanishes. In the same way, the RHS of Eq. \eqref{LBTE_steady_state_short} has to be equal to zero as well. Eq. \eqref{LBTE_steady_state_short}allows us to identify the following expression
\begin{equation}
\sum_{\nu}\omega_{\nu}{\rm{CT}}_{\nu}=0
\end{equation}
as the condition under which the system's energy is conserved and a local temperature can be defined. This is exactly the condition given in Eq. \eqref{energy_cons_condition} of the main text. As it will be clear from the next Section, the energy conservation condition \eqref{energy_cons_condition} mirrors the requirement of having the Bose-Einstein eigenvectors being a left eigenvector of the scattering matrix with zero eigenvalue \cite{allen2018temperature}.

\subsection{Scattering matrix symmetrization, Bose-Einstein eigenvector and relation between depumping and repumping matrices} \label{bose_einstein_Section}

The Bose-Einstein eigenvector representing conservation of energy is usually introduced using a standard symmetrized form of the scattering matrix. As we will see, such transformation is not general, in the sense that it does not lead to a symmetric scattering matrix in the case of Lorentzian broadening. This condition has in turn crucial consequences with respect to the energy conservation condition \eqref{energy_cons_condition}. We start by recalling that we can write the collision term using either the matrices $A_{\nu\nu'}$ or ${\rm{\Omega}}_{\nu\nu'}$:
\begin{equation}
\sum_{\nu'}A_{\nu\nu'}h_{\nu'}=\sum_{\nu'}{\rm{\Omega}}_{\nu\nu'}\Delta n_{\nu'},
\end{equation}
where
\begin{equation}
{\rm{\Omega}}_{\nu\nu'}=A_{\nu\nu'}\frac{1}{\bar{n}_{\nu'}(\bar{n}_{\nu'}+1)}.
\end{equation}
By defining the conventional symmetrized phonon scattering matrix as \cite{hardy1970phonon,chaput2013direct,cepellotti2016thermal}:
\begin{equation} \label{symmetric_matrix}
\tilde{{\rm{\Omega}}}_{\nu\nu'}={\rm{\Omega}}_{\nu\nu'}\sqrt{\frac{\bar{n}_{\nu'}(\bar{n}_{\nu'}+1)}{\bar{n}_{\nu}(\bar{n}_{\nu}+1)}}=\frac{1}{\sqrt{\bar{n}_{\nu'}(\bar{n}_{\nu'}+1)}}A_{\nu\nu'}\frac{1}{\sqrt{\bar{n}_{\nu}(\bar{n}_{\nu}+1)}},
\end{equation}
which is clearly symmetric upon interchanging $\nu$ and $\nu'$. Similarly, one can apply the same transformation to the phonon distribution
\begin{equation}
\tilde{\Delta n}_{\nu}=\frac{\Delta n_{\nu}}{\bar{n}_{\nu}(\bar{n}_{\nu}+1)}.
\end{equation}
By evaluating the collision term of the LBTE \eqref{LBTE_steady_state} using the symmetrized notation and in the case where the solution is $h_{\nu}=h_{\nu}^{T}$ (or $\Delta n_{\nu}=\Delta 
 n_{\nu}^{T}$) we have \cite{simoncelli2020generalization}
\begin{equation} \label{LBTE_with_n_T}
\begin{split}
\sum_{\nu'}\tilde{{\rm{\Omega}}}_{\nu\nu'}\tilde{\Delta n}_{\nu'}^{T}&=\sum_{\nu'}\tilde{{\rm{\Omega}}}_{\nu\nu'}\frac{1}{\sqrt{\bar{n}_{\nu}[\bar{n}_{\nu}+1]}}\left.\frac{\partial\bar{n}_{\nu}}{\partial T}\right|_{T=\bar{T}}\cdot\left[T(\boldsymbol{r},t)-\bar{T}\right]=\\
&=\sum_{\nu'}\tilde{{\rm{\Omega}}}_{\nu\nu'}\sqrt{\bar{n}_{\nu}[\bar{n}_{\nu}+1]}\frac{\hbar\omega_{\nu'}}{k_{\text{B}}\bar{T}^{2}}\cdot\left[T(\boldsymbol{r},t)-\bar{T}\right]=0.
\end{split}
\end{equation}
Eq. \eqref{LBTE_with_n_T} has to be equal to zero because of energy conservation (see also Eq. 13.8 of Ref. \cite{spohn2006phonon}); in fact, one can easily see that $\tilde{\Delta n}_{\nu}^{T}$ is nothing but phonon energy. From which the Bose-Einstein eigenvector can be identified:
\begin{equation}
\theta_{\nu}^{0}=\frac{\sqrt{\bar{n}_{\nu}[\bar{n}_{\nu}+1]}}{k_{\text{B}}\bar{T}^{2}}\hbar\omega_{\nu}.
\end{equation}
This is an eigenvector of the scattering matrix with zero eigenvalue. One should also note that the result of Eq. \eqref{LBTE_with_n_T} is entirely general in the sense that it is valid for any kind of hierarchy of the scattering matrix, that is, for both quasiparticle and Lorentzian broadening ansatze.
Finally, from Eq. \eqref{LBTE_with_n_T} it is easy to separate the scattering-out and scattering-in matrixec:
\begin{equation} \label{omega_autovettore_destro}
-{\rm{\Omega}}^{\text{out}}_{\nu\nu}\bar{n}_{\nu}(\bar{n}_{\nu}+1)\omega_{\nu}+\sum_{\nu'\ne\nu}{\rm{\Omega}}^{\text{in}}_{\nu\nu'}\bar{n}_{\nu'}(\bar{n}_{\nu'}+1)\omega_{\nu'}=0,
\end{equation}
leaving us with 
\begin{equation} \label{global_scattering_energy_cons_appendix}
{\rm{\Omega}}^{\text{out}}_{\nu\nu}=\frac{A^{\text{out}}_{\nu\nu}}{\bar{n}_{\nu}(\bar{n}_{\nu}+1)}=\Gamma_{\nu}=\sum_{\nu'\ne\nu}{\rm{\Omega}}^{\text{in}}_{\nu\nu'}\frac{\bar{n}_{\nu'}(\bar{n}_{\nu'}+1)\omega_{\nu'}}{\bar{n}_{\nu}(\bar{n}_{\nu}+1)\omega_{\nu}}=\sum_{\nu'\ne\nu}A^{\text{in}}_{\nu\nu'}\frac{\omega_{\nu'}}{\bar{n}_{\nu}(\bar{n}_{\nu}+1)\omega_{\nu}},
\end{equation}
which is Eq. \eqref{global_scattering_energy_cons} of the main text and coincides with Eq. 39 of Ref. \cite{simoncelli2022wigner}. Using the symmetrized matrices, we clearly see that Eq. \eqref{global_scattering_energy_cons_appendix} has the exact same form:
\begin{equation} \label{global_scattering_energy_cons_sym_appendix}
\tilde{{\rm{\Omega}}}^{\text{out}}_{\nu\nu}=\Gamma_{\nu}=\sum_{\nu'\ne\nu}\tilde{{\rm{\Omega}}}^{\text{in}}_{\nu\nu'}\frac{\bar{n}_{\nu'}(\bar{n}_{\nu'}+1)\omega_{\nu'}}{\bar{n}_{\nu}(\bar{n}_{\nu}+1)\omega_{\nu}}
\end{equation}
Eq. \eqref{global_scattering_energy_cons_appendix} (or Eq. \eqref{global_scattering_energy_cons_sym_appendix}) provides a way to compute the linewidths from the repumping term and is exact within the quasiparticle ansatz. As can be seen, this is closely related to the concept of energy conservation in the system, which, in the quasiparticle ansatz, holds by construction since each individual scattering event exactly conserves energy. More precisely, Eq. \eqref{global_scattering_energy_cons_appendix} arises from the fact that the Bose-Einstein eigenvector is a right eigenvector of the scattering matrix with zero eigenvalue. However, for the Bose-Einstein eigenvector to be well defined, it must also be a left eigenvector of the scattering matrix with zero eigenvalue \cite{allen2018temperature}, that is, condition \eqref{energy_cons_condition} has to be verified, and so temperature is well-defined. This condition holds in the case of the quasiparticle ansatz, as explicitly verified in the SM \cite{supplementary}. As we shall see, when using Eq. \eqref{global_scattering_energy_cons_sym_appendix} directly to compute the depumping part of the scattering matrix in the case of Lorentzian broadening, this condition no longer holds \cite{supplementary}. Therefore, in order to impose a well-defined Bose-Einstein eigenvector, and thus ensure energy conservation in the system, we will need to modify the form of the matrix.

\subsection{Adaptive smearing approach} \label{symmetrization_with_adaptive_smearing}

In this Section, we examine how the symmetry of the scattering matrix is affected when the broadening is treated using an adaptive smearing technique, as proposed in Refs. \cite{ShengBTE_2014,li2012thermal}. In this approach, the smearing applied to the Dirac delta function $\delta(\omega_{\nu}-\omega_{\nu'}-\omega_{\nu''})$, which enforces energy conservation for the phonon decay process, is chosen to be on the order of the energy mismatch $\Delta\epsilon_{\nu'\nu''}=|\boldsymbol{v}_{\nu'}-\boldsymbol{v}_{\nu''}|\,\Delta\boldsymbol{Q}$ defined in Eq. \eqref{energy_mismatch}. This comes from the idea of linearizing the detuning in the vicinity of the energy-conserving manifold, which defines a surface in $\boldsymbol{q}$-space. To first order we have
\begin{equation}
\omega_{s'}(\boldsymbol{q}'-\Delta\boldsymbol{Q})\simeq\omega_{s'}(\boldsymbol{q}')-\boldsymbol{v}_{\nu'}\cdot\Delta\boldsymbol{Q}
\end{equation}
and 
\begin{equation}
\omega_{s''}(\boldsymbol{q}''+\Delta\boldsymbol{Q})\simeq\omega_{s''}(\boldsymbol{q}'')+\boldsymbol{v}_{\nu''}\cdot\Delta\boldsymbol{Q}
\end{equation}
because momentum conservation links the two shifts. Therefore the detuning varies as
\begin{equation} \label{detuning_adaptive_smearing}
\omega_{\nu}-\omega_{\nu'}-\omega_{\nu''}\simeq(\boldsymbol{v}_{\nu'}-\boldsymbol{v}_{\nu''})\cdot\Delta\boldsymbol{Q}.
\end{equation}
With this choice of smearing, the Dirac delta function can be effectively replaced by $\delta(\omega_{\nu}-\omega_{\nu'}-\omega_{\nu''}-|\boldsymbol{v}_{\nu'}-\boldsymbol{v}_{\nu''}|\,\Delta\boldsymbol{Q})$. This allows us to write:
\begin{equation} \label{extended_detailed_balance_relations_1_adaptive}
\begin{split}
&\,(\bar{n}_{\nu''}+1)(\bar{n}_{\nu'}+1)\bar{n}_{\nu}=\frac{e^{\omega_{\nu'}+\omega_{\nu''}}}{(e^{\omega_{\nu}}-1)(e^{\omega_{\nu'}}-1)(e^{\omega_{\nu''}}-1)}=\frac{e^{\omega_{\nu}-|\boldsymbol{v}_{\nu'}-\boldsymbol{v}_{\nu''}|\,\Delta\boldsymbol{Q}}}{(e^{\omega_{\nu}}-1)(e^{\omega_{\nu'}}-1)(e^{\omega_{\nu''}}-1)}=\\
=&\,e^{-|\boldsymbol{v}_{\nu'}-\boldsymbol{v}_{\nu''}|\,\Delta\boldsymbol{Q}}\bar{n}_{\nu''}\bar{n}_{\nu'}(\bar{n}_{\nu}+1)=\big(1-|\boldsymbol{v}_{\nu'}-\boldsymbol{v}_{\nu''}|\,\Delta\boldsymbol{Q}\big)\bar{n}_{\nu''}\bar{n}_{\nu'}(\bar{n}_{\nu}+1).
\end{split}
\end{equation}
The scattering-out matrix is diagonal in $\nu$ and therefore symmetric by construction. We can thus focus exclusively on the scattering-in matrix in Eq. \eqref{final_A_in_A_out_qp_main} of the main text. Rewriting it using the conventional symmetrization transformation in Eq. \eqref{symmetric_matrix} and incorporating the adaptive smearing, we obtain:
\begin{equation} \label{sym_ext_matrix_1_adaptive}
\begin{split}
\tilde{\Omega}^{\text{in}}_{\nu\nu'}=8\pi\hbar\sum_{\nu''}\Bigg[&\mathcal{L}_{\nu\nu''}^{\nu'}\frac{\bar{n}_{\nu''}}{1\mp|\boldsymbol{v}_{\nu''}-\boldsymbol{v}_{\nu'}|\,\Delta\boldsymbol{Q}(\bar{n}_{\nu'}+1)}\sqrt{\frac{\bar{n}_{\nu}(\bar{n}_{\nu'}+1)}{\bar{n}_{\nu'}(\bar{n}_{\nu}+1)}}-\\
&-\mathcal{L}_{\nu\nu'}^{\nu''}\big[1\mp|\boldsymbol{v}_{\nu'}-\boldsymbol{v}_{\nu''}|\,\Delta\boldsymbol{Q}(\bar{n}_{\nu''}+1)\big]\bar{n}_{\nu''}
\sqrt{\frac{(\bar{n}_{\nu}+1)(\bar{n}_{\nu'}+1)}{\bar{n}_{\nu}\bar{n}_{\nu'}}}+\\
&+\mathcal{L}_{\nu}^{\nu'\nu''}\big[1\mp|\boldsymbol{v}_{\nu'}-\boldsymbol{v}_{\nu''}|\,\Delta\boldsymbol{Q}(\bar{n}_{\nu}+1)\big](\bar{n}_{\nu''}+1)\sqrt{\frac{\bar{n}_{\nu}(\bar{n}_{\nu'}+1)}{\bar{n}_{\nu'}(\bar{n}_{\nu}+1)}}\Bigg].
\end{split}
\end{equation}
Using the extended detailed balance relations as shown in Eq. \eqref{extended_detailed_balance_relations_1_adaptive} to incorporate adaptive smearing, and expanding for small smearing (i.e., to linear order in the group velocity differences), we obtain:
\begin{equation}  \label{final_matrix_sym_nunu'_adaptive}
\begin{split}
\tilde{\Omega}^{\text{in}}_{\nu\nu'}=8\pi\hbar\sum_{\nu''}\Bigg\{\Bigg(&\mathcal{L}_{\nu\nu''}^{\nu'}\left[1\pm\bar{n}_{\nu'}|\boldsymbol{v}_{\nu''}-\boldsymbol{v}_{\nu'}|\,\Delta\boldsymbol{Q}\right]-\mathcal{L}_{\nu\nu'}^{\nu''}\left[1\pm\bar{n}_{\nu''}|\boldsymbol{v}_{\nu'}-\boldsymbol{v}_{\nu''}|\,\Delta\boldsymbol{Q}\right]+\\
&+\mathcal{L}_{\nu}^{\nu'\nu''}\left[1\pm\bar{n}_{\nu}|\boldsymbol{v}_{\nu'}-\boldsymbol{v}_{\nu''}|\,\Delta\boldsymbol{Q}\right]\Bigg)\sqrt{\bar{n}_{\nu''}(\bar{n}_{\nu''}+1)}\Bigg\},
\end{split}
\end{equation}
where we also considered the limit of large $\bar{n}_{\nu^{i}}$. If we evaluate $\tilde{\Omega}^{\text{in}}_{\nu'\nu}$ by interchanging $\nu$ and $\nu'$ we get
\begin{equation}  \label{final_matrix_sym_nu'nu_adaptive}
\begin{split}
\tilde{\Omega}^{\text{in}}_{\nu'\nu}=8\pi\hbar\sum_{\nu''}\Bigg\{\Bigg(&\mathcal{L}_{\nu'\nu''}^{\nu}\left[1\pm\bar{n}_{\nu}|\boldsymbol{v}_{\nu''}-\boldsymbol{v}_{\nu}|\,\Delta\boldsymbol{Q}\right]-\mathcal{L}_{\nu'\nu}^{\nu''}\left[1\pm\bar{n}_{\nu''}|\boldsymbol{v}_{\nu}-\boldsymbol{v}_{\nu''}|\,\Delta\boldsymbol{Q}\right]+\\
&+\mathcal{L}_{\nu'}^{\nu\nu''}\left[1\pm\bar{n}_{\nu'}|\boldsymbol{v}_{\nu}-\boldsymbol{v}_{\nu''}|\,\Delta\boldsymbol{Q}\right]\Bigg)\sqrt{\bar{n}_{\nu''}(\bar{n}_{\nu''}+1)}\Bigg\},
\end{split}
\end{equation}
from which we clearly see that due to the adaptive smearing technique $\tilde{\Omega}^{\text{in}}_{\nu'\nu}\ne\tilde{\Omega}^{\text{in}}_{\nu\nu'}$. This demonstrates that collisional broadening not only surpasses adaptive smearing in terms of physical motivation, is anharmonic rather than harmonic like adaptive smearing, is mode-resolved, and is self-consistent, but also performs better in preserving the symmetry of the scattering matrix. In particular, when extending the detailed balance relations to phonon frequencies and phonon occupations, adaptive smearing breaks the symmetrization condition coming from the standard transformation \eqref{symmetric_matrix}, making it insufficient to symmetrize the scattering matrix. As a consequence, adaptive smearing does not allow for a well-defined Bose-Einstein eigenvector. This implies that total energy conservation is no longer guaranteed, leading to ill-defined temperature and thermal conductivity.

\section{Analytical solution to overdamped (long wavelenght) phonon lifetimes in 2D systems via self-consistent collisional broadening} \label{solution_overdamped_section}

\begin{figure}[!htb]
\centering
\includegraphics[width=0.66\textwidth]{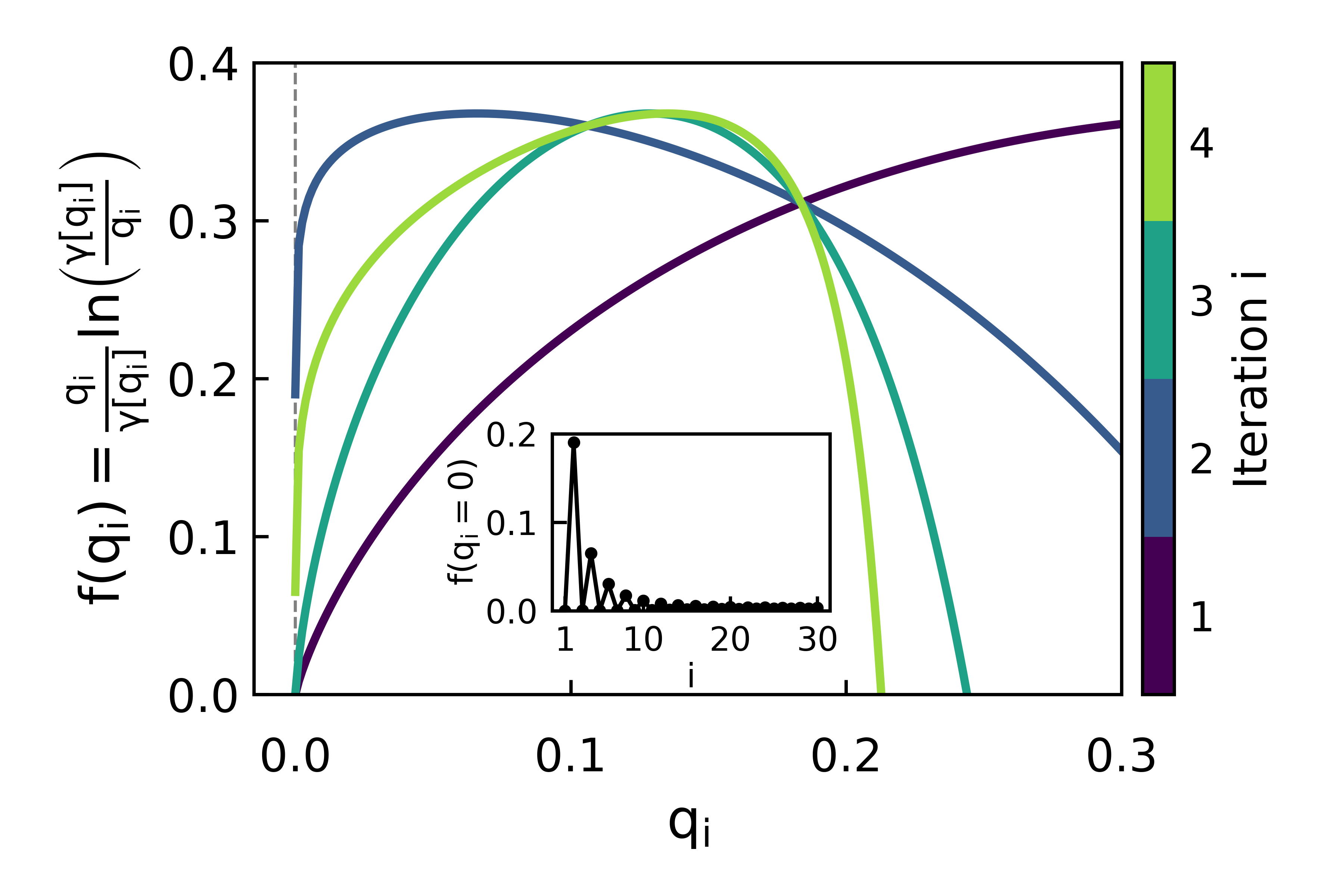}
\caption{Evolution of the first four iterations of the self-consistent cycle for the function $f(q_i)$ defining the analytical behavior of the scattering rate ($\tau_i(\boldsymbol q_i)^{-1}\sim f(q_i)$) associated with the LA (TA) $\rightarrow$ ZA + ZA scattering channel in the long-wavelength limit $q_i \to 0$. As the iterations proceed, the value at $q_i=0$ oscillates between zero and a finite value; however, this finite value decreases monotonically with increasing iteration number, ultimately tending to zero. The inset shows explicitly how $f(q_i=0)$ converges to zero as the number of iterations increases.}
\label{fig:analytic_scf_broadening}
\end{figure}

In this section, we derive the analytic long-wavelength limit of the scattering rate for in-plane acoustic modes in generic two-dimensional crystals. In the SM \cite{supplementary}, we revisit the analysis of Ref. \cite{bonini2012acoustic}, considering phonon–phonon scattering within the Dirac-delta approach of FGR. Here, we extend it by incorporating self-consistent Lorentzian collisional broadening. This approach provides an analytical resolution of the overdamped phonon lifetime problem associated with scattering processes involving out-of-plane flexural (ZA) modes. Numerical validation of this result is presented in Section \ref{result_section}.\\
We start from the three-phonon Normal processes of the type LA (TA) $\to$ ZA + ZA. The scattering rate for this process is given by
\begin{equation} \label{scattering_rate}
\begin{split}
\tau_{i}(\boldsymbol{q}_{i})^{-1}\propto\sum_{\boldsymbol{q}_{f}}F(\boldsymbol{q}_{i},\boldsymbol{q}_{f})\mathrm{d}\left(\Delta\omega(\boldsymbol{q}_{i},\boldsymbol{q}_{f})\right),
\end{split}
\end{equation}
where
\begin{equation} \label{delta_omega_initial_lor}
\Delta\omega(\boldsymbol{q}_{i},\boldsymbol{q}_{f})=\omega_{i}(\boldsymbol{q}_{i}) - \omega_{f}(\boldsymbol{q}_{f}) - \omega_{f}(\boldsymbol{q}_{i}+\boldsymbol{q}_{f})
\end{equation}
and
\begin{equation} \label{F_expression_lor}
F(\boldsymbol{q}_{i},\boldsymbol{q}_{f})=|\uppsi_{\boldsymbol{q}_{i}\boldsymbol{q}_{f}}|^{2}\big[n\big(\omega_{f}(\boldsymbol{q}_{f})\big) + n\big(\omega_{f}(\boldsymbol{q}_{i} + \boldsymbol{q}_{f})\big) + 1\big].
\end{equation}
For 2D materials in the long-wavelength limit ($\boldsymbol{q}_{i}\to0$) we assume
\begin{equation}
\begin{split}
&\omega_{i}(\boldsymbol{q})=A|\boldsymbol{q}|\\
&\omega_{f}(\boldsymbol{q})=B|\boldsymbol{q}|^{2}
\end{split}
\end{equation}
In the case of exact energy conservation, enforced by the Dirac delta function, there exists a value $q_{f}=q_{f,0}$ such that $\Delta\omega(q_{i},q_{f},\theta)=0$ \cite{supplementary}. The scattering rate in Eq. \eqref{scattering_rate} can be written as
\begin{equation} \label{polar_coordinate_int}
\tau_{i}(\boldsymbol{q}_{i})^{-1} = \int_{0}^{2\pi} d\theta \int_{0}^{\infty} dq_{f}\, q_{f}\, F(q_{i}, q_{f}, \theta) \, \mathrm{d}\big(\Delta\omega(q_{i}, q_{f}, \theta)\big),
\end{equation}
where $q_{f}$ appears due to the 2D Jacobian of the coordinate change. The Dirac delta identity
\begin{equation} \label{dirac_delta_identity}
\delta\!\big(\Delta\omega(q_{i},q_{f},\theta)\big)=\frac{\delta(q_{f}-q_{f,0})}{\big|\nabla_{q_{f}}\Delta\omega(q_{i},q_{f},\theta)\big|_{q_{f}=q_{f,0}}}
\end{equation}
does not have an exact counterpart for a Lorentzian distribution. Nevertheless $\Delta\omega(q_{i},q_{f},\theta)$ is smooth and assuming that it has a simple zero at $q_{f,0}$, one may linearize
\begin{equation} \label{Delta_omega_linearization}
\Delta\omega(q_{i},q_{f},\theta)\simeq\underbrace{\left.\Delta\omega(q_{i},q_{f},\theta)\right|_{q_{f}=q_{f,0}}}_{=0}+(q_f-q_{f,0})\left.
\nabla_{q_f}\Delta\omega(q_{i},q_{f},\theta)\right|_{q_{f}=q_{f,0}}+\cdots
\end{equation}
This yields the approximate identity (for compactness, the explicit dependence of $\Delta\omega(q_{i}, q_{f}, \theta)$ is omitted)
\begin{equation} \label{expandend_lorentzian}
\mathrm{d}\big(\Delta\omega\big)=\frac{1}{\pi}\frac{\gamma}{\Delta\omega^2+\gamma^2}\simeq\frac{1}{\pi}\frac{\gamma}{\left[(q_f-q_{f,0})\left.\nabla_{q_f}\Delta\omega\right|_{q_{f}=q_{f,0}}\right]^2+\gamma^2}=\frac{1}{\big.\nabla_{q_f}\Delta\omega\big|_{q_{f,0}}}\frac{1}{\pi}\frac{\frac{\gamma}{\big.\nabla_{q_f}\Delta\omega\big|_{q_{f,0}}}}{(q_f-q_{f,0})^2+\Big(\frac{\gamma}{\big.\nabla_{q_f}\Delta\omega\big|_{q_{f,0}}}\Big)^2}.
\end{equation}
The choice of a constant broadening parameter $\gamma$ is fully consistent with the hierarchy of ansätze underlying the present framework. Within the Lorentzian GKBA, the broadening is provided by the phonon linewidth evaluated at the quasiparticle ansatz, i.e., using a Dirac delta spectral function. As shown in the SM \cite{supplementary}, for the scattering channel considered here this quasiparticle linewidth approaches a constant in the long-wavelength limit \cite{bonini2012acoustic}. The use of a constant $\gamma$ in the present analytical treatment therefore follows directly from this self-consistent construction. In the limit $\gamma\to 0$, the Lorentzian reduces to the Dirac delta and the exact identity \eqref{dirac_delta_identity} is recovered. Substituting Eq. \eqref{expandend_lorentzian} into Eq. \eqref{polar_coordinate_int} yields
\begin{equation} \label{scattering_rate_2}
\tau_{i}(\boldsymbol{q}_{i})^{-1}=\int_{0}^{2\pi} d\theta \int_{0}^{\infty} dq_{f}\,q_{f} F(q_{i}, q_{f}, \theta)\frac{1}{\big.\nabla_{q_f}\Delta\omega\big|_{q_{f,0}}}\,\frac{1}{\pi}\frac{\frac{\gamma}{\big.\nabla_{q_f}\Delta\omega\big|_{q_{f,0}}}}{(q_f-q_{f,0})^2+\Big(\frac{\gamma}{\big.\nabla_{q_f}\Delta\omega\big|_{q_{f,0}}}\Big)^2}.
\end{equation}
Then, as in the case of Dirac delta spectral function, we have:
\begin{equation} \label{limit_q_f_zero}
q_{f,0}\sim\sqrt{q_{i}}\quad\text{for}\,q_{i}\to0,
\end{equation}
and 
\begin{equation} \label{limit_nabla_Delta_omega}
\left.\nabla_{q_{f}} \Delta\omega(q_{i},q_{f},\theta)\right|_{q_{f}=q_{f,0}}=\left. \big[-4Bq_{f}-2Bq_{i}\cos(\theta)\big]\right|_{q_{f}=q_{f,0}}\sim\sqrt{q_{i}}\quad\text{for}\,q_{i}\to0.
\end{equation}
So we are finally left with 
\begin{equation} \label{on_going_linewidth_q_i}
\tau_{i}(\boldsymbol{q}_{i})^{-1}\sim\int_{0}^{2\pi} d\theta\int_{0}^{\infty} dq_{f}\,q_{f}F(q_{i},q_{f,0},\theta)\frac{1}{\pi}\frac{\frac{\gamma}{\sqrt{q_{i}}}}{(q_f-\sqrt{q_{i}})^2+\left(\frac{\gamma}{\sqrt{q_{i}}}\right)^2}.
\end{equation}
We now analyze the behavior of $F(q_{i},q_{f},\theta)$ in the limit $q_{i}\to0$. This follows the same procedure as in the Dirac-delta case: considering the LA mode polarized along the $x$ direction, we obtain \cite{supplementary}:
\begin{equation} \label{V_final}
\uppsi_{\boldsymbol{q}_{i},\boldsymbol{q}_{f}}\propto\sum_{\substack{\bm{R}'\bm{R}''\\bb'b''\\ \alpha\beta\eta}}\uppsi_{b\alpha;\boldsymbol{q},b'\beta;\boldsymbol{q}',b''\eta;\boldsymbol{q}''}\frac{e^{i\boldsymbol{q}_{i}\cdot(\bm{r}_{b}-\bm{R}'')}e^{i\boldsymbol{q}_{f}\cdot(\bm{R}'-\bm{R}'')}}{\sqrt{q_{i}q_{f}^{2}|\boldsymbol{q}_{i}+\boldsymbol{q}_{f}|^{2}}}.
\end{equation}
Although in the present case the exact power-law dependence of $q_f$ on $q_i$ in the $q_i\to0$ limit is unknown, we can nevertheless consider the joint limit $q_f\to0$ as $q_i\to0$ and analyze the denominator of Eq. \eqref{V_final}:
\begin{equation} \label{understand_root}
\begin{split}
\sqrt{q_{i}q_{f}^{2}|\boldsymbol{q}_{i}+\boldsymbol{q}_{f}|^{2}}=\sqrt{q_{i}q_{f}^{2}(q_{i}^{2}+q_{f}^{2}+2q_{i}q_{f}\cos(\theta))}.
\end{split}
\end{equation}
The asymptotic behavior of this should be studied in the joint limit $q_i \to 0$, $q_f \to 0$, distinguishing two different scaling regimes: $q_i > q_f$ and $q_i < q_f$. This means that the dispersion relation \eqref{delta_omega_initial_lor} can be still assumed to be vanishing:
\begin{equation} \label{condition_disp_rel}
\Delta\omega(q_i,q_f)=A q_i-2B q_f^{2}-B q_i^{2}-2B\cos(\theta) q_i q_f \simeq 0,
\end{equation}
Then one has to analyze the solutions of Eq. \eqref{condition_disp_rel} in the joint small-momentum limit $q_i,q_f \to 0$, distinguishing the two possible hierarchies $q_i < q_f$ and $q_i > q_f$. Eq. \eqref{condition_disp_rel} can be further rewritten as a quadratic equation for $q_i$,
\begin{equation}
B q_i^{2} -\bigl(A-2B\cos(\theta) q_f\bigr) q_i +2B q_f^{2}=0,
\end{equation}
whose solutions are
\begin{equation} \label{qi_branches}
q_i(q_f)=\frac{A-2B\cos(\theta)\, q_f\pm\sqrt{\bigl(A-2B\cos(\theta)\, q_f\bigr)^{2}-8B^{2}q_f^{2}}}{2B}.
\end{equation}
The choice of branch is fixed by continuity in the $q_f \to 0$ limit:
the minus (plus) branch is the only solution that vanishes (remains finite)
as $q_f \to 0$, and is therefore consistent with the hierarchy
$q_i < q_f$ ($q_i > q_f$).
Expanding Eq. \eqref{qi_branches} for small  $q_f$, one finds
\begin{equation}
q_i(q_f)=\begin{cases}
q_i^{-}(q_f)=\frac{2B}{A}\,q_f^{2}+\mathcal O(q_f^{3}),\\
q_i^{+}(q_f)=\frac{A}{B}-2\cos(\theta)\, q_f+\mathcal O(q_f^{2}).
\end{cases}
\end{equation}
The minus branch satisfies  $q_i \sim q_f^{2} < q_f$ and approaches the origin  $(q_i,q_f)=(0,0)$, remaining consistent with the condition $\Delta\omega(q_i,q_f)\simeq 0$ in the joint limit  $q_i,q_f \to 0$. By contrast, the plus branch approaches a finite value $q_i \to A/B$ as  $q_f \to 0$ and therefore violates the requirement  $q_i \to 0$. Consequently, although mathematically admissible, the regime  $q_i > q_f$ is incompatible with the joint small-momentum limit and must be discarded. In the following, the long-wavelength limit $q_i,q_f\to0$ is thus understood as a joint limit taken with the constraint $q_i<q_f$. In this case we can factor out $q_f^{2}$ from the term in parentheses in Eq. \eqref{understand_root}:
\begin{equation} \label{psi_den_contr}
\begin{split}
\sqrt{q_{i}q_{f}^{2}|\boldsymbol{q}_{i}+\boldsymbol{q}_{f}|^{2}}&=\sqrt{q_{i}q_{f}^{2}q_f^{2}\left[1+2\left(\frac{q_i}{q_f}\right)\cos(\theta)+\left(\frac{q_i}{q_f}\right)^{2}\right]}\sim q_f^{2}q_i^{\frac{1}{2}}\quad q_i,q_f \to 0.
\end{split}
\end{equation}
Then, the exponentials in the numerator of Eq. \eqref{V_final} can be expanded for  as $q_i\to0$ and $q_f\to0$ respectively. We find that the terms proportional to $1$, $q_f$, $q_i$, $q_i q_f$, $q_f^2$, and $q_f^3$ vanish due to the same symmetry and acoustic sum-rule arguments that render them zero in the case of a Dirac-delta spectral function. The leading nonvanishing contributions are those quadratic in $q_i$ and proportional to $q_i q_f^2$.
\begin{equation} \label{leading_psi_num_contr}
\begin{split}
e^{iq_{i}|\bm{r}_{b}-\bm{R}''|\cos(\theta)}e^{iq_{f}|\bm{R}'-\bm{R}''|cos(\alpha)}\sim-\frac{1}{2}|\bm{r}_{b}-\bm{R}''|^{2}\cos^{2}(\theta)q_{i}^{2}-\frac{1}{2}i|\bm{r}_{b}-\bm{R}''||\bm{R}'-\bm{R}''|^{2}\cos(\theta)\cos^{2}(\alpha)q_{i}q_{f}^{2}.
\end{split}
\end{equation}
So, putting together Eqs. \eqref{psi_den_contr} and \eqref{leading_psi_num_contr} yields
\begin{equation} \label{V_final_limit}
\uppsi_{\boldsymbol{q}_{i},\boldsymbol{q}_{f}}\sim\frac{q_{i}^{2}+q_{i}q_{f}^{2}}{q_f^{2}q_i^{\frac{1}{2}}}=\frac{q_i^{3/2}}{q_f^{2}}+\sqrt{q_i}\sim\sqrt{q_i}\quad q_i,q_f \to 0.
\end{equation}
Indeed, for $q_i < q_f$ one has $q_i^{3/2}/q_f^{2}<\sqrt{q_i}$. Now we can evaluate the populations term in Eq. \eqref{F_expression_lor}. By using the simplify notation,
\begin{equation}
a \equiv \beta\hbar B,\qquad X \equiv q_i^2 + 2 q_i q_f \cos(\theta) + q_f^2,
\end{equation}
this reads
\begin{equation} 
\begin{split}
n\big(\omega_{f}(\boldsymbol{q}_{f})\big) + n\big(\omega_{f}(\boldsymbol{q}_{i} + \boldsymbol{q}_{f})\big)+ 1=\left(e^{a q_f^{2}}-1\right)^{-1}+\left(e^{a X}-1\right)^{-1}+1.
\end{split}
\end{equation}
Then we can exploit the small-$aq^{2}$ expansion:
\begin{equation}
\frac{1}{e^{a q^2}-1}=\frac{1}{a q^2}-\frac12+\mathcal O(a q^2),\qquad (a q^2< 1).
\end{equation}
In this way, in the regime of $q_i< q_f$, we get
\begin{equation} \label{populations_limit}
n\big(\omega_{f}(\boldsymbol{q}_{f})\big) + n\big(\omega_{f}(\boldsymbol{q}_{i} + \boldsymbol{q}_{f})\big)+ 1\sim\frac{1}{q_f^2}+\mathcal O(1)\qquad q_i,q_f\to 0,
\end{equation}
where $q_{f}^{-2}$ dominates over the other terms in the regime $q_i< q_f,\; q_i,q_f\to 0$.
So merging Eqs. \eqref{V_final_limit} and \eqref{populations_limit} we globally have
\begin{equation} \label{F_limit}
F(q_{i},q_{f,0})\sim\sqrt{q_i}\left[\frac{1}{q_f^2}+\mathcal O(1)\right]\sim\frac{\sqrt{q_i}}{q_f^{2}}\qquad q_i,q_f\to 0,
\end{equation}
where the $\mathcal O(1)$ contribution is finite in the limit $q_f \to 0,\;q_i< q_f$, whereas the first term diverges as $q_f^{-2}$ and, as a consequence, $q_f^{-2}>\mathcal O(1)$ for $q_f \to 0$. Substituting Eq. \eqref{F_limit} into the expression of the phonon linewidth \eqref{on_going_linewidth_q_i} finally yields
\begin{equation} \label{partial_tau_final_lorentz}
\tau_i(\boldsymbol q_i)^{-1}\sim2\pi\sqrt{q_i}\int_0^{\infty} dq_f\,W(q_f)L(q_f),\quad q_i,q_f\to 0,
\end{equation}
where the Lorentzian
\begin{equation}
L(q_f)=\frac{1}{\pi}\frac{\frac{\gamma}{\sqrt{q_i}}}{(q_f-\sqrt{q_i})^2+\left(\frac{\gamma}{\sqrt{q_i}}\right)^2}
\end{equation}
is peaked at $q_f\simeq\sqrt{q_i}$ and the integral weight is $W(q_f)=q_{f}^{-1}$. To analyze the integral we perform the substitutions $q_{f}=\frac{\gamma}{\sqrt{q_{i}}}s$, $dq_{f}=\frac{\gamma}{\sqrt{q_{i}}}ds$ and $c=\frac{q_{i}}{\gamma}$. In this way Eq. \eqref{partial_tau_final_lorentz} becomes:
\begin{equation}
\begin{split}
\tau_i(\boldsymbol q_i)^{-1}&\sim2\pi\sqrt{q_i}\int_0^{\infty} ds\,\frac{1}{s}\frac{1}{\pi}\frac{\frac{\gamma}{\sqrt{q_i}}}{\left(\frac{\gamma}{\sqrt{q_{i}}}s-\sqrt{q_i}\frac{\gamma}{\sqrt{q_i}}\frac{\sqrt{q_i}}{\gamma}\right)^2+\left(\frac{\gamma}{\sqrt{q_i}}\right)^2}=\\
&=2\pi\sqrt{q_i}\int_0^{\infty} ds\,\frac{1}{s}\frac{1}{\pi}\frac{\frac{\gamma}{\sqrt{q_i}}}{\left[\frac{\gamma}{\sqrt{q_{i}}}\left(s-\sqrt{q_i}\frac{\sqrt{q_i}}{\gamma}\right)\right]^2+\left(\frac{\gamma}{\sqrt{q_i}}\right)^2}=2\frac{q_{i}}{\gamma}\int_0^{\infty} ds\,\frac{1}{s\left[\left(s-\frac{q_i}{\gamma}\right)^2+1\right]}.
\end{split}
\end{equation}
The integral over $s$ goes as $\ln\big(\frac{\gamma}{q_{i}}\big)$ for $q_{i}\to0$, so we finally have
\begin{equation}
\tau_i(\boldsymbol q_i)^{-1}\sim2\frac{q_{i}}{\gamma}\ln\left(\frac{\gamma}{q_{i}}\right).
\end{equation}
At the first iteration, $\gamma$ is treated as a constant, since it corresponds to the result of the Dirac delta regime \cite{supplementary,bonini2012acoustic}. Given the first iteration (with iteration zero corresponding to the quasiparticle ansatz), we have
\begin{equation} \label{first_analytic_iteration}
\tau_i(\boldsymbol q_i)^{-1}\sim f_{0}(q_{i})=\frac{q_{i}}{\gamma}\ln\left(\frac{\gamma}{q_{i}}\right).
\end{equation}
Then, by proceeding iteratively in this way, $\gamma$ progressively acquires an explicit dependence on $q_i$. Using Eq. \eqref{first_analytic_iteration} as the mode-resolved broadening, i.e. setting $\gamma\to\gamma[q_i]=f_0(q_i)$ in the next iteration, we obtain
\begin{equation}
f_{1}(q_{i})=\frac{q_{i}}{\frac{q_{i}}{\gamma}\ln\left(\frac{\gamma}{q_{i}}\right)}\ln\left(\frac{\frac{q_{i}}{\gamma}\ln\left(\frac{\gamma}{q_{i}}\right)}{q_{i}}\right)=\frac{\gamma}{\ln\left(\frac{\gamma}{q_{i}}\right)}\ln\left(\frac{1}{\gamma}\ln\left(\frac{\gamma}{q_{i}}\right)\right),
\end{equation}
and repeating the procedure once more yields
\begin{equation}
f_{2}(q_{i})=\frac{\frac{q_{i}}{\gamma}\ln\left(\frac{\gamma}{q_{i}}\right)}{\ln\left(\frac{1}{\gamma}\ln\left(\frac{\gamma}{q_{i}}\right)\right)}\ln\left(\frac{1}{\frac{q_{i}}{\gamma}\ln\left(\frac{\gamma}{q_{i}}\right)}\ln\left(\frac{1}{\gamma}\ln\left(\frac{\gamma}{q_{i}}\right)\right)\right)
\end{equation}
and so on. This recursive procedure generates increasingly nested logarithmic corrections to the small-$q_i$ behavior of the scattering rate. To understand how this approach effectively resolves the problem of overdamped phonon lifetimes, it is useful to examine graphically the analytical behavior of the scattering rate obtained from the first few iterations shown in Fig. \ref{fig:analytic_scf_broadening}.
We observe that, as the iterations proceed, the long-wavelength limit of the scattering rate at $q_i=0$ oscillates between zero and a finite value. Importantly, this finite value becomes progressively smaller with each iteration, ultimately vanishing in the self-consistent limit. This demonstrates that the self-consistent collisional broadening corrects the overdamping of the lifetime associated with the LA (TA) $\rightarrow$ ZA + ZA scattering process in two-dimensional materials. It is worth emphasizing that, as shown in Fig. \ref{fig:analytic_scf_broadening}, already at the first iteration one obtains $\tau_i(\boldsymbol q_i)^{-1} \to 0$ as $q_i \to 0$. However, a single iteration may be numerically unstable if the chosen Lorentzian broadening is very sharp and approaches a Dirac delta distribution. The fully self-consistent procedure, which is physically motivated, ensures stability and guarantees the absence of overdamping at convergence.\\
This demonstrates that the inclusion of a Lorentzian collisional broadening regularizes the scattering channel involving ZA acoustic modes: the corresponding phonon lifetimes are no longer overdamped and, as physically expected, the associated linewidth tends to zero in the long-wavelength limit $q_i \to 0$. In particular, taking the small-broadening limit restores the Dirac delta function behavior of the Lorentzian. In this case, the integral in Eq. \eqref{partial_tau_final_lorentz} has to be evaluated exactly at $q_{f}=q_{f,0}\sim\sqrt{q_{i}}$ and so it reduces to
\begin{equation}
\begin{split}
\tau_{i}(\boldsymbol{q}_{i})^{-1}&\sim2\pi\sqrt{q_i}\int dq_{f}\,\frac{1}{q_f}\,\delta(q_f-\sqrt{q_{i}})\sim2\pi\sqrt{q_i}\,\frac{1}{\sqrt{q_{i}}}=\text{const.},\quad q_i,q_f \to 0,
\end{split}
\end{equation}
yielding a finite linewidth in the long-wavelength limit and thus allowing for overdamped phonon lifetimes. Such limiting behavior is consistent with Eq. \eqref{bonini_res} of the main text, which reproduces the result originally reported in Ref. \cite{bonini2012acoustic} and revised in the SM \cite{supplementary}.

\bibliographystyle{apsrev4-1}
%\bibliography{bib}

%

\section*{Supplemental Material}

\section{Computational details and supporting data} \label{computational_details}

\begin{figure}[!htb]
\centering
\includegraphics[width=0.62\textwidth]{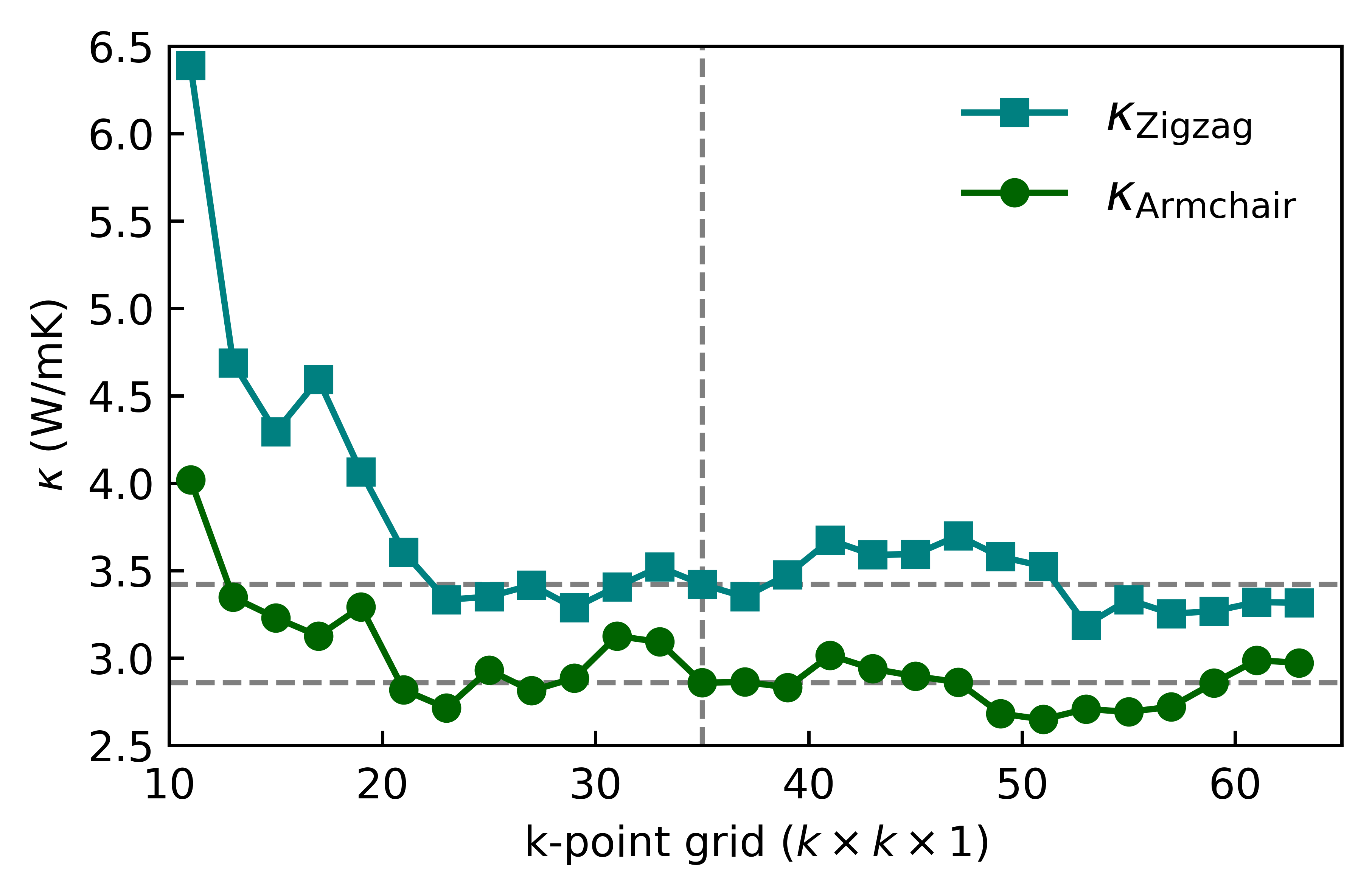}
\caption{Convergence of the zigzag (teal) and armchair (green) components of the room-temperature lattice thermal conductivity of the $\alpha$-GeSe monolayer with respect to the $k$-point grid used to solve the LEBTE. The Lorentzian numerical smearing is fixed at 0.1 cm$^{-1}$. The vertical dashed line marks the grid adopted in the simulations.}
\label{fig:k_grid_convergence}
\end{figure}

\begin{figure}[!htb]
\centering
\includegraphics[width=0.66\textwidth]{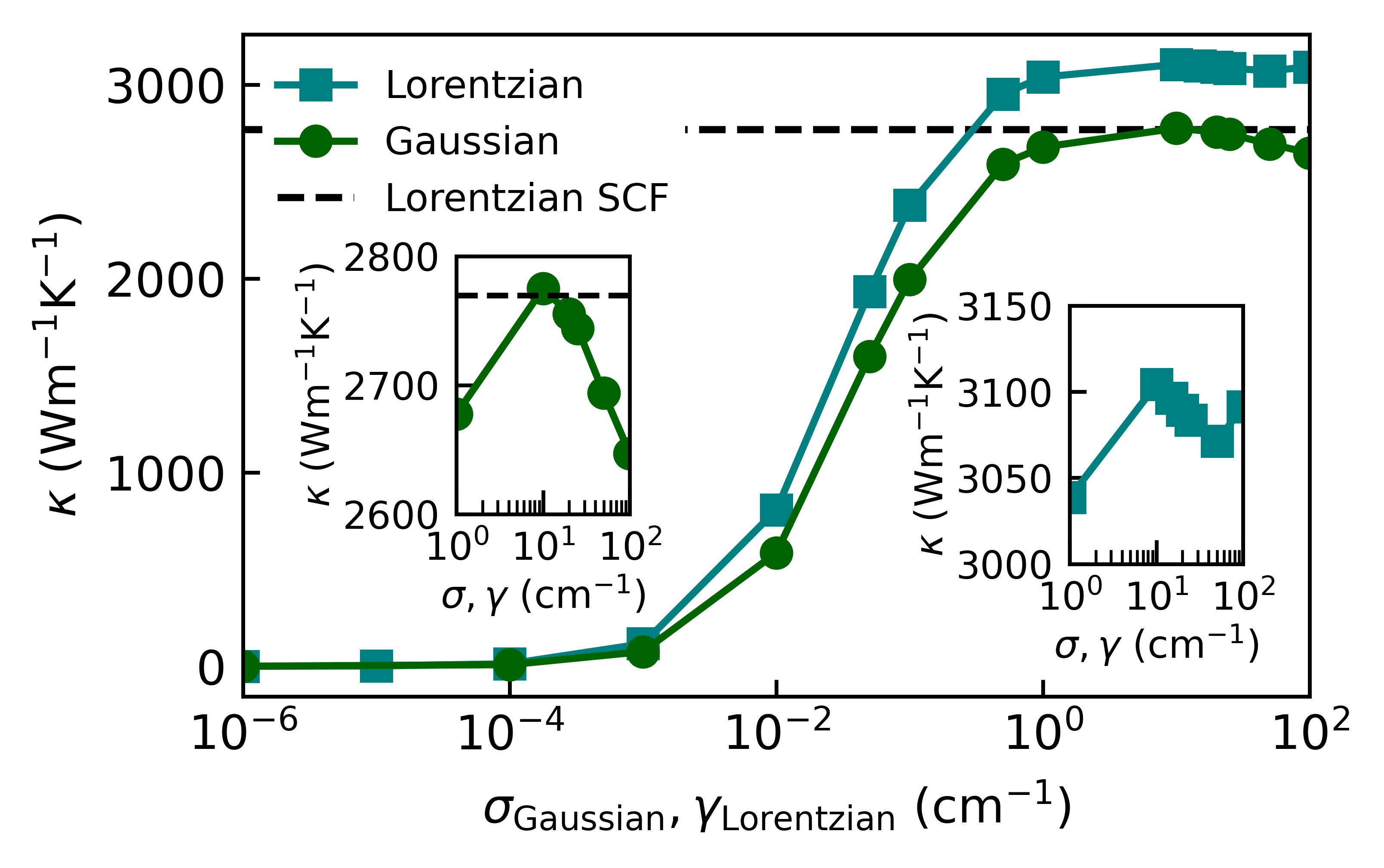}
\caption{Room-temperature lattice thermal conductivity of bulk diamond as a function of the smearing parameter for Gaussian (green) and Lorentzian (teal) numerical broadening schemes. This figure extends panel (b) of Fig. \ref{fig:kappa_vs_smearing} by including larger smearing values up to $10^{2}$ cm$^{-1}$. Convergence with respect to the smearing parameter is not achieved over the entire explored range. In particular, after reaching a maximum around $\sim$10 cm$^{-1}$, the conductivity decreases for larger smearing values. This qualitative behavior is consistent with the trend reported in Ref. \cite{fugallo2013ab} (see Fig. 6 therein), which refers to a temperature of 100 K. The result obtained with the present Lorentzian self-consistent broadening (SCF) approach is shown as a black dashed line and is independent of the smearing parameter.}
\label{fig:kappa_vs_smearing_diamond}
\end{figure}

\begin{figure}[!htb]
\centering
\includegraphics[width=0.66\textwidth]{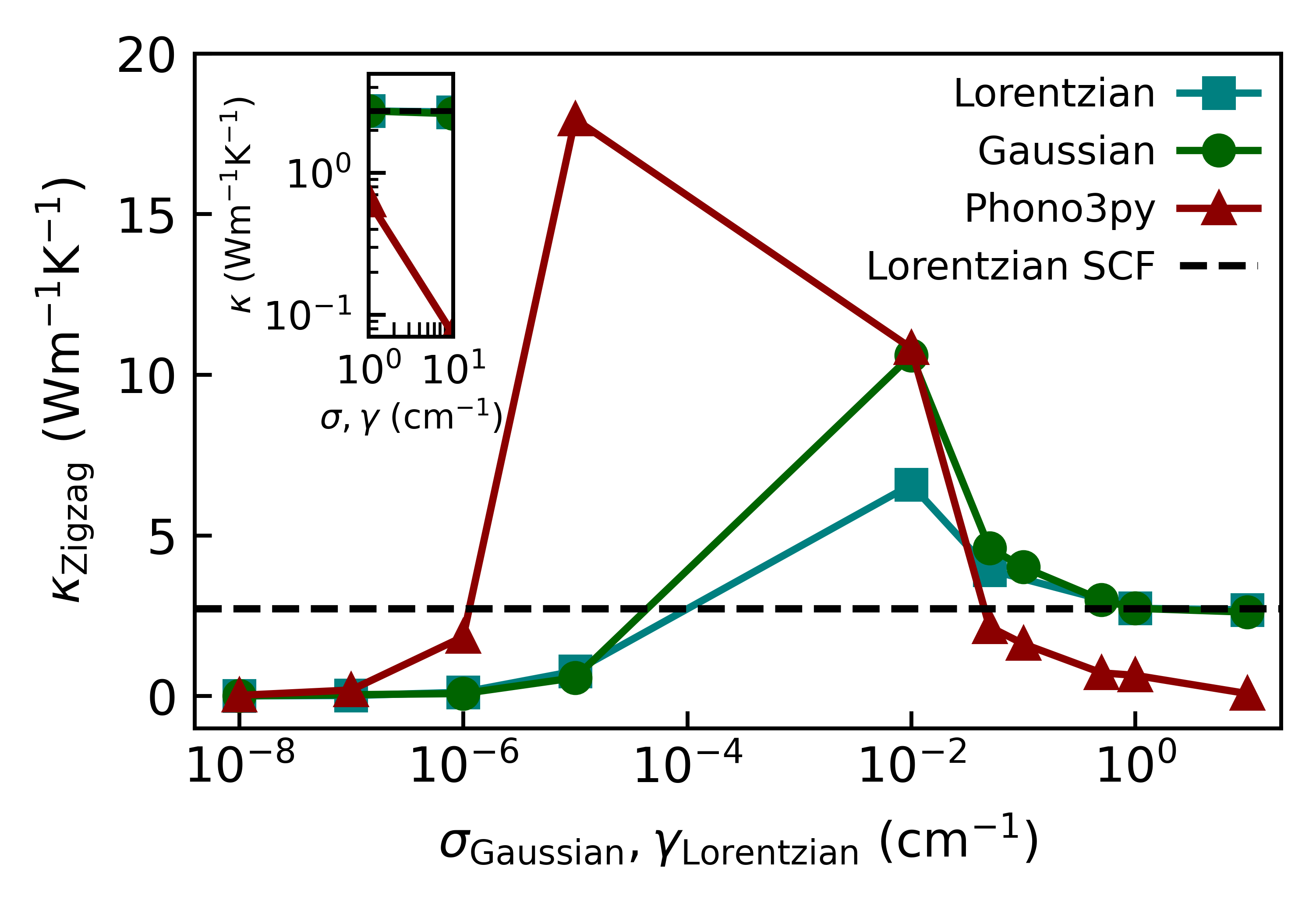}
\caption{Room-temperature in-plane lattice thermal conductivity along the zigzag direction of monolayer $\alpha$-GeSe as a function of the smearing parameter. Results are shown for Gaussian (green) and Lorentzian (teal) numerical broadening schemes obtained using our in-house implementation of the \texttt{D3Q} code, together with Gaussian smearing results from \texttt{Phono3py} (bordeaux). The inset shows a zoom of the 1–10 cm$^{-1}$ range, highlighting that the \texttt{Phono3py} calculations do not exhibit convergence in this commonly adopted smearing interval.}
\label{fig:kappa_vs_smearing_vs_phono3py}
\end{figure}

The equilibrium crystal configuration of $\alpha$-GeSe monolayer is obtained using the \texttt{Quantum ESPRESSO} distribution \cite{quantum_espresso} through Kohn-Sham density-functional theory (DFT) \cite{DFT_1,DFT_2} calculations with ultrasoft pseudopotentials from the SSSP library \cite{SSSP,SSSP2,SSSP3}. The Perdew-Burke-Emzerhof approximation for solids (PBEsol) is employed for the exchange-correlation functional \cite{perdew2008restoring}. Convergence of total energy, forces, and pressure with respect to $k$-mesh and kinetic energy cutoff is achieved using a $11\times11\times1$ $k$-points grid, and 40\,Ry plane wave cutoff respectively.
\begin{figure}[!htb]
\centering
\includegraphics[width=0.66\textwidth]{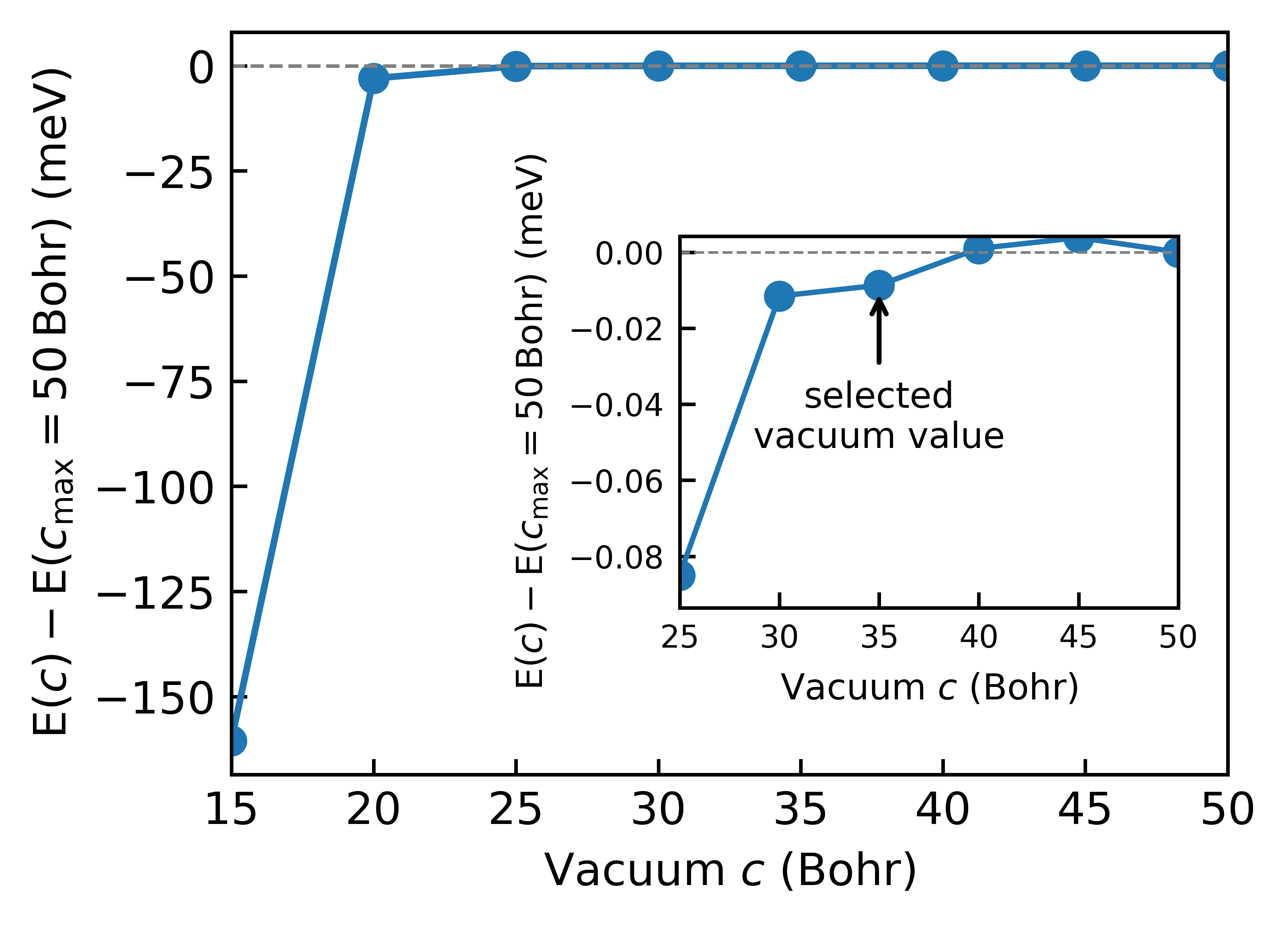}
\caption{Energy difference with respect to the reference calculation performed using the largest vacuum spacing along the $z$ direction, quantified by the lattice parameter $c$ of the monolayer $\alpha$-GeSe cell with vacuum. The energy of the same structure is reported as a function of the vacuum parameter $c$ in the range 15--50 Bohr. Inset: zoom of the 25--50 Bohr region. As shown, the total energy converges rapidly with increasing vacuum spacing; all results presented in this work were obtained using $c = 35$ Bohr.}
\label{fig:E_vs_vacuum_c}
\end{figure}
To model the material as an isolated two-dimensional system, we carefully examined the convergence of the total energy with respect to the vacuum thickness along the out-of-plane direction. In Fig. \ref{fig:E_vs_vacuum_c}, we plot the energy difference between the system with the largest considered cell thickness (50 Bohr) and that obtained for several smaller values of the lattice parameter $c$. As shown, the total energy is effectively converged already for vacuum thicknesses of approximately 30 Bohr. In the same figure, we indicate the vacuum thickness adopted in the present calculations, namely 35 Bohr. The optimized lattice parameters of the $\alpha$-GeSe monolayer are $a = 3.941$ \AA and $b = 4.064$ \AA. The optimized orthorhombic bulk GeSe structure was also evaluated, yielding lattice parameters $a_{\mathrm{bulk}} = 3.852$ \AA, $b_{\mathrm{bulk}} = 4.262$ \AA, and $c_{\mathrm{bulk}} = 10.693$ \AA.\\
Phonon properties are computed within density-functional perturbation theory (DFPT) using an $8\times8\times1$ $k$-point grid.
\begin{figure}[!htb]
\centering
\includegraphics[width=0.66\textwidth]{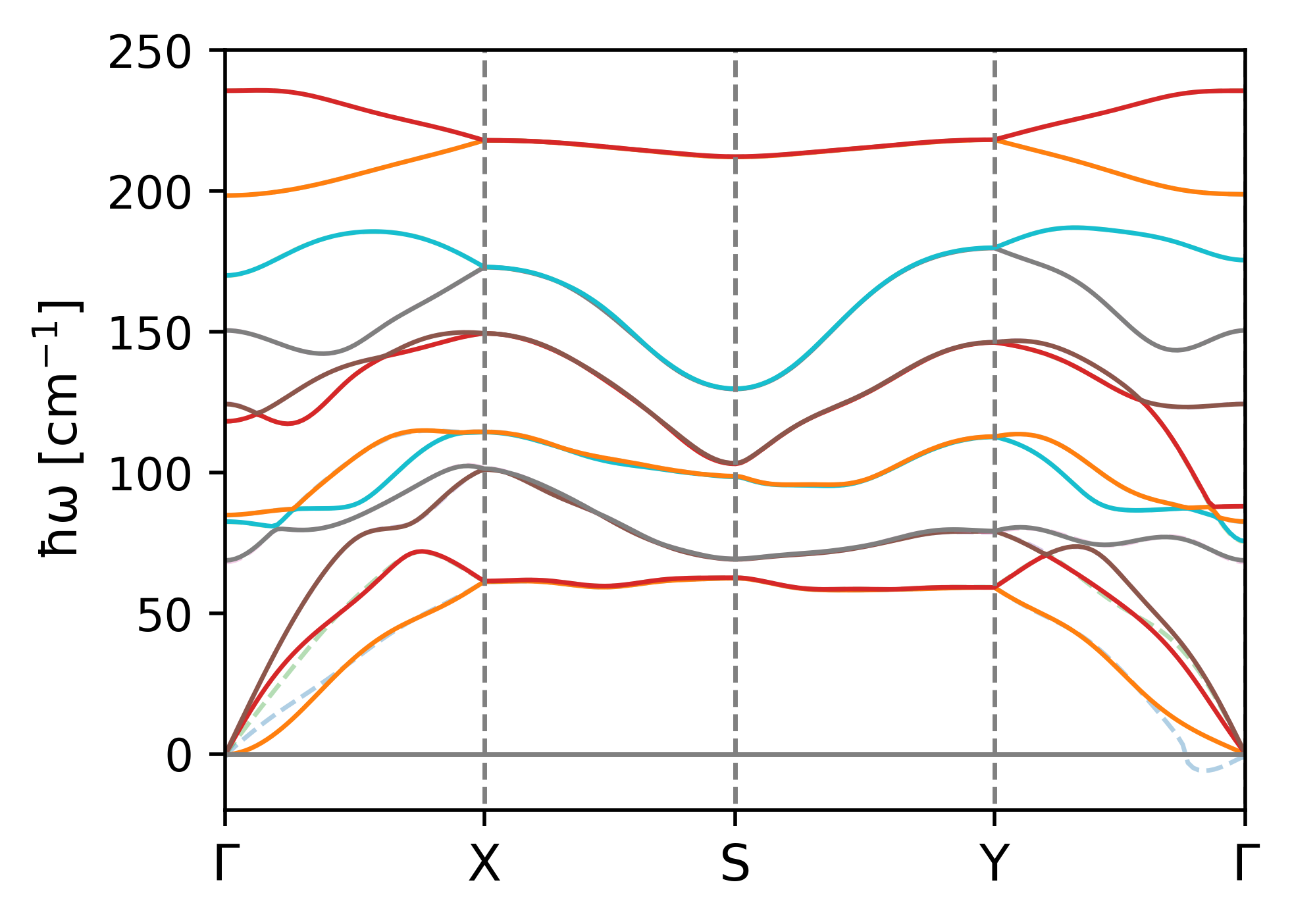}
\caption{Phonon dispersion of monolayer $\alpha$-GeSe calculated along the high-symmetry directions of the Brillouin zone. The dashed shaded curves represent the dispersion prior to enforcing the three rotational sum rules, whereas the solid lines correspond to the corrected dispersion obtained by imposing the 15 Huang conditions, which ensure a vanishing stress tensor through optimized correction of the force constants, as described in Ref. \cite{lin2022general}.}
\label{fig:phonons_GeSe}
\end{figure}
The dielectric tensor and the Born effective charge tensors are included in the calculations in order to account for long-range electrostatic interactions. In order to correctly describe the flexural (out-of-plane) acoustic mode, which exhibits a quadratic dispersion near the $\Gamma$ point of the Brillouin zone, rotational sum rules are imposed through the 15 Born-Huang conditions \cite{begbie1947thermal,born1996dynamical,huang1951lattice}. These conditions enforce the vanishing of the residual stress tensor by applying an optimized correction to the interatomic force constants, as described in Ref. \cite{lin2022general}. A comparison of the phonon dispersion obtained with and without the enforcement of rotational sum rules is shown in Fig. \ref{fig:phonons_GeSe}. \\
The LEBTE was solved using the variational method \cite{fugallo2013ab} using an in-house version of the \texttt{D3Q} code \cite{paulatto2013anharmonic,paulatto2015first} as done in Ref. \cite{simoncelli2020generalization}
To express the lattice thermal conductivity of the 2D system in W m$^{-1}$ K$^{-1}$ and enable comparison with 3D materials, the value obtained from simulations including vacuum was rescaled. Because the thermal conductivity expression is normalized by the simulation cell volume—while a strictly 2D system would require normalization by surface area—we define an effective volume. This volume is taken as the in-plane unit cell area multiplied by half of the bulk interlayer spacing, $c_{\mathrm{bulk}}/2$. Accordingly, the monolayer thickness is chosen as half the out-of-plane lattice parameter of the layered bulk counterpart, whose conventional cell contains two layers \cite{qin2016diverse}. The convergence with respect to the $k$-point sampling in the solution of the LEBTE is shown in Fig. \ref{fig:k_grid_convergence}.
For diamond, we adopt the computational parameters reported in Ref. \cite{simoncelli2020generalization}, including a $27 \times 27 \times 27$ $k$-point grid for solving the LEBTE, as in our previous work. In Fig. \ref{fig:kappa_vs_smearing_diamond}, we reproduce the same analysis shown in panel (b) of Fig. \ref{fig:kappa_vs_smearing}, but extending the $x$-axis to a numerical smearing value of 100 cm$^{-1}$ for the room-temperature lattice thermal conductivity of bulk diamond. The results show that convergence is not achieved even at such large smearing values—permissible in bulk diamond due to its maximum phonon frequency of approximately 1400 cm$^{-1}$—thereby confirming that a self-consistent collisional broadening is required.
Finally, in Fig. \ref{fig:kappa_vs_smearing_vs_phono3py} we present the analogue of panel (a) of Fig. \ref{fig:kappa_vs_smearing} for monolayer $\alpha$-GeSe, including a comparison with the lattice thermal conductivity along the zigzag direction obtained from simulations performed with \texttt{Phono3py} \cite{phono3py} at different smearing values. As can be seen, \texttt{Phono3py} also exhibits a pronounced lack of convergence with respect to the smearing parameter, even more marked than in our implementation. This behavior is accompanied, at small smearing values, by strongly overdamped phonon modes.

\end{document}